\documentclass[]{jfm}

\usepackage{lmodern}
\usepackage[T1]{fontenc}
\usepackage{newtxmath, bm, natbib, comment, graphicx, subfigure}

\usepackage{hyperref}
\hypersetup{
    colorlinks = true,
    citecolor  = blue,
}

\newcommand{\RomanNumeralCaps}[1]
\linenumbers

\renewcommand{\vec}[1]{\bm{#1}}
\newcommand{\e}{{\mathrm e}}
\renewcommand{\i}{{\mathrm i}}

\definecolor{orange}{rgb}{1,0.3,0}
\definecolor{darkgreen}{rgb}{0,0.5,0}

\usepackage{graphicx,epstopdf}
\DeclareGraphicsExtensions{.ps}
\title{Large-scale multifractality and lack of self-similar decay for Burgers and 3D Navier-Stokes turbulence}

\author{
    Takeshi Matsumoto\aff{1},
    Dipankar Roy\aff{2},
    Konstantin Khanin\aff{3},
    Rahul Pandit\aff{4},
    \and 
    Uriel Frisch\aff{5}
     \corresp{\email{uriel@oca.eu}}
}

\affiliation{
    \aff{1}Division of Physics and Astronomy, Kyoto University, Kyoto, Japan
    \aff{2}Universit\'{e} C\^{o}te d’Azur, CNRS UMR 7351, Nice, France
    \aff{3} BIMSA, Beijing, China; University of Toronto, Toronto, Canada
    \aff{4}Department of Physics, Indian Institute of Science, Bangalore, India
    \aff{5} Observatoire de la  C\^{o}te d’Azur, CNRS Laboratoire J.-L. Lagrange, 06300 Nice, France
}

\begin{document}
\maketitle

\begin{abstract}

We study decaying turbulence in the 1D Burgers equation
(Burgulence) and 3D Navier-Stokes (NS) turbulence.
We first investigate the decay in time $t$ of the energy $E(t)$
in Burgulence, for a fractional Brownian initial potential, with Hurst
exponent $H$, and demonstrate rigorously a self-similar time-decay of $E(t)$, previously determined heuristically. This is a consequence
of the nontrivial boundedness of the energy for any positive
time. We define a spatially forgetful
\textit{oblivious fractional Brownian motion} (OFBM), with Hurst exponent $H$,
and prove that Burgulence, with an OFBM as initial potential
$\varphi_0(x)$, is not only intermittent, but it also displays, a
hitherto unanticipated, large-scale bifractality or multifractality;
the latter occurs if we combine OFBMs, with different values of $H$.
This is the first rigorous proof of genuine multifractality
for turbulence in a nonlinear hydrodynamical partial differential
equation.
We then present direct numerical simulations (DNSs) of freely decaying
turbulence, capturing some aspects of this multifractality. For
Burgulence, we investigate such decay for two cases: (A) $\varphi_0(x)$ a
multifractal random walk that crosses over to a fractional Brownian
motion beyond a crossover scale $\mathcal{L}$, tuned to
go from small- to large-scale multifractality; (B) initial energy
spectra $E_0(k)$, with wavenumber $k$, having one or more power-law
regions, which lead, respectively, to self-similar and
non-self-similar energy decay. Our analogous DNSs of the 3D NS equations also uncover self-similar and non-self-similar energy decay. Challenges confronting the detection of genuine large-scale multifractality, in numerical and experimental studies of NS and MHD turbulence, are highlighted. 
\end{abstract}

\begin{keywords}
Decaying turbulence. 
\end{keywords}

{\bf MSC Codes }  {\it(Optional)} Please enter your MSC Codes here

\tableofcontents

\section{Introduction}\label{sec:intro}

The decay of homogeneous, isotropic fluid turbulence is a problem of fundamental significance in fluid dynamics. Not surprisingly, studies of this problem have a long history, which we outline below.
The large-scale dynamics of three-dimensional (3D) fluid turbulence has been investigated since Leonardo da Vinci (1505), who wondered why vortices (which he called \textit{turbulences}), generated at the pillars of a bridge in the Arno river in Florence, tended to endure for a long time. The exact text of Leonardo's three lines on hydrodynamics, in his \textit{Codex Atlanticus}, can be found in \cite{1995-frisch-book} on p. 112. The date of publication, within the more than one thousand pages of \textit{Codex Atlanticus}, was at first set in the late 16$^{\rm{th}}$ century by Pompeo Leoni. He was no specialist of the evolution of Leonardo's hand-writing and thus wrongly positioned Leonardo's text on ``turbulences'' to around 1470s, during Leonardo's roughly 30-year initial stay in Florence. But Leonardo was not really interested in hydrodynamics at that time.  During his second stay, in the early 1500s, Leonardo had become strongly interested in hydrodynamics and seriously considered advanced fluvial engineering, to divert the Arno's path. The (probably correct) dating, 1505, was
made recently by Augusto Marinoni  and may be found in his book ``Il codice Atlantico di Leonardo da Vinci. Indici per materie e alfabetico'' (Giunti
Editore, Milan, 2017) [see ~\cite{marinoni2004codice}].

More than four centuries later, \cite{1938-karman-howarth} and \cite{1941-kolmogorov-b} speculated that, in the limit of vanishing viscosity and in the absence of forces, the energy of 3D incompressible turbulence would decline, at long time $t$, as the power-law $E(t) \propto t^{-n} $.  Following the work of \cite{1938-karman-howarth}\footnote{de K\'arm\'an in the original}  on velocity correlations, \cite{1941-kolmogorov-b}  [for an English translation see \cite{sinai2003russian}, pages 332-336] arrived at a decay law $E(t) \propto t^{-10/7} $ using the Loitsiansky invariant [see \cite{1939-loitsiansky}]. However, in 1954 \cite{1954-proudman-reid} discovered that the Loitsiansky invariant is typically infinite, thereby bringing Kolmogorov's result into question. A subsequent direct numerical simulation (DNS) by \cite{ishida2006decay}  was designed to investigate freely decaying fluid turbulence with an initial energy spectrum $\sim k^4$; over the duration of their DNS, \cite{ishida2006decay} found 
that the Loitsiansky invariant remained approximately constant and the decay of the energy could possibly be consistent with the Kolmogorov result $E(t) \sim t^{-10/7}$. A subsequent DNS by \cite{davidson2012freely} investigated such decay  with an initial energy spectrum $\sim k^2$ and obtained results consistent with the Saffman suggestion $E(t) \sim t^{-6/5}$ for turbulence with a Saffmann invariant\footnote{Some authors prefer to call this the Birkhoff-Saffman invariant [see \cite{panickacheril2022laws}].} [see \cite{birkhoff1954fourier,saffman1967large}]. 
The free decay of 3D NS turbulence has been studied with other types of initial data also; e.g., \cite{biferale2003decay} have examined such decay with initial velocity fields taken from a simulation of forced, statistically steady turbulence; and \cite{krstulovic2024initial} have studied 
the initial evolution of 3D NS turbulence as it evolves towards a spectrum \textit{\`a la} Kolmogorov.
A recent overview and results from high-resolution DNSs is contained in \cite{panickacheril2022laws}. Experimental studies of such energy decay have a long history [see, e.g., \cite{batchelor1947decay,comte-bellot_use_1966,2012-meldi-sagaut,panickacheril2022laws}]; 
decay data from such experiments are often fit to the form $E(t) \sim t^{-n}$, but \cite{2012-meldi-sagaut} and \cite{panickacheril2022laws} note that the values reported for the exponent $n$ are spread over a wide range $\simeq 1.4 -2$. \cite{2012-meldi-sagaut} use the eddy-damped-quasi-normal-Markovian (EDQNM) closure to suggest that this range of exponents may be understood because of non-self-similar energy decay that occurs if the initial energy spectrum has three power-law regions; \cite{eyink2000free} also employ the EDQNM to discuss non-self-similar energy decay of 
the type uncovered by \cite{1997-gurbatov--toth} in the context of the 1D Burgers equation (see below).

On theoretical grounds, it is often said that the self-similar power-law decay of the energy arises from \textit{the principle of the permanence of large eddies} [see Section 7.8 in \cite{1995-frisch-book}]. This principle builds upon results of~\cite{1954-proudman-reid}, \cite{1978-tatsumi--mizushima}, and \cite{1980-frisch--schertzer}, which show that the beating interaction of two nearly opposite wavenumbers $k$, whose absolute values are near the integral-scale wavenumber $K(t)= L^{-1}(t)$, contributes to low-wavenumber dynamics and a (transfer) input $T(k) \propto k ^4$ (in dimension $d=3$). As a consequence, if the low-wavenumber initial data have an energy spectrum $E(k, t=0) \sim k^n$, with $n > 4$, then, for wavenumbers $k \ll K(t)$ and to leading order, the beating interaction leads to $E(k,t) \sim k^4$ and thence a power-law decay of $E(t)$.

Starting in the late 1970s, energy decay was studied in the context of \textit{Burgulence}, i.e., for random solutions to the Burgers equation arising from randomness in the initial conditions [see, e.g., \cite{1979-kida,1992-she--frisch,1997-gurbatov--toth,frisch2002burgulence}]. Such studies shed light on the principle of the permanence of large eddies. In particular, \cite{1979-kida} and \cite{1997-gurbatov--toth} showed that, for the one-dimensional (1D) Burgers equation, initial data with single-power-law energy spectra lead to energy decay with an inverse power of time $t$, sometimes modified by a logarithmic prefactor [see \cite{1997-gurbatov--toth}], which is now referred to as the Gurbatov phenomenon. 

Given this historical background, we decided, at first, to focus mostly on {\it self-similar} and {\it non-self-similar} decay of the energy $E(t)$. This was done here for:
\begin{itemize}
\item the 1D Burgers equation;
\item and the 3D viscous and hyperviscous Navier--Stokes equations.
\end{itemize}
In the process, we discovered a novel type of \textit{large-scale multifractality}, which is intimately connected to the lack
of self-similar decay.

We carry out two types of studies of freely decaying 1D Burgulence: in the first type, we specify initial data in physical space, via the initial potential  $\varphi_0(x)$, which is related to the velocity by $u_0(x) = -  \partial_{x} \varphi_0(x)$; in the second type, we start with an initial energy spectrum $E_0(k)$, which is chosen to have one or more power-law regions as a function of the wavenumber $k$.
For the first type of initial data, we obtain both rigorous and numerical results for the decay of the total energy $E(t)$; these studies are designed to explore signatures of large-scale multifractality and the crossover from small-scale to large-scale multifractality. With the second type of initial data we quantify, via direct numerical simulations, non-self-similar decay of $E(t)$, the associated temporal evolution of the energy spectrum $E(k,t)$, and the Gurbatov phenomenon [see ~\cite{1997-gurbatov--toth}] for initial energy spectra $E_0(k)$ that have more than one power-law region.

We perform two types of studies of freely decaying 3D NS turbulence: in the first type, we use the viscous 3D NS equation; in the second type, we use the hyperviscous 3D NS equation,
because this allows us to carry out long direct numerical simulations (DNSs) with enough spatial resolution to examine the temporal evolution of the energy spectrum $E(k,t)$. In both these types of DNSs, 
we start with an initial energy spectrum $E_0(k)$, which is chosen to have one or more power-law regions as a function of the wavenumber $k$. With two power-law regimes, we obtain non-self-similar decay of $E(t)$ %
and, with certain initial-power-law exponents, the 3D NS counterpart of the Gurbatov phenomenon [see~\cite{1997-gurbatov--toth} and \cite{frisch2002burgulence}]. 

The remaining part of this paper is organised as follows. In Section~\ref{sec:burg-model} we discuss the 1D Burgers equation and the rigorous results that we obtain for freely decaying 1D Burgulence. Section~\ref{sec:burg-num} contains the results of our direct numerical simulations (DNSs) for such decay in the 1D Burgers equation with different types of initial potentials $\varphi_0(x)$ or different initial energy spectra $E_0(k)$. In Section~\ref{sec:ns-hyper-vis} we generalise these DNSs to studies of freely decaying turbulence in the three-dimensional (3D) viscous and hyperviscous Navier-Stokes equations. Section~\ref{sec:conc} is devoted to a discussion of the theoretical and experimental implications of our work. Technical details, both mathematical and numerical, are dicussed in Appendices~\ref{sec:AppA} - \ref{sec:app_ns}.

\section{The Burgers PDE in 1D: models and methods} 
\label{sec:burg-model}

The one-dimensional Burgers PDE without any forcing is given by
\begin{equation}
    \partial_{t} u(x,t) + u(x,t) \partial_{x} u(x,t) = \nu \partial_{xx} u(x,t)\,,   
    \label{eq:burg}
\end{equation}
where $\nu>0$ is the kinematic viscosity. The initial velocity is denoted by $u_{0}(x)$. It is frequently convenient to work with the potential $\varphi(x,t)$, related to the velocity by
\begin{equation}
    u(x,t) = -  \partial_{x} \varphi(x,t)\,.
    \label{eq:u-pot}
\end{equation}
The potential satisfies
\begin{equation}
    \partial_{t} \varphi(x,t) = \frac{1}{2} \left( \partial_{x} \varphi(x,t) \right)^2  + \nu \partial_{xx} \varphi(x,t)\,.
    \label{eq:burg-pot}
\end{equation}
As is well known, the Burgers equation~(\ref{eq:burg}) can be mapped into the heat equation by a nonlinear transformation, introduced by \cite{1948-hopf, 1950-hopf} and \cite{1951-cole}. One consequence, strongly emphasized by~\cite{1974-burgers-book}, is that the zero-viscosity limit of the potential has a very simple explicit representation in terms of the initial potential $\varphi_0(x)$:
\begin{equation}
    \varphi(x,t)  = \max_a\left(\varphi_0(a) - \frac{(x -a)^2}{2t}\right)\,,
    \label{eq:max}
\end{equation}
where $\underset{a}{\max}$ is the maximum over all initial fluid particle positions $a$. We shall refer to Eq.~\eqref{eq:max} as the ``max formula'', which is essentially a Legendre transform. Indeed, $\varphi(xt,t) + (xt)^2/(2t)$ is a Legendre transform of $\psi_0(a) := \varphi_0(a) - a^2/(2t)$ [see, e.g., \cite{1992-she--frisch}];
numerically, we can move from the initial data to the solution at any time $t>0$, directly, without having to consider any intermediate times. This Legendre transform has an implementation whose spatial complexity is $\mathcal{O}(N \ln N)$, where $N$ is the number of equally spaced collocation points [see, e.g., \cite{1992-she--frisch}].

There is an important difference between the Burgers equation and the Navier-Stokes equation: The unforced Burgers equation has no mechanism allowing for the generation of stochastic solutions unless the initial conditions are random [see, e.g., \cite{1992-she--frisch} and \cite{1997-gurbatov--toth}]. Given that the Burgers equation is translationally invariant, we are particularly interested in stochastic solutions whose statistical properties have \textit{translational invariance} (i.e., are homogeneous) or have \textit{translationally invariant increments} (i.e., whose space derivatives are homogeneous). 

\subsection{Energy decay with a fractional Brownian initial potential}
\label{sec:EdecayFBM}

\label{subsec:brownian-motion}

We work with the initial potential $\varphi_0(a)$ that we take to be a fractional Brownian
motion with a Hurst exponent $0<H<1$.
This means that the initial potential $\varphi_0(a)$ is Gaussian, it vanishes at the origin,  and its
second-order structure function is given by 
\begin{equation}
  \mathcal{S}_0(b) \equiv \langle (\varphi_0(a+b) -\varphi_0(a))^2 \rangle_a = C |b|^{2H}\,,
  \label{eq:2nd-structure-function-scales_A}
\end{equation}
where the subscript $a$ denotes an average\footnote{The only source of randomness in the setting of decaying Burgulence is provided by the random initial conditions.
In the case of the random Burgers equation the random initial condition is determined by the random initial potential $\varphi_0(a)$.
Since $\varphi_0(a)$ is a process with stationary increments, the averaging of increments $\varphi_0(a+b) -\varphi_0(a)$ with respect to $\varphi_0$ can be replaced by averaging with respect to the Lagrangian coordinate $a$. For a fixed time $t>0$, stationarity of increments of the process $\varphi_0$ is reflected in translation invariance of the velocity field $u(x,t)$. Hence, averaging with respect to $\varphi_0$ can be replaced by averaging with respect to the Eulerian coordinate $x$.} over the (initial) Lagrangian coordinate $a$ and $C$ is a positive constant\footnote{We recall that genuine Brownian motion ($H=1/2$) is not only a Gaussian process, but it is also a Markov process with no memory. In contrast, if $H\neq 1/2$, it is
not a Markov process
[see, e.g., \cite{molchan1997burgers,molchan2000maximum}].}. Note that Eq.~\eqref{eq:2nd-structure-function-scales_A} implies that the initial energy spectrum has the following power-law dependence on the wavenumber $k$:
\begin{equation}
  E_0(k) \sim |k|^{1-2H}\,.
  \label{eq:E0kFBM}
\end{equation}
From a fluid-mechanical point of view, all these processes are self-similar, namely, for any $\lambda >0$ and any real $a$ and $b$, the increments of the initial potential $\varphi_0(a)$ are scale invariant in the
following  sense:
\begin{equation}
    \varphi_0\left(a+\lambda (b-a)\right) - \varphi_0(a) \stackrel{{\rm law}}{=} 
    \lambda ^H \left(\varphi_0(b) -\varphi_0(a)\right)\,,
    \label{eq:lambda-invariance}
\end{equation}
where ``$\stackrel{{\rm law}}{=}$'' is read as ``have the same (probabilistic) law  as''.
Taking $a=0$ and using $\varphi_0(0)=0$, the scale invariance~\eqref{eq:lambda-invariance} becomes
\begin{equation}
    \varphi_0 (\lambda b) \stackrel{{\rm law}}{=} 
    \lambda ^H \varphi_0(b)\;\; \implies \lambda ^{-H}\varphi_0 (\lambda b) \stackrel{{\rm law}}{=} \varphi_0(b)\,,
    \label{eq:lambda-invariance-short}
\end{equation}
which implies 
\begin{equation}
 \varphi_0(\lambda a_1)\varphi_0(\lambda a_2)\ldots \varphi_0(\lambda a_p) \stackrel{{\rm law}}{=} \lambda ^{pH} \varphi_0(a_1)\varphi_0(a_2)\ldots \varphi_0(a_p)\,.
    \label{eq:lambda-invariance-p}
\end{equation}

We now look at the finite-time evolution given by the max formula~\eqref{eq:max} and show rigorously below, for all $0 < H < 1$, that the average kinetic energy decays self-similarly as a power of the time $t$:
\begin{equation}
E(t) \equiv \langle u^2(x,t)\rangle_x = \langle(-\partial_x\varphi(x,t))^2\rangle_x\,;\;\; E(t) = \left(\frac{t}{t_*}\right)^{-\frac{2-2H}{2 - H}}\,\langle u^2\left(x, t_*\right) \rangle_x\,;
\label{eq:energy-exact-power-law}
\end{equation}
here, the subscript $x$ denotes an average over the Eulerian coordinate $x$ and $t_*$ is a non-vanishing reference time. This result was obtained as an asymptotic formula, based on the permanence of large eddies and the long-distance behaviour of the velocity correlation function, by \cite{1997-gurbatov--toth}.\footnote{Rigorous results are more easy to obtain for the Brownian case $H=1/2$, because it is a Markov process [see, e.g., \cite{1960-girsanov,1983-pitman,1983-groeneboom,1995-avellaneda-e-oct}].}

To prove the law of energy decay~\eqref{eq:energy-exact-power-law}, we change $x$ into $\lambda x$ and $a$ into $\lambda a$ in~\eqref{eq:max}. Note that, because $\lambda >0$, the maximum over $a$ is also the maximum over $\lambda a$, so using~\eqref{eq:lambda-invariance-short} we obtain
\begin{equation}
    \varphi(\lambda  x, t) 
    \stackrel{{\rm law}}{=}
    \max_{ a} \left(\lambda^{H}\varphi_0({a}) - \frac{\lambda^2}{2t}(x - {a})^2\right).
    \label{eq:getting-there-minus-1}
\end{equation}
\textit{Here comes the essential step:} In the right-hand side (RHS) of \eqref{eq:getting-there-minus-1} 
the coefficients of $\varphi_0({a})$ and of $- (x-a)^2/2$ have the ratio $\lambda ^{H}/\left(\lambda ^2/t\right)$. 
Therefore, it is natural to demand that the scale factor $\lambda$ be chosen in such a way that this ratio be unity. However, $\lambda$ is dimensionless but $t$ is not, so it is convenient to rewrite Eq.~\eqref{eq:getting-there-minus-1} in terms of the dimensionless ratio $t/t_*$ as follows\footnote{For a discussion of scaling functions in the context of the statistical mechanics of critical phenomena, see, e.g., Chapter 11 of \cite{stanley1971phase}.}:
\begin{equation}
    \varphi(\lambda  x, t) 
    \stackrel{{\rm law}}{=}
    \max_{a} \left(\lambda^{H}\varphi_0({a}) - \frac{\lambda^2}{2(t/t_*) t_*}(x - {a})^2\right)\,,
    \label{eq:getting-there-minus-0}
\end{equation}
where $t_*$ is an arbitrary positive, non-vanishing reference time. This requires 
\begin{equation}
    \lambda(t) = \left(\frac{t}{t_*}\right)^\frac{1}{2-H},
    \label{eq:lambda-becomes-a-function-of-t}
\end{equation}
so Eq. \eqref{eq:getting-there-minus-1} can be rewritten as the scale-invariant equation
\begin{equation}
    \varphi(\lambda(t) x, t)
    \stackrel{{\rm law}}{=}
    \frac{\lambda ^2(t)}{(t/t_*)}  \varphi(x,t_*)\,.
    \label{eq:potential-at-time-t-related-to-potential-at-time-unity}
\end{equation}
By combining Eqs. ~\eqref{eq:u-pot} and \eqref{eq:potential-at-time-t-related-to-potential-at-time-unity} we obtain 
\begin{eqnarray}
    u(\lambda (t) x, t)  &\stackrel{{\rm law}}{=}&
    \frac{\lambda (t)}{t/t_*}  u(x,t_*)\;\; \rm{and} \nonumber \\
    u ^2(\lambda (t) x, t)  &\stackrel{{\rm law}}{=}&  \frac{\lambda ^2(t)}{(t/t_*)^2} u ^2(x,t_*)\,,
    \label{eq:velocity-at-time-t-in-terms-of-velocity-at-time-unity}
\end{eqnarray}
which we use to obtain the decay law of the (mean) energy. We show below that the average energy is finite; therefore, we can use 
\begin{equation}
    \langle u ^2(\lambda (t) x, t) \rangle_x =\frac{\lambda ^2(t)}{(t/t_*)^2}
    \langle u ^2(x, t_*) \rangle_x\,,
    \label{eq:mean-energy-at-time-t-in-terms-of-mean-energy-at-time-unity}
\end{equation} 
whence we obtain the following law for the temporal variation of the energy:
\begin{equation}
    E(t) = \langle u ^2(\lambda (t) x, t) \rangle_x = \left(\frac{t}{t_*}\right)^\frac{2H
    -2}{2-H} \langle u ^2(x, t_*)\rangle_x \,.
    \label{eq:lo-the-law-of-energy-decay}
\end{equation}
If $E(t)$ is finite at $t_* =1$, as we prove below, then Eq.~\eqref{eq:velocity-at-time-t-in-terms-of-velocity-at-time-unity} implies that the energy will be finite at \textit{any positive time}. Indeed, with Brownian initial data (ordinary or fractional) for the potential $\varphi_0$, the initial energy is infinite. We emphasize that scaling arguments [see, e.g.,~\cite{1997-gurbatov--toth}] cannot be used to prove that $E(t)$ is finite at $t_* =1$, because, as $t \to 0$, Eq.~\eqref{eq:lo-the-law-of-energy-decay} yields $E(t) \to \infty$. To handle this, we need special tools that we forge hereafter.

\subsection{An initial potential with a Fractional Brownian Motion: boundedness of the energy}
\label{sec:finite_energy} 

We now demonstrate the boundedness of $E(t)$ at any finite time, e.g., at $t_*=1$.
We use the rigorous asymptotic relation for large deviations of the maximum of Fractional Brownian Motion [see Eq.~\eqref{eq:finite-energy-4} in Appendix~\ref{sec:AppA}] of~\cite{1978-piterbarg-prisyazhnyuk}. Let us first make a general remark
about \textit{Fractional} Brownian Motions $W_H(x)$ with Hurst exponent $H$. Many important properties of the processes $W_H(x)$ are very similar to the properties of the standard Brownian Motion for which $H=1/2$. However, rigorous mathematical analysis for other values of $H$ is very challenging because of the non-Markovian character of the process [for Burgulence, see, e.g., ~\cite{molchan1997burgers,2017-molchan}] when $H\neq 1/2$. Although increments of $W_H(x)$ are stationary, they are not independent anymore if $H\neq 1/2$. In fact, such increments are positively correlated, for $H>1/2$, and negatively correlated, for $H<1/2$. These correlations create mathematical difficulties in the analysis of $W_H(x)$.

For an arbitrary fixed  $U\ge 0$, we have
\begin{equation}
\langle u^2(0,t_*=1)\rangle\leq U^2 {\mathcal P}\left( |u(0,1)|\leq U \right) + \sum_{k=1}^\infty{(2^kU)^2\mathcal P\left(2^{k-1}U < |u(0,1)|\leq 2^kU\right)}\,.
\label{eq:finite-energy-1}    
\end{equation}
Here, $\mathcal P$ is the probability distribution corresponding to the random initial condition $\varphi_0(a)$ given by the Fractional Brownian Motion with the Hurst exponent $H$, i.e., $\varphi_0(a)=W_H(a)$; furthermore,
we have divided the range of possible values of $|u(0,1)|$ into the initial interval $[0,U]$ and a sequence of intervals $[2^{k-1}U,\; 2^kU],\; {\rm{for}} \; k~\geq~1$. Within each one of these intervals, we have used, for $|u(0,1)|$, an upper bound $U$, in the initial interval, and the bounds $2^kU$, in the intervals with the integer $ k \geq 1$ . The squares of these upper bounds, $U^2$ and $(2^kU)^2$, respectively, have been used in the estimate~\eqref {eq:finite-energy-1}, as an obvious upper bound for $|u(0,1)|^2$ within the corresponding interval.
The Lagrangian coordinate $a=a_{x,t}$, corresponding to the location $x$ at time $t$, is the location at time $t=0$ corresponding to the maximum  value of $\left(\varphi_0(a) - \frac{(x-a)^2}{2t}\right)$ [see the max formula~\eqref{eq:max}]. Given that $u(x,t)$ is the velocity, $a=x-tu(x,t)$, which can be interpreted, for a fixed $t$, 
as the inverse of the Lagrangian map from $a$ to $x$. 
In particular, if $t=1$, then the estimate $2^{k-1}U < |u(0,1)|\leq 2^kU$ implies that the Lagrangian coordinate $a$,
corresponding to $x=0$ at time $t=1$, satisfies the same estimate $2^{k-1}U <|a|=|u(0,1)| \leq 2^kU$. Since $\varphi(0,1)= \varphi_0(a) - a^2/2$, and $\varphi(0,1)$ corresponds to maximizing over all $a$, we have $\varphi_0(a) - a^2/2\geq \varphi(0,0)=0$. Hence, $\varphi_0(a)\geq a^2/2> (2^{k-1}U)^2/2$. 
From here on, using standard inequalities, the scaling invariance of $\varphi_0(x)$ and a bound for the probability distribution of $\varphi_0(a)$, obtained by ~\cite{1978-piterbarg-prisyazhnyuk} (see Appendix~\ref{sec:AppA} for details), we have for $U$ large enough
\begin{equation}
 \langle u^2(0,t_*\rangle\leq U^2 + \sum_{k=1}^\infty{(2^kU)^2 C_1M_k^\mathfrak{h}e^{-M_k^2/4}}\,,
 \label{eq:finite-energy-6}
\end{equation}
where $M_k=U^{2-H}2^{(2-H)k}/8$ and $\mathfrak{h}=\max\{(1/H-2)\,, 0\}$.
It follows that the series in Eq.~\eqref {eq:finite-energy-6} converges,
given that the term $e^{-M_k^2/4}$ dominates $2^{2k}M_k^\mathfrak{h}$.
Hence, boundedness of energy at time $t_*=1$ is established.

\subsection{Oblivious Fractional Brownian Motion and Large-scale Multifractality}
\label{subsec:OBHM}
In this Section we construct an initial potential that exhibits large-scale multifractality.
In Sections~\ref{sec:EdecayFBM} and \ref{sec:finite_energy} we proved that, when the initial potential $\varphi_0(a)$ is a fractional Brownian motion with a Hurst exponent $H$, between $0$ and $1$, the  potential $\varphi(x,t)$ evolves in time in a self-similar way. This implies that the mean kinetic energy decays like a negative power of the time $t$.
We now show how to avoid the pitfall of self-similar evolution by making the initial potential a variant of the fractional Brownian motion [see ~\cite{levy1953random,mandelbrot1968fractional}],
\textit {but with long-time forgetfulness}. As we have stated, the standard  Brownian motion, with $H=1/2$, is a Markov process: if the initial potential is known for some Lagrangian coordinate $a$, then its (spatial) future, for $b>a$, is independent of its  (spatial) past, for $c<a$. A fractional Brownian motion with a Hurst index $H \neq 1/2$ has a lot of (spatial) memory. How can we make it somewhat (spatially) forgetful or oblivious (from Latin \textit{obliviosus})?

In brief, the idea of making an oblivious version of fractional Brownian motion, without breaking the homogeneity, is the following. For the initial potential $\varphi(a, 0)$, generate a realization of a fractional Brownian motion with Hurst exponent $H$. Take any initial Lagrangian point (for example $a=0$). Let $l_1$ be positive and random (its probabilistic law will be specified in a moment). In the Lagrangian interval $[0, l_1]$, let the initial potential be one realization of a fractional Brownian motion with exponent $H$. Pick another positive random $l_2$. In the Lagrangian interval $[l_1 , l_2]$, the initial potential will be essentially another realization of the same fractional Brownian motion. By “essentially” we mean that the initial potential should be continuous at $a=l_1$ (an obvious way to achieve this is to generate the potential between $l_1$ and $l_2$ first and then to perform a vertical translation, which ensures the continuity at $l_1$). Now we extend the definition in the Lagrangian space to  $l_3$, … We proceed similarly to the left of the Lagrangian origin $a=0$. Finally, we specify the probability laws of the “oblivious” intervals of lengths $l_1$, $l_2$, $l_3$. We demand that their PDF should have “heavy tails”, i.e., $p(l) \sim l^{-\gamma}$ and that the oblivious intervals be independent. We emphasize that the condition of heavy power-law tails for the PDF of $l$ is crucially important for the large-scale-multifractality analysis that we present below.  The above construction 
provides a good description of the main idea. However, the resulting point field generated by the end points of the intervals is not spatially homogeneous because it always contains the origin.
The correct procedure, described below, starts with first selecting the interval containing the origin and then extending it by independently sampling intervals to the right and to the left of it. Translation invariance is achieved if the PDF for the initial interval is
different from those of the rest of the intervals. It is shown in  
Appendix~\ref{sec:AppB} that the PDF for the initial interval should be proportional to $lp(l)$. 

Our aim is to demonstrate that non-trivial scaling behaviour may occur at large times $t$, and arises from fluctuations of the initial potential at large distances. More precisely,  multifractality should manifest itself in the statistical behaviour of the averaged powers of the speed $|u(0,t)|$ at large times $t$.

As we have noted above, a fractional Brownian motion with Hurst exponent $H$ is non-Markovian
if $H \neq 1/2$ and, therefore, it retains memory of all past steps. We now define an \textit{Oblivious Fractional Brownian Motion} OFBM$_H$, which is \textit{continuous} by construction and comprises a random translationally invariant sequence of intervals, over which memory is present. The initial potential restricted to these intervals will be given by the Fractional Brownian Motions with the Hurst exponent $H$. However, the increments of the Fractional Brownian Motions inside a particular interval will be statistically independent 
from the pieces in other intervals. One can say that the new piece does not remember the behaviour in the previous (spatial) pieces.  We will call the processes with such loss of memory Oblivious Fractional Brownian Motions (OFBMs). We shall assume that the probability density
for the length of the intervals, where memory is present, has heavy tails. Hence, the probability of having very long intervals cannot be ignored. As a result, such large deviation events will give dominant contributions to the scaling behaviour of $\langle |u(0,t)|^{m}\rangle$ in the case of large enough values of $|m|$.
Below we consider two cases. In the first one, when $0<H<1/2$, this dominant contributions will correspond to negative values of the exponent $m$ such that $-1<m<-1 + \epsilon$ for $\epsilon$ small enough. In the second case, when $1/2<H<1$,  contributions corresponding to long intervals will determine the power-law behaviour of $|u(0,t)|^{m}$ for all large-enough positive values of $m$. In both cases, the scaling behaviour for small values of $|m|$ will be determined by the events of high probability, i.e., by the typical behaviour in terms of the lengths of the intervals used in the construction of the $\text{OFBM}_H$. Below we provide the detailed analysis in both cases.

We start with the construction of a spatially homogeneous sequence
of random intervals.  
Consider a sequence of independent identically distributed intervals (iid) in the Lagrange variable $a$. We shall assume that the length $l$ of the intervals has the probability density function (PDF) $p(l)$, where 
$p(l)\sim l^{-\gamma}$ as $l\to \infty$, with the \textit{tail exponent} $\gamma>1$. We are interested in random initial potentials with stationary increments, so we need to ensure that the point process, corresponding to the end points of the intervals, has translational invariance. This can be achieved by the following procedure. 
We start with some large negative $a=-L$, and then begin adding iid intervals, sampled according to the PDF $p(l)$ in the positive direction. 
In the limit $L\to \infty$, the starting point $a=-L$ plays no role, so, in this limit, we will obtain 
a translationally invariant point field of the endpoints of the intervals. Although the above construction is conceptually correct,
it is better to achieve our goal of constructing a
translationally invariant point field of the endpoints of the intervals as follows. It is easy to see that, for a fixed non-random point, the distribution of the length of the interval containing this point is different from the PDF $p(l)$. Indeed, it is more probable that long intervals will contain a given point.
It is not difficult to show that the corresponding PDF is proportional to $lp(l)$. In Appendix~\ref{sec:AppB} we explain the appearance of this extra factor $l$.
Now, the construction of the translationally invariant point field can be described as follows.
We first sample the length $l_0$ of the interval $\Delta_0$ containing the origin $a=0$ using its PDF which is proportional to $lp(l)$. Then we sample the location of the origin uniformly within the interval of length $l_0$. In other words,
we choose $\Delta_0=[-\epsilon, l_0-\epsilon]$, where $\epsilon$ is uniformly distributed in $[0, l_0]$. 
Next, we add intervals $\Delta_{-i}, \; i>0$ and $\Delta_i, \; i>0$ to the left and to the right of $\Delta_0$. The length of each interval $\Delta_i, \; i\neq 0$ is an independent
random variable with the distribution given by the PDF $p(l)$. 
Note that the above construction can be carried out only if the exponent $\gamma>2$. Otherwise, $\int_0^\infty{lp(l)dl}=+\infty$ and a probability distribution with the PDF proportional to $lp(l)$
does not exist.

\begin{center}
    \begin{figure}
        \centering
        \includegraphics[width=0.8\linewidth]{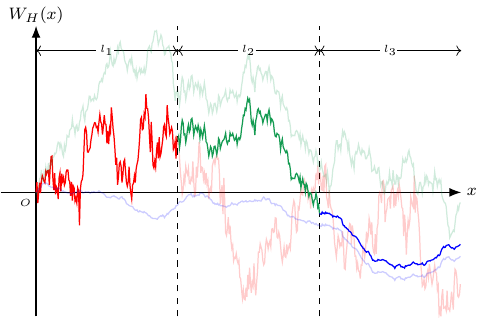}
        \caption{A schematic plot showing the construction of an Oblivious Fractional Brownian Motion [OFBM$_H$] with Hurst exponent $H \in (0,1)$: We consider intervals of length $l_i$, with $i\in[1,2, \ldots]$; for this illustration we consider three contiguous segments, of lengths $l_1$, $l_2$, and $l_3$. To obtain an OFBM$_H$, we start with three independent realizations (light red, light green, and light blue) of a fractional Brownian motion [FBM$_H$] with a given Hurst exponent $H$; in the first, second, and third segments, we use the dark red, dark green, and dark blue FBM$_H$, respectively, after a vertical translation of the starting point of the FBM$_H$, in segment labelled by $i$, so that it meets the end point of the FBM$_H$, in segment $(i-1)$. The OFBM$_H$ is shown in dark red, dark green, and dark blue (for our three-segment illustration).  The lengths of the segments are random variables that are distributed according to the pdf $p(l)$; here we choose a power-law tail for this pdf: $p(l) \sim l^{-\gamma}, \;\; l \to \infty$, with $\gamma$ the tail exponent (see text).}
        \label{fig:orw_schematic}
    \end{figure}    
\end{center}

We next construct the initial potential $W(a), \, a\in \mathbb R^1$. 
In each of the intervals $\Delta_i, \; i \in \mathbb{Z}$, we choose an independent realization of  Fractional Brownian Motion with the Hurst exponent $H$. Notice that these Fractional Brownian Motions are not extended beyond $\Delta_i$. For the interval $\Delta_0$ we assume that the FBM starts at the origin. For all other $\Delta_i$ we shall assume that it starts at the leftmost point of $\Delta_i$ for positive $i$, and at the rightmost point of $\Delta_i$ for negative $i$. 
Since we need our potential $W(a)$ to be continuous, we next move Fractional Brownian Motions inside $\Delta_{-1}$ and $\Delta_1$ vertically,
so that the values at the end points of $\Delta_0$ are matched.
We repeat this matching process for intervals $\Delta_{-i}$
and $\Delta_i$ consequently for $i= 2, \, 3, \dots$. The process constructed above is exactly the process which we call Oblivious Fractional Brownian Motions  with the Hurst exponent $H$ ($\text{OFBM}_H$).

\subsubsection{Large-scale Bifractality}\label{subsec:bifrac}

{\bf {Case A ($0<H<1/2$).}} We have mentioned above that, in the case $0<H<1/2$,  we will be interested in the averages of the inverse powers of the speed, i.e., $\langle|u(x,t)|^{m}\rangle$ for negative values of $m$. We will show that there are two important contributions, the first from intervals $\Delta_i$, which are not anomalously long,  and the second from the interval $\Delta_0$ when it is so long that the velocity at the origin will be determined by the FBM with fixed Hurst exponent $H$ inside this interval. This leads to 
\begin{eqnarray}
    \langle|u(0,t)|^{m}\rangle &\sim& t^{s(m)}\,,\;\; {\rm{with}} \nonumber \\
    \label{eq:smdef}
    s(m)&=&-m/3\;\; {\rm{for}} \;\; m\geq m_A(H,\gamma)\equiv-\frac{3(\gamma-2)}{1-2H}\,,\nonumber\\
    s(m)&=&-\alpha(H)m-(\gamma-2)/(2-H) \;\; {\rm{for}} -1<m<m_A(H,\gamma)\,, 
    \label{eq:smbifractal1}
\end{eqnarray}
where $\alpha(H)= (1-H)/(2-H)$. This is an example of \textit{bifractal} scaling, insofar as $s(m)$ is a piecewise linear function of $m$.
The detailed derivation of Eq.~\eqref{eq:smbifractal1} is given in Appendix~\ref{sec:AppB}.

{\bf Case B ($1/2<H<1,\; \tau= (\gamma-1)/2H >1$).} The analysis proceeds as in Case A above and we get
\begin{eqnarray}
    s(m)&=&-m/3\;\; {\rm{for}} \;\; -1 < m \leq m_B(H, \gamma) \,,\nonumber\\
    s(m)&=&-\alpha(H)m-(\gamma-2)/(2-H) \;\; {\rm{for}} \;\; m > m_B(H, \gamma)\,, 
    \label{eq:smbifractal2}
\end{eqnarray}
where $\alpha(H) < 1/3$ and $m_B(h,\gamma)=\frac{3(\gamma-2)}{2H-1}$; given that $\gamma > 1+2H$, the exponent $m_B(h,\gamma)$ is greater than $3$ [see Appendix~\ref{sec:AppB} for details]; again this is an example of bifractal scaling.

{\bf Case C ($1/2<H<1;\; \tau= (\gamma-1)/2H <1$).} In Case B above, with $1/2<H<1$, we had assumed that $\tau>1$. We now consider the last possible case with $1/2<H<1$ and $\tau<1$, so $2<\gamma<1+2H$. The analysis in the case of the long interval $\Delta_0$ remains unchanged, but the analysis for the other intervals $\Delta_i$ has to be modified, as we discuss in detail in Appendix~\ref{sec:AppB}. Finally, we obtain
\begin{eqnarray}
    s(m)&=&-m \frac{\gamma-1-H}{2\gamma-2-H}\;\; {\rm{for}} \;\; -1 < m \leq m_C(H, \gamma) \,,\nonumber\\
    s(m)&=&-\alpha(H)m-(\gamma-2)/(2-H) \;\; {\rm{for}} \;\; m > m_C(H, \gamma)\,, 
    \label{eq:smbifractal3}
\end{eqnarray}
where $m_C(H,\gamma)=\frac{2\gamma-2-H}{H}$. 

\begin{figure}
    \centering
    \includegraphics[width=0.3\linewidth]{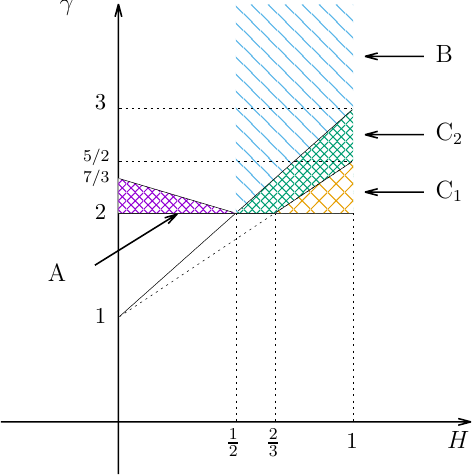}
    \caption{The regions A, B, and C, in the $H-\gamma$ parameter space, in which Cases (A), (B), and (C) [see text] are applicable, respectively.}
    \label{fig:CasesABC}
\end{figure}

Note that the energy corresponds to the exponent $m=2$. For this value of $m$, in the first two cases considered above, the dominant contribution to $s(m)$ comes from the term $-\frac{m}{3}$. Hence, the energy decays as $t^{-2/3}$. The threshold $m_C(H,\gamma)$ is:
\begin{eqnarray}
m_C(H,\gamma)&\leq& 2 \;\; {\rm{ in}} \;\; C_1\equiv \{(H,\gamma): 2/3<H<1,\;\; 2<\gamma\leq 3H/2 +1\}\,;\nonumber \\
m_C(H,\gamma) &>& 2 \;\; {\rm{ in}} \;\; C_2 \equiv \{(H,\gamma): 1/2<H<1, \;\max{\{2, 3H/2+1\}}<\gamma < 1+2H\}.\nonumber\\ 
\end{eqnarray}
The areas $C_1$ and $C_2 = C\setminus C_1$ are shown in Fig.~\ref{fig:CasesABC}. 
Therefore, the energy decays as follows: 
\begin{eqnarray}
E(t)&\sim& t^{-(\gamma-2H)/(2-H)}\,,\;\; {\rm{for}}\;\; (H,\gamma) \in C_1\,; \\
E(t)&\sim& t^{-(1-H/(2\gamma -2-H))}\,,\;\;{\rm{for}}\;\; (H,\gamma) \in C_2\,.
\end{eqnarray}
The exponent $(\gamma-2H)/(2-H)\leq 1/2$, when $(H,\gamma) \in C_1$, whereas, if $(H,\gamma) \in C_2$, the exponent $1/2<1-H/(2\gamma -2-H)<2/3$. Note that in all three cases there is also a subdominant contribution to $E(t)$ with a faster decay in the limit $t\to \infty$.

In all three Cases A, B, and C considered above, the exponent $s(m)$ consists of two different pieces that are linear in $m$, so this can again be viewed as bifractal behavior. 
\

\subsubsection{Genuine large-scale multifractality}\label{subsec:genmultifrac}

\begin{center}
    \begin{figure}
        \centering
        \includegraphics[width=1.0\linewidth]{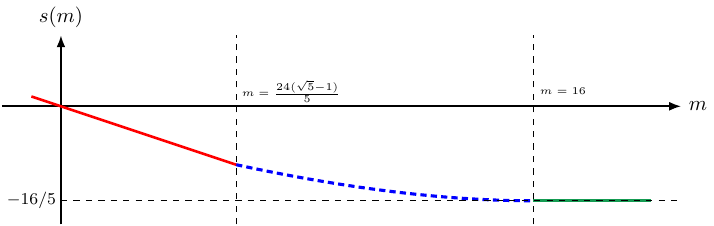}
        \caption{In Eq.~\eqref{eq:smdef} we define the exponent $s(m)$; here, we plot this exponent versus the order $m$ using Eq.~\eqref{s(m)_final} for case B and the Hurst exponent $H$ uniformly distributed in the interval $[1/2,1]$; the three parts of $s(m)$ in Eq.~\eqref{s(m)_final} are shown by full-red, dashed-blue, and full-green curves.}
        \label{fig:sm_plot}
    \end{figure}
\end{center}

We now generalize the OFBM$_H$, which we used in Cases A-C above, to build an initial condition that leads to genuine large-scale multifractality. The crucial idea is to allow the Hurst exponent $H$ to vary, over different oblivious intervals, and then use the construction of the OFBM$_H$ with an $H$-dependent tail exponent $\gamma=\gamma(H)$. 
We provide the details, for Case B, of such a construction in Appendix~\ref{sec:AppB}; we outline the essential steps below. We proceed as in Case B above by choosing $H_0$, $\gamma_0$, and $m_0$ such that  $1/2 < H_0 <1, \; \gamma_0 > (1+2H_0), \; {\rm{and}}\; m_0 > m_B(H,\gamma)$. We then sample
$H$ uniformly from the interval $[H_0-\epsilon,H_0+\epsilon]$, where $\epsilon$ is small and positive.
Furthermore, we use the tail exponent $\gamma=\gamma(H)$, with $\gamma(H_0)=\gamma_0$, and then show that $s(m)$ is related to  $[-\alpha(H)m-\frac{\gamma(H)-2}{2-H}]$ by a Legendre transformation. For example, if we use the values $H_0=3/4, \, \gamma_0=3, \, m_0=7$, we get
\begin{equation}
    \label{s(m)_final}
s(m)=\begin{cases}
			-\frac{m}{3}, & \, -1<m\leq \frac{24(\sqrt{5}-1)}{5}\,; \\
           \frac{224}{5}-m -8\sqrt{32-m}, & \, \frac{24(\sqrt{5}-1)}{5}<m<16\,; \\
           -\frac{16}{5}, &\, m\geq 16\,.
		 \end{cases}
\end{equation}
We plot $s(m)$ versus $m$ in Fig.~\ref{fig:sm_plot}. Clearly, Eq.~\eqref{s(m)_final} implies genuine multifractality because $s(m)$ has truly nonlinear dependence on $m$, and not just a combination of different linear functions of $m$ [as, e.g., in Eq.~\eqref{eq:smdef}].

\section{Numerical results for energy decay in 1D Burgulence} 
\label{sec:burg-num}

We now present the results from our direct numerical simulations (DNSs) for the decay of energy in the 1D Burgers equation. 
We then consider multifractal initial conditions for the initial potential $\varphi_0(x)$, which are constructed differently from the OFBM$_H$, as we describe in detail in Section~\ref{subsec:mfracinit}. In Section~\ref{subsec:res-burg-power-law} we use initial energy spectra $E_0(k)$ that have multiple ranges characterized by power laws that are distinct from each other.

\subsection{Burgers equation in 1D}
\label{subsec:res-burg}

We have introduced the 1D Burgers equation~\eqref{eq:burg} in Section~\ref{sec:burg-model}. Here, we consider the case in which the velocity field $u(x,t)$ is defined on the periodic interval $\left[ 0, L_{s}\right] $ and the kinematic viscosity $\nu \in \vvmathbb{R}_{+}$. 
We relate the velocity to the potential $\varphi(x,t)$ [Eq.~\eqref{eq:u-pot}] that satisfies the Eq.~\eqref{eq:burg-pot}. In the limit of $\nu \rightarrow 0^{+}$, the solution $\varphi(x,t)$ is given by the max formula~\eqref{eq:max}. The \emph{inverse Lagrangian function} $a(x,t)$ gives the (initial) position at time $t_{0}$ of a fluid particle that is at $x$ at time $t$. Thus the velocity $u(x,t)$ is found to be [see \cite{1992-she--frisch,vergassola1994burgers}]
\begin{equation}
    u[x, t] = u [a(x, t), t_{0}] = \frac{x - a(x, t)}{ t - t_{0}}\,,
    \label{eq:burgvel}
\end{equation} 
where $x$ is the Eulerian position (coordinate).

\subsection{Multifractal initial conditions with periodicity}
\label{subsec:mfracinit}

We briefly describe the algorithm that we have developed for generating multifractal initial data, whose spatiotemporal evolution we then monitor using the max formula~\eqref{eq:max} and Eq.~\eqref{eq:burgvel} for the 1D inviscid Burgers equation. Multifractal random walks, which were studied by \cite{2001-bacry--muzy},
take the form:
\begin{equation}
    X_n = \sum_{k=1}^{n} \xi_{k} e^{\Omega_k}, \quad n = 1, 2, \ldots
     \label{eq:mfin1}
\end{equation}
Here, the sequence $\xi_k$ is Gaussian white noise and $e^{\Omega_k}$ is a log-normal variable. In addition, $\Omega_k$'s are correlated, with the covariance matrix
\begin{equation}
    \text{cov}\big( \Omega_{k_1} , \Omega_{k_2} \big) \propto \log  \frac{\mathcal{L}}{|k_1 -k_2| +1}, \quad |k_1 -k_2| \leq \mathcal{L}\,,
    \label{eq:omega-cor}
\end{equation}
where $\mathcal{L}$ is the length scale below which the walk displays multifractality [see \cite{bacry2001multifractal}] and above which the walk is a fractional Brownian motion. Note that by tuning $\mathcal{L}$, \textit{we can go from small-scale to large-scale multifractality.}
However, we cannot use this numerical scheme of \cite{bacry2001multifractal} directly because we here impose periodic boundary conditions in our system. 
Therefore, to generate a multifractal random walk with periodic boundary conditions, we generalise the method of \cite{2001-bacry--muzy} by combining it with the technique of \cite{1997-dietrich-newsam} as follows. We consider the sequence
\begin{equation}
	A_n \equiv B_n e^{ \Omega_n}, \quad 1 \leqslant n \leqslant N\,,
 \label{eq:mfin2}
\end{equation}
where  $B_n$ and $\Omega_n$ are random numbers with the following statistics: 
We choose $B_n$ to be Gaussian random numbers, but with the additional restriction $\sum_{i=1}^N B_n = 0$. Thus, the random walk $W_n = \sum_{k=1}^{n-1} B_n$, constructed using $B_n$, is periodic and approximately Brownian ($H=1/2$) at scales much smaller than the length of the system; i.e., the $B_n$'s are increments of the random walk $W_n$, for which $H=1/2$ at small scales. Later, we will consider increments of a random walk for which $0 < H <1$.
Specifically, we use the procedure prescribed by \cite{1997-dietrich-newsam} to compute the sequence $\Omega_n$. Then, in terms of $A_n$, we define 
\begin{equation}
	A'_n = A_n - \frac{1}{N}\sum_{i=1}^{N} A_i\,.
 \label{eq:mfin3}
\end{equation}
This ensures that $A'_n$ is periodic. 
Then we consider the following sequence of random numbers, which are computed using $A'_n$:
\begin{equation}
	M_n = 
	\begin{cases}
	0, & \quad  n=1 \\
	\sum_{i=1}^{n-1} A'_i,  & \quad n > 1 \,.
 \label{eq:mfin4}
	\end{cases}
\end{equation}
The random numbers $M_n$ show multifractal properties\footnote{Note that if $\Omega_n \rightarrow 0$, we recover a simple random walk with $H=1/2$ at small scales (by construction as explained above).}, because of the factors $e^{\Omega_n}$.

We now consider freely decaying turbulence in the 1D inviscid Burgers equation~\eqref{eq:burg}, with the multifractal initial condition that we have obtained using Eqs.~\eqref{eq:mfin1}-\eqref{eq:mfin4} with $H=1/2$
for the initial potential $\varphi_0$. [We describe our results for a multifractal random walk (MRW) with $H=0.75$ for the initial potential $\varphi_0$ in Appendix~\ref{subsec:app_burg1}, where we also discuss decaying 1D Burgulence with such MRWs for the \textit{initial velocity}.] 
In our numerical studies, which use Eqs.~\eqref{eq:max} and ~\eqref{eq:burgvel}, we discretize the system with $N=2^{14}$ points. 

\begin{figure} 
    \centerline{
        \includegraphics[width=0.5\columnwidth]{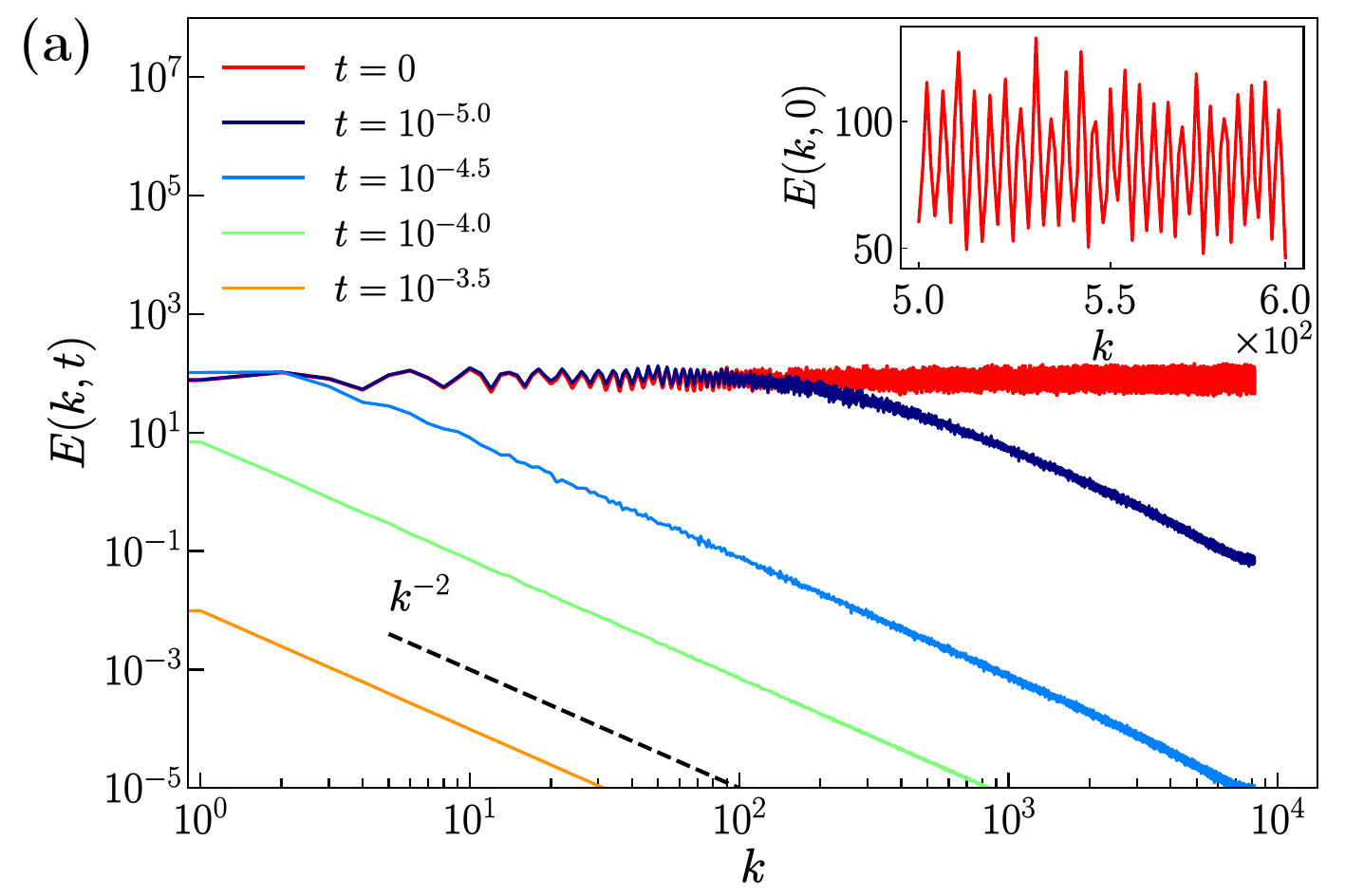} 
        \includegraphics[width=0.5\columnwidth]{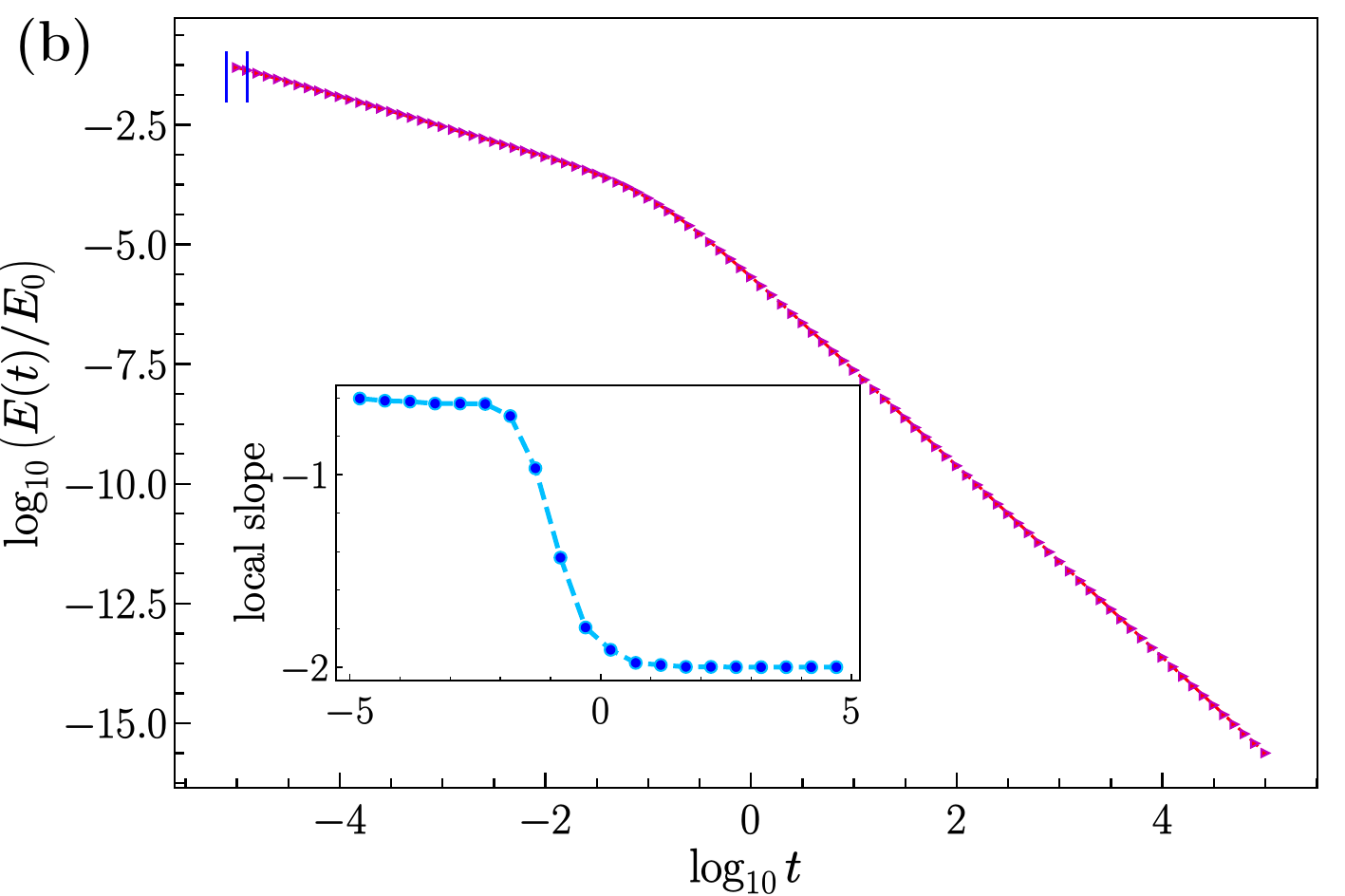}
    }
    \caption{Log-log plots of (a) the energy spectrum $E(k,t)$ versus the wavenumber $k$, at some representative values of $t$ [the inset shows the initial spectrum $E(k,0)$ in detail in the wavenumber range $5\times 10^2\leq k\leq 6 \times 10^2$] and (b) the scaled total energy $E(t)/E(t=0)$ versus the time $t$ for the  multifractal initial condition [Eqs.~\eqref{eq:mfin1}-~\eqref{eq:mfin4}], for the potential $\varphi_0$, with Hurst exponent $H=1/2$; the inset shows a plot of the local slope. We compute structure functions in Fig.~\ref{fig:sfunmfrwphi} at the point in time that lies at the centre of the interval indicated by the blue vertical lines. }
    \label{fig:mrw-a1-ende-phi}
\end{figure}

\begin{center}
	\begin{figure}
            \centering
		\includegraphics[width=0.9\linewidth]{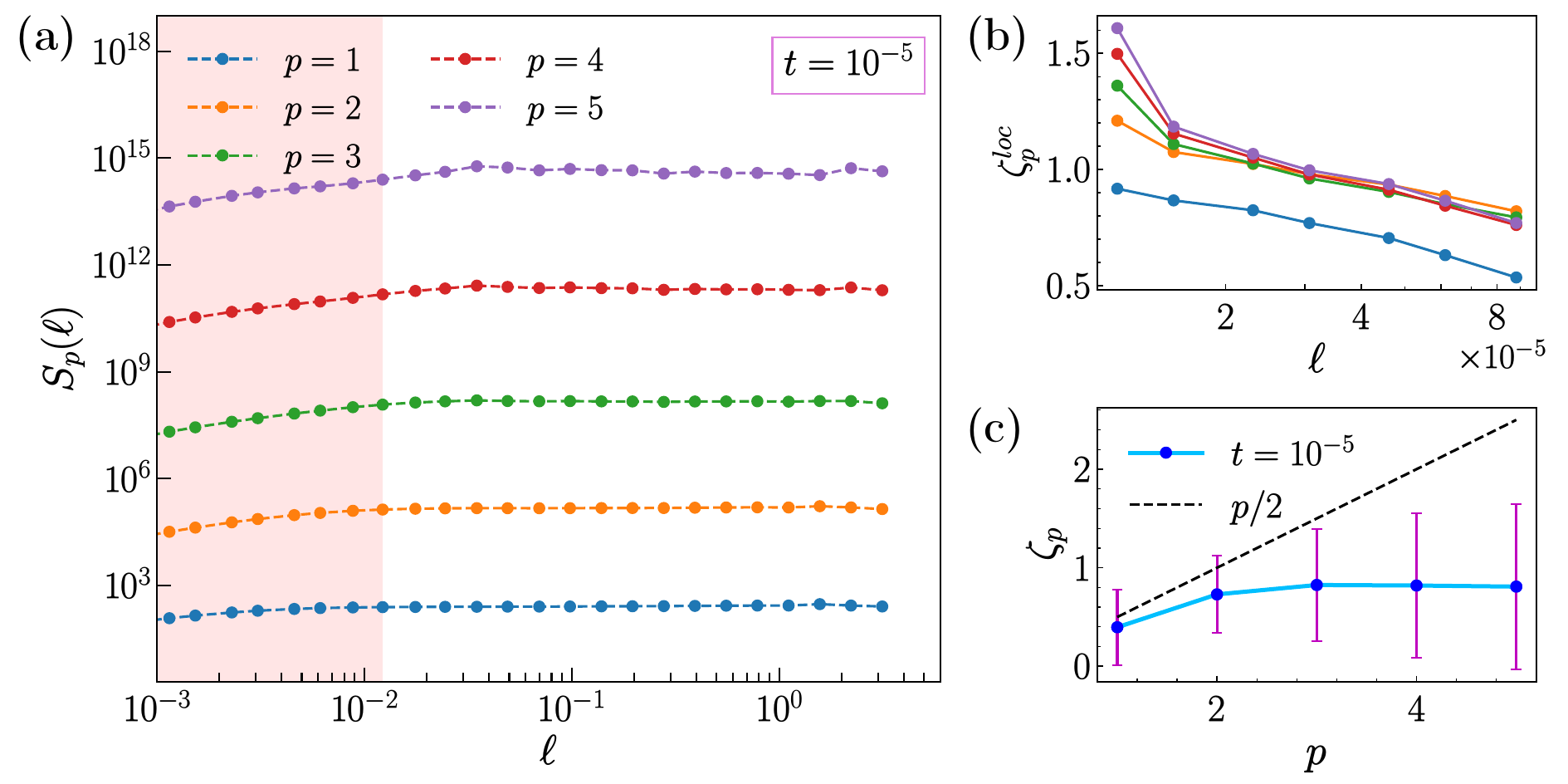}
		\caption {Plots for the  multifractal initial condition, for the potential $\varphi_0$, with Hurst exponent $H=1/2$ at $t=10^{-5}$: (a) Log-log plots versus $\ell$ of the structure functions of order $p=1, \ldots, 5$. (b) Plots of $\zeta_p^{loc}$, obtained from local slopes of the structure functions in (a), versus $\ell$. (c) Plots of $\zeta_p$ versus $p$ (in blue) at $t=10^{-5}$.}
   \label{fig:sfunmfrwphi}
	\end{figure}
\end{center}

In Fig.~\ref{fig:mrw-a1-ende-phi} (a) we show log-log plots of the energy specturm $E(k,t)$ versus the wave number $k$ at different representative times $t$; at early times 
$0 \leq t \lesssim 10^{-5}$, this spectrum is not of a simple, power-law form because of the multifractal initial condition for $\varphi_0(x)$; however, for $10^{-4.5} \lesssim t$, the spectrum has the power-law form $E(k,t) \sim k^{-2}$ because of the formation of shocks. The decay of the total energy $E(t)$ is shown in the log-log plot of Fig.~\ref{fig:mrw-a1-ende-phi} (b); the temporal decay does not have a single-exponent, power-law form for $0 \leq t \lesssim 1$; however, at later times, it shows the power-law decay $E(t) \sim t^{-2}$, once the integral length scale becomes comparable to the system size.  We compute the order-$p$ velocity structure functions $S_p(\ell,t) \equiv [u(x+\ell,t) - u(x,t)]^p$ and plot it versus the separation $\ell$ [see the log-log plot in Fig.~\ref{fig:sfunmfrwphi} (a)]; we obtain the multiscaling exponents $\zeta_p$, which follow from the power-law form $S_p(\ell,t) \sim \ell^{\zeta_p}$ for $\ell$ in the pink-shaded region in  Fig.~\ref{fig:sfunmfrwphi} (a). We use a local-slope analysis [Fig.~\ref{fig:sfunmfrwphi} (b)] to extract these exponents, which we plot versus the order $p$ in  Fig.~\ref{fig:sfunmfrwphi} (c) at  $t=0$ (red curve)  and $t=10^{-3}$ (blue curve). We observe that multifractality is present at $t=10^{-5}$ (see Fig.~\ref{fig:sfunmfrwphi}), in so far as $\zeta_p$ is a nonlinear function of $p$. 


\subsection{Burgers equation in 1D: Power-law Initial Data}
\label{subsec:res-burg-power-law} 

We have seen in Section~\ref{sec:burg-model} that, if the initial potential $\varphi_0$ is a fractional Brownian motion, with Hurst exponent $0 < H < 1$, then the initial energy spectrum has the power-law form $E_0(k) \sim |k|^{(1-2H)}$ given in Eq.~\eqref{eq:E0kFBM}. Therefore, in our direct numerical simulations (DNSs), we examine various types of initial conditions whose Fourier transforms lead to power-law regions in the initial energy spectrum $E_0(k)$. To obtain a single-power-law regime, as in Eq.~\eqref{eq:E0kFBM}, we use a Gaussian random initial velocity profile $u(x,0)$ for which the Fourier modes $\widetilde{u}_{k}(0)$ for the wavenumber $k \geq 0$ take the following form: 
\begin{eqnarray}
    \widetilde{u}_{k}(0) 
    &=& 
    \sqrt{ A  \, \mathcal{E}(k) }  \, \exp \! \left( - k^{2}/k^{2}_{c} \right) X_{k}\,; \nonumber\\
{\rm{here,}}\;\;    \mathcal{E}(k) &\equiv& k^n\,,\;\; {\rm{with}} \;\; -1< n < 2\,;
    \label{eq:burg_one_power}
\end{eqnarray}  
$A$ is a positive constant, with the cutoff wavenumber $k_{c} \gg 1$, and $X_{k}$ is a standard complex Gaussian random variable. With these initial data, the energy decay is self-similar, as in Eq.~\eqref{eq:energy-exact-power-law}
for $-1< n<1$, which is associated with the permanence of large eddies, with $E(t)  \sim  t^{\frac{-2(n+1)}{(n+3)}}$
and $L(t) \sim t^{\frac{2}{(n+3)}}$ [see, e.g., \cite{1992-she--frisch}, \cite{1997-gurbatov--toth}, and page 114 of \cite{roy2021steady}]. If $1 < n < 2$, we encounter the Gurbatov phenomenon,  namely, $E(k,t) > E_0(k)$ for wavenumbers $k \leq K(t) \sim 1/L(t)$. This leads to non-self-similar decay [growth] of the $E(t)$ [$L(t)$] because of logarithmic corrections [see \cite{1997-gurbatov--toth} and \cite{roy2021steady}].

\subsubsection{Case I: Two-power-law initial energy spectrum}
\label{subsubsec:tworange}

We consider next the case in which the initial energy spectrum $E_{0}(k)$ has two spectral ranges with different power laws,
specifically,
\begin{equation}
	E_{0}(k) = 
	\begin{cases}
		A_{1} \mathcal{E}_{1}(k)  & \qquad \text{ for } k < k_{1} \ ,\\
		A_{2}  \mathcal{E}_{2}(k) \exp \! \left[ - 2 k^{2}/k^{2}_{c} \right]  & \qquad \text{ for } k \geq k_{1} \,,
	\end{cases}
    \label{eq:E0k2p1}
\end{equation}  
where the constants $A_{1}$ and $A_{2}$ are chosen such that $E_{0}(k)$ is continuous and the functions
\begin{equation}
	\mathcal{E}_{i}(k) = k^{n_{i}}\,, \;\; i = 1,2\,;
	\label{eq:power-law}
\end{equation}
so, in this case, the initial spectrum $E_{0}(k)$ depends on the pair of integers $\overline{n} = (n_{1}, n_{2})$. We consider the following four pairs:
I (a): $\overline{n} = (0.25, 0.75)$; I (b): $\overline{n} = (0.5, 1.5)$; I (c): $\overline{n} = (1.5, 0.5)$; and I (d):  $\overline{n} = (1.25, 1.75)$.
We describe our results for case I (a) in detail below and discuss the other cases in Appendix~\ref{subsec:app_burg3}.

For case I(a), the energy spectrum $E(k,t)$ at time $t>0$ has one peak at $k_{p}(t)$, so we describe its features as follows: 
\begin{equation}
	E(k,t) =
	\begin{cases}
		E_{0}(k) 	&\qquad \text{ for } k < k'(t) < k_{p}(t) \ , \\
		J(k, t) 		&\qquad \text{ for } k'(t) \leq k \leq k''(t) \ , \\
		J(k''(t),t) (k''(t)/ k)^{2} 	&\qquad \text{ for } k > k''(t) > k_{p}(t) \ .
	\end{cases}
\end{equation}
The function $J(k,t)$ is defined on the interval $\left[ k'(t), k''(t)\right] $. $J(k,t)$ describes the smooth portion of the continuous part of the spectrum that includes the peak at $k_{p}(t)$. In this case I (a),  $\overline{n} = (0.25, 0.75)$, both $n_1$ and $n_2$ are less than $1$, so $E(k,t)$ is bounded above by $E(k,0)$ for all $t$, i.e., there is no Gurbatov effect [see Fig. 5 in \cite{1997-gurbatov--toth}]. By contrast, cases  I (b), $\overline{n} = (0.5, 1.5)$, I (c), $\overline{n} = (1.5, 0.5)$, and I (d), $\overline{n} = (1.25, 1.75)$, show the Gurbatov effect with regions where $E(k,t)$ is not bounded above by $E(k,0)$ for all $t$ [see Appendix~\ref{subsec:app_burg3}].
Furthermore, at large times, $E(k,t) \sim k^{-2}$ because of the formation of shocks.

\begin{figure}
    \centerline{
        \includegraphics[width=0.333\columnwidth]{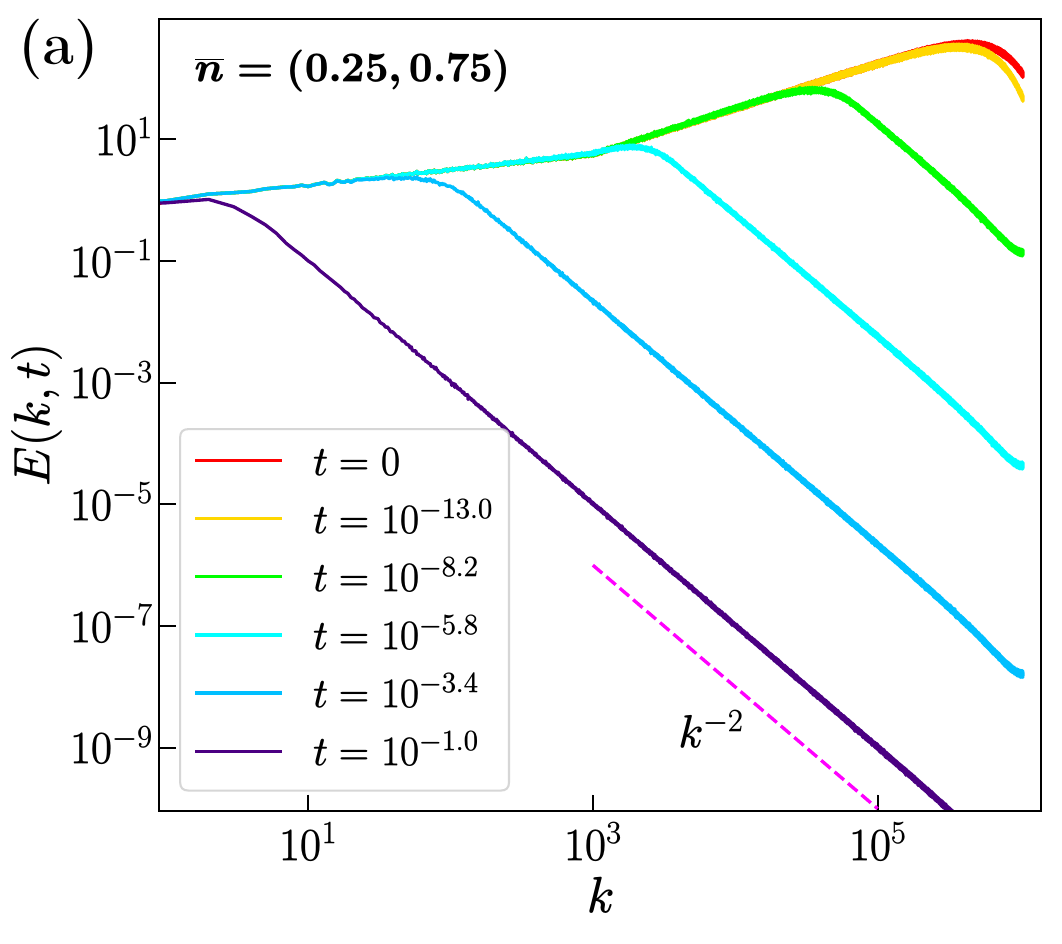} 
        \includegraphics[width=0.333\columnwidth]{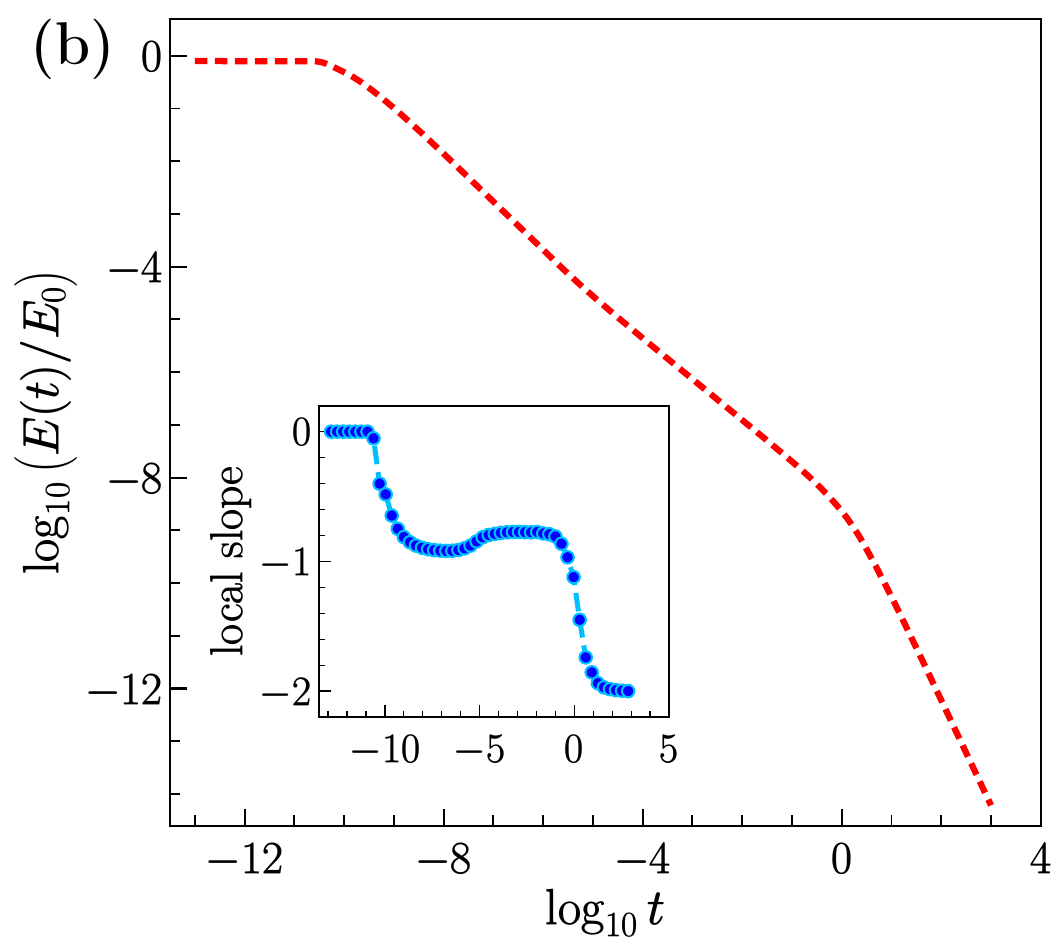}
        \includegraphics[width=0.333\columnwidth]{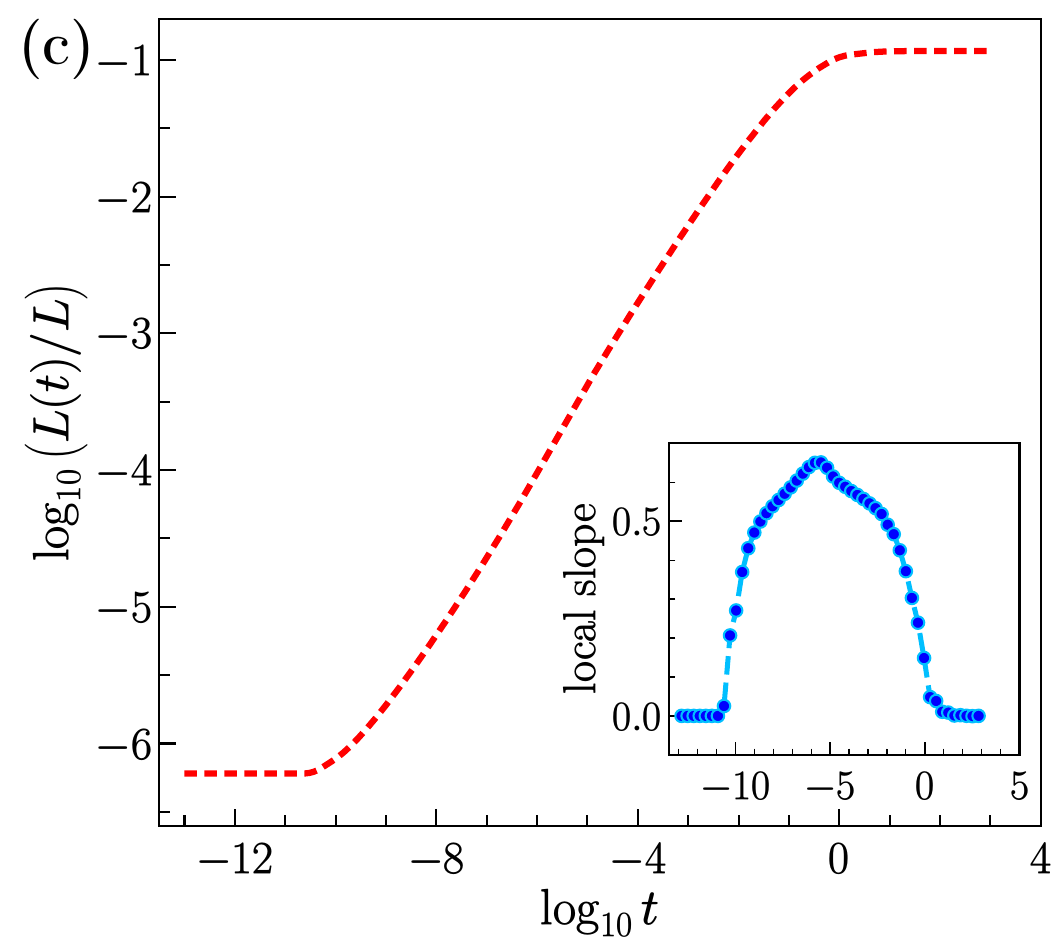}
    }
    \caption{Log-log plots for (a) the energy spectrum $E(k,t)$ versus the wavenumber $k$ at representative times $t$, (b) the decay of the total energy $E(t)$ with time $t$, and (c) the growth of integral length scale  $L(t)$ with time $t$ for case I (a), with a two-power-law initial energy spectrum [see Section~\ref{subsubsec:fourrange}]. The insets show local slopes that can be used to
    estimate the decay and growth exponents in (b) and (c), respectively.
    }
    \label{fig:caseI_burg}
\end{figure}

The temporal evolution of the energy spectrum is shown in Fig.~\ref{fig:caseI_burg}(a) by log-log plots of $E(k,t)$ versus $k$ at some representative times.
The decay of the total energy [Fig.~\ref{fig:caseI_burg}(b)] and the growth of the integral length scale [Fig.~\ref{fig:caseI_burg}(c)] clearly show two temporal regimes: $E(t)$ decays as $\sim t^{-0.9}$ (resp. $\sim t^{-0.8}$) for $t \in [10^{-7}, 10^{-2}]$ (resp. $t \in [10^{-2}, 10^{2}] $). The integral scale $L(t)$ grows with an exponent greater than $ 0.5$ throughout these two regimes. The variation in the local slopes are shown insets of the plots of $E(t)$ versus $t$ and $L(t)$ versus $t$ in Figs.~\ref{fig:caseI_burg}(b) and (c), respectively. The exponents for the energy decay, $\simeq -0.9$ and $\simeq -0.8$, compare well with the values $-0.93$ and $-0.77$, respectively, which are computed using the formula for the single-power-law case [see Eq.~\eqref{eq:expe} in Appendix~\ref{subsec:app_burg2}] and taking into consideration account the peak position $k_{p}(t)$. A single exponent, which characterises the growth $L(t)$, cannot be extracted from  Fig.~\ref{fig:caseI_burg}(c). Thus, Figs.~\ref{fig:caseI_burg}(b) and (c) provide clear evidence for \textit{non-self-similar} decay of $E(t)$ and growth of $L(t)$, respectively.

\subsubsection{Case II: Four-power-law initial energy spectrum}
\label{subsubsec:fourrange}
\begin{figure}
    \centerline{
        \includegraphics[width=0.333\textwidth]{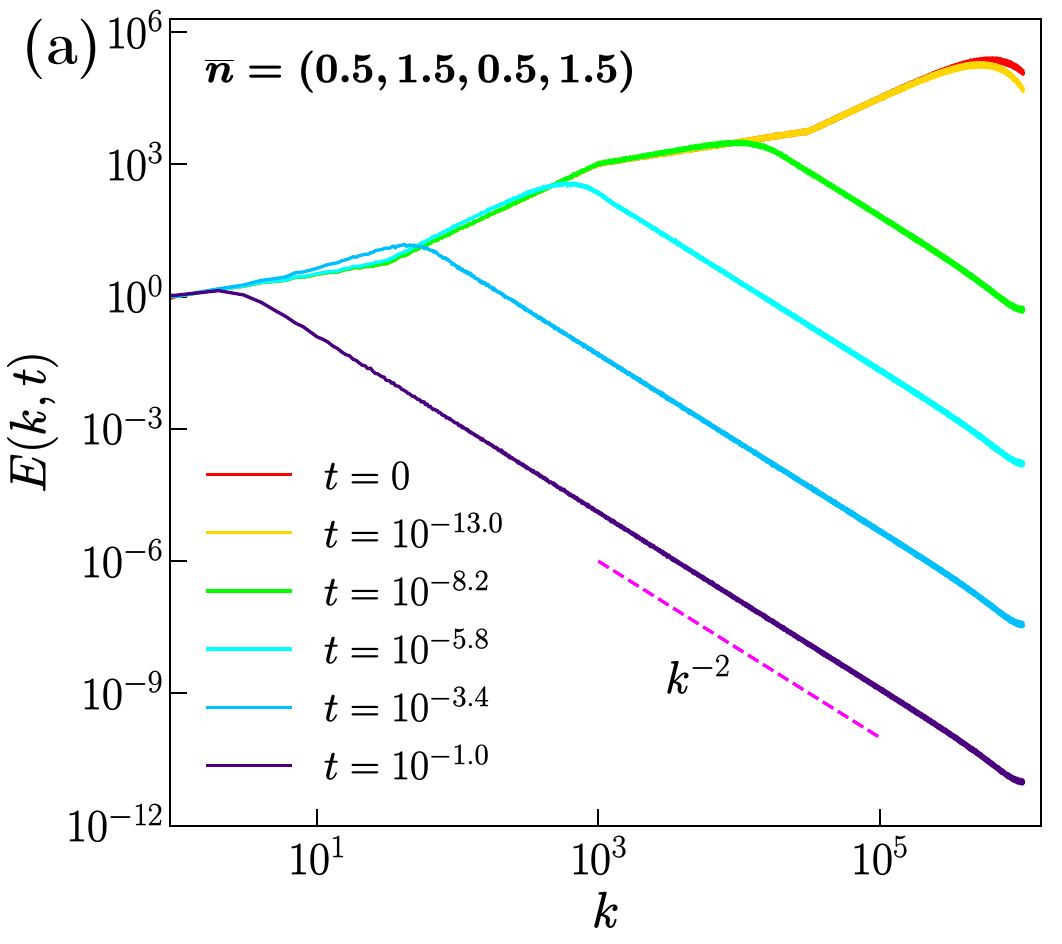}
        \includegraphics[width=0.333\textwidth]{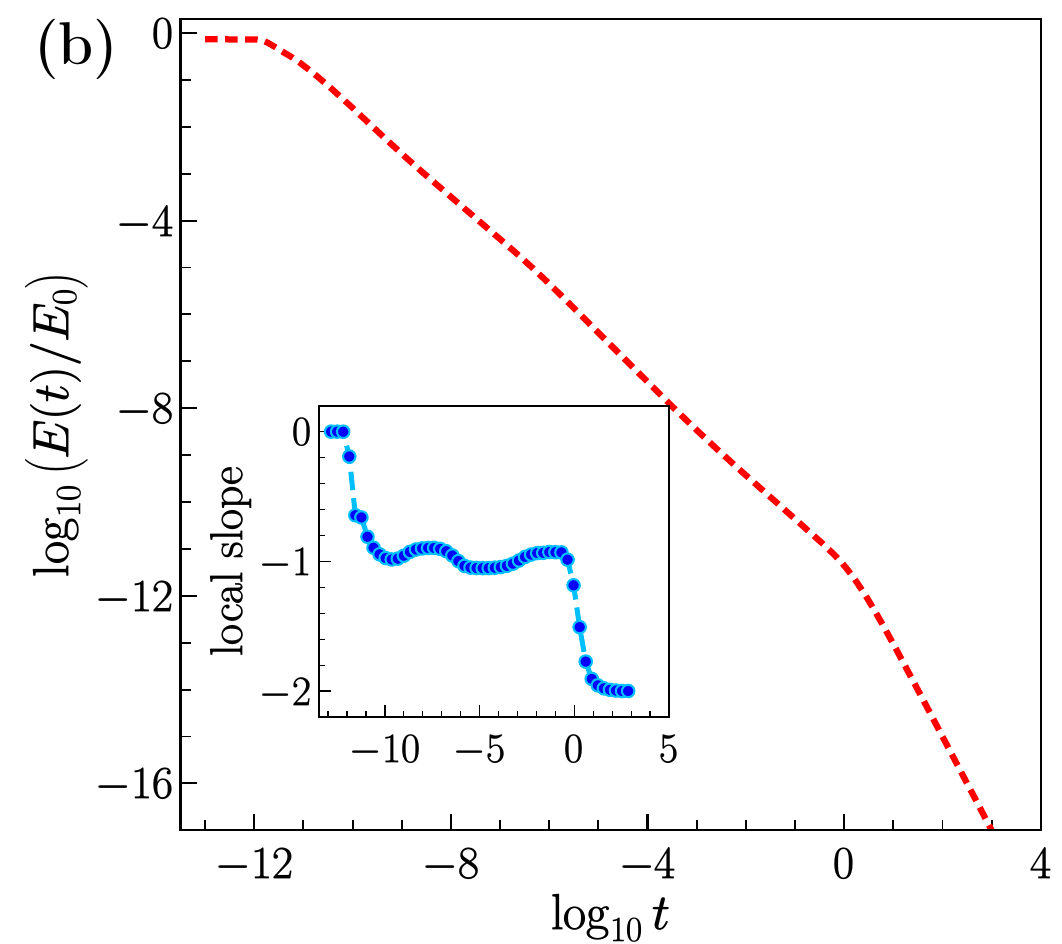}
        \includegraphics[width=0.333\textwidth]{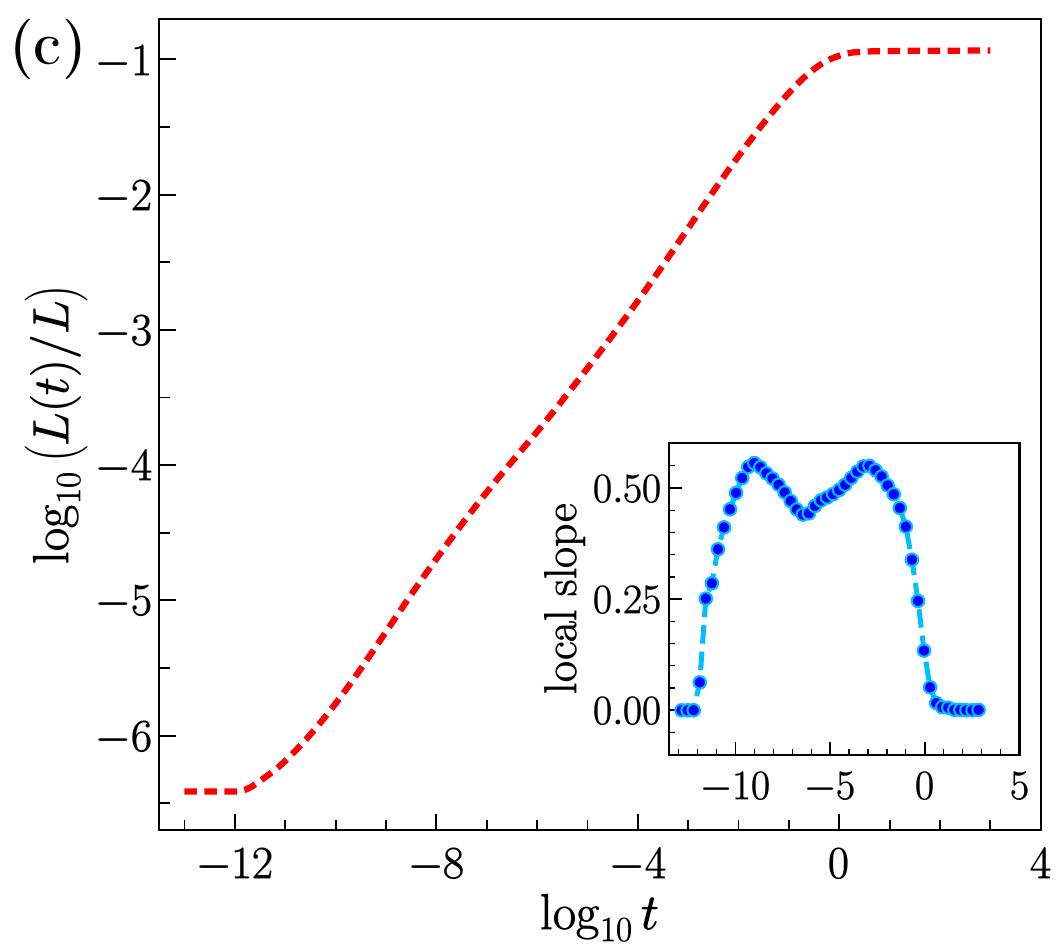}
    }
    \caption{Log-log plots for (a) the energy spectrum $E(k,t)$ versus the wavenumber $k$ at representative times $t$, (b) the decay of the total energy $E(t)$ with time $t$, and (c) the growth of integral length scale  $L(t)$ with time $t$ for case II (a), with a four-power-law initial energy spectrum [see Section~\ref{subsubsec:fourrange}]. The insets show local slopes that can be used to
    estimate the decay and growth exponents in (b) and (c), respectively.}
    \label{fig:caseII_burg}
\end{figure}

The initial energy spectrum, involving four main spectral ranges with power-law dependences on $k$, is given by 
\begin{equation}
	E_{0}(k) = 
	\begin{cases}
		A_{1} \mathcal{E}_{1}(k)  & \qquad \text{ for } k < k_{1} \ ,\\
		A_{2} \mathcal{E}_{2}(k)  & \qquad \text{ for } k_{1} \leq k < k^{2}_{1} \ ,\\
		A_{3} \mathcal{E}_{3}(k)  & \qquad \text{ for } k^{2}_{1} \leq k < k^{3}_{1} \ ,\\
		A_{4} \mathcal{E}_{4}(k) \exp \! \left[ - 2 k^{2}/k^{2}_{c} \right]  & \qquad \text{ for } k \geq k^{3}_{1} \ , \\
        {\rm{with}}\;\; \mathcal{E}_{i} \sim k^{n_i}\ , & \qquad {\rm{and}} \; i = 1,\,2,\,3, 4 \ ,
	\end{cases}
	\label{eq:four-power}
\end{equation} 
and power-law regions specified by the quartet of exponents $\overline{n}= (n_1,\,n_2,\,n_3,\,n_4)$ [cf. Eq.~\eqref{eq:power-law}].
We consider the following two examples: II (a) $\overline{n} = (0.5,1.5,0.5,1.5)$; and II (b) $\overline{n} = (1.5,-1.5,1.5,-1.5)$.
We describe  below our results for the choice of exponents  II (a) [for the choice  II (b) see Appendix~\ref{subsec:app_burg4}].
The temporal evolution of $E(k,t)$ for case IIa is shown in Fig.~\ref{fig:caseII_burg}(a) by log-log plots of $E(k,t)$ versus $k$ at some representative times. This is similar to the spectral evloution in case I (a) [if, roughly speaking, we consider the first two and the last two spectral ranges independently]. However, there is one important difference: Because $n_2$ and $n_4$ are both greater than $1$, we obtain a Gurbatov effect [see \cite{1997-gurbatov--toth}] so $E(k,t)$ rises above $E(k,0)$ in some ranges of $k$ and $t$.  
The decay of the total energy $E(t)$ and the growth of the integral length $L(t)$, which can be surmised from the log-log plots in Figs.~\ref{fig:caseII_burg}(b) and (c), respectively, are \textit{non-self-similar}.

\section{Energy decay in incompressible Navier--Stokes turbulence}
\label{sec:ns-hyper-vis}

In Sections~\ref{sec:burg-model} and ~\ref{sec:burg-num} we have shown, theoretically and numerically, respectively, that for 1D Burgulence, in the limit of vanishing viscosity and without forcing, we do not necessarily have a self-similar power-law decay of the energy $E(t)$. 
To what extent can we obtain analogous non-self-similar decay of $E(t)$ for the \textit{three-dimensional} (3D)  Navier--Stokes equation (3DNSE) in the limit of vanishing viscosity? 

From a mathematical point of view, this is a rather difficult question, because we do not know under what conditions these equations, with smooth initial conditions, possess unique solutions, devoid of singularities for all positive times. Let us leave such mathematical concerns aside for the moment and recaptitulate, briefly, some results for the decay of $E(t)$ in the 3DNSE. 
\cite{1941-kolmogorov-b} shed some light on this problem by deriving the following equation:
\begin{equation} 
	\partial_t E(t) = -C E^{3/2}(t)L^{-1}(t)\,,
	\label{eq:k41b-eq-22}
\end{equation}
where $C$ is a positive dimensionless constant; this is equation (22) in \cite{1941-kolmogorov-b}, which makes a scaling assumption [Eq.~(19) in \cite{1941-kolmogorov-b}]; we note, with hindsight, that this assumption implicitly excludes multifractality. However, Eq.~\eqref{eq:k41b-eq-22} can still not be solved because it contains two unknown functions, $E(t)$ and $L(t)$. To overcome this problem, Kolmogorov then used the Loitsiansky invariant, whose invariance was later called into question by \cite{1954-proudman-reid}, who used the quasi-normal closure. Subsequent studies [see, e.g., \cite{1978-tatsumi--mizushima}, \cite{1980-frisch--schertzer}, \cite{1997-gurbatov--toth}] argued that these results  of \cite{1954-proudman-reid} are robust and that they are valid even if we do not employ the quasi-normal closure. Furthermore, they formulated the principle of permanence of large eddies [see Section 7.8 in \cite{1995-frisch-book}] by building upon the key result by \cite{1954-proudman-reid} that the beating interaction of two nearly opposite wavenumbers $k$, whose absolute values are near the integral-scale wavenumber $\sim L^{-1}(t)$, contributes to low-wavenumber dynamics a (transfer) input $T(k) \propto k ^4$ in 3D. As a consequence, if the low-wavenumber initial energy has a spectrum much steeper than $k^4$, the low-$k$ energy spectrum develops a $k^4$ regime.

In Sections~\ref{subsec:ns} we present the results of our direct numerical simulations (DNSs)
of freely decaying turbulence in the viscous or hyperviscous 3D incompressible Navier--Stokes equations; DNSs with
hyperviscosity allow us to overcome the limited-resolution problems that beset their viscous counterparts. 
In order to alleviate this problem, we then move to the hyperviscous incompressible Navier--Stokes
equations [for a precise definition see Section~\ref{subsec:ns}]. 

\subsection{Numerical simulation of the Navier--Stokes case\label{ssec:ns}}
\label{subsec:ns}

We study freely decaying turbulence in the incompressible 3D Navier--Stokes equations (3DNSE): 
\begin{eqnarray}
  \partial_t {\bm u} + ({\bm u}\cdot\nabla){\bm u} &=& -\nabla p - \nu_\upbeta [-\triangle]^\upbeta {\bm u}\,; \nonumber \\
 \nabla \cdot \bm u &=& 0\,;
 \label{eq:ns}
\end{eqnarray}
${\bm u}({\bm x}, t)$, $p({\bm x}, t)$, and $\triangle = \nabla^2$ denote the velocity, pressure, and Laplacian, respectively; we set the constant density $\rho=1$; and we consider the viscous and hyperviscous cases $\upbeta = 1$ and $\upbeta = 2$ with kinematic viscosity $\nu_1$ and kinematic hyperviscosity $\nu_2$, respectively. 

We use a cubical domain with sides of length $2\pi$ and periodic boundary conditions and a standard pseudospectral method [see, e.g., \cite{canuto2007spectral}] for our DNSs with isotropic truncation for dealiasing, i.e., the Fourier modes with wavevector $|{\bm k}| \ge \sqrt{2}N/3$ are set to zero; $N^3$ is the number of collocation points; we use $N = 512$.
To obtain reliable data for the decay of turbulence at long times, it is important to use a sophisticated time-stepping scheme, so we employ the high-order Runge--Kutta method, known as the Dormand-Prince 853 scheme, accompanied with adaptive step-size control [see \cite{HNW}], so that we can increase the time-step $\delta t$ as the flow decays.
We use dense output [see ~\cite{HNW}] in order to have the data output at regulary spaced intervals in time~\footnote{Dense output is an interpolation method for data that are obtained with irregular time steps; the  Dormand-Prince 853 scheme used in conjunction with dense-output data has an accuracy of $\mathcal{O}[(\delta t)^7]$.}. If we set the order of the Laplacian $\upbeta$ larger than $2$, the 3DNS equations become so stiff that this adaptive control is not efficient.

\subsubsection{One-power-law initial energy spectrum for Navier-Stokes turbulence}
\label{subsubsec:1PLNS}

\begin{figure}
    \centerline{
    \includegraphics[width=0.35\columnwidth]{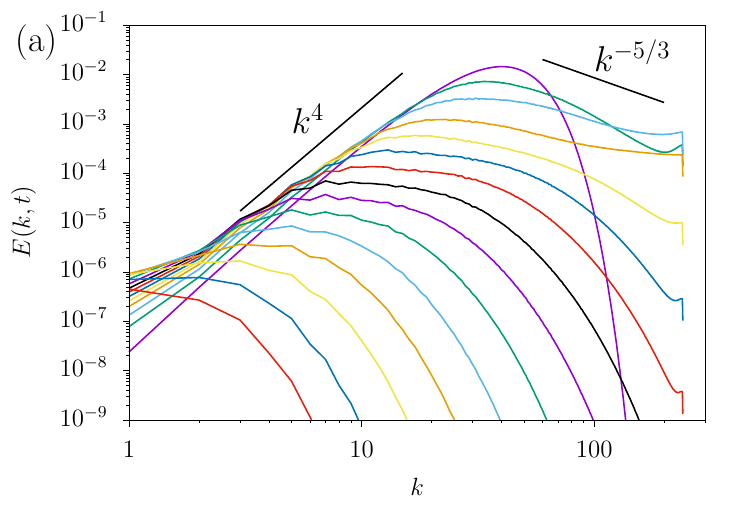} 
    \includegraphics[width=0.35\columnwidth]{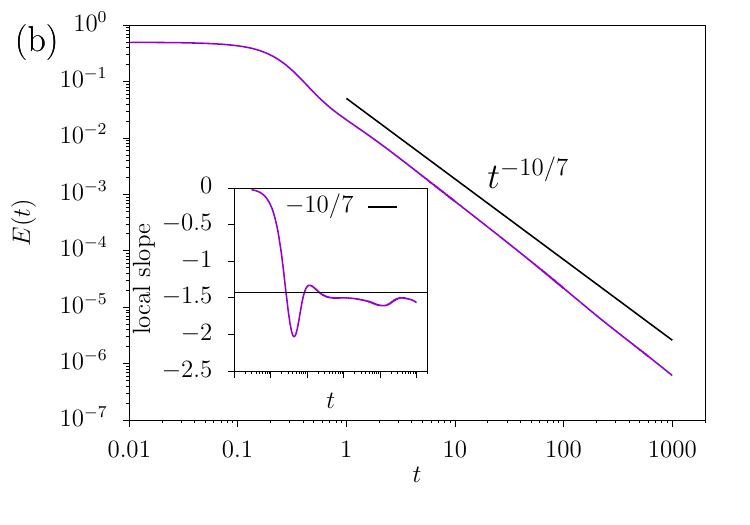}
    \includegraphics[width=0.35\columnwidth]{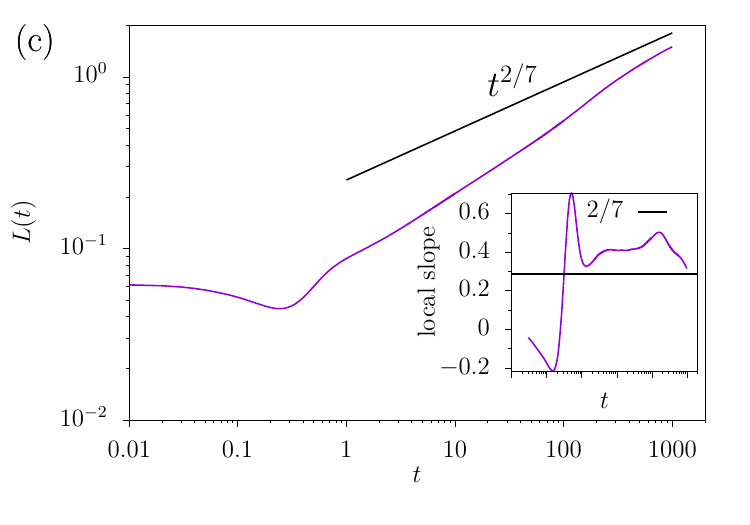}
    }
    \centerline{
    \includegraphics[width=0.35\columnwidth]{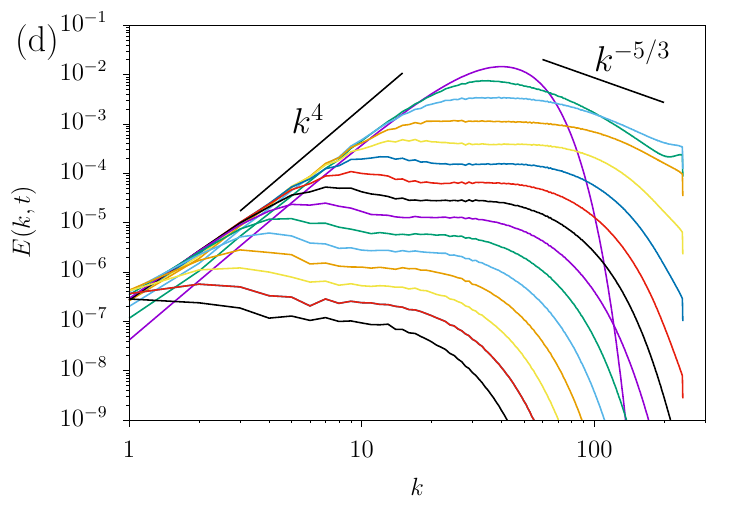} 
    \includegraphics[width=0.35\columnwidth]{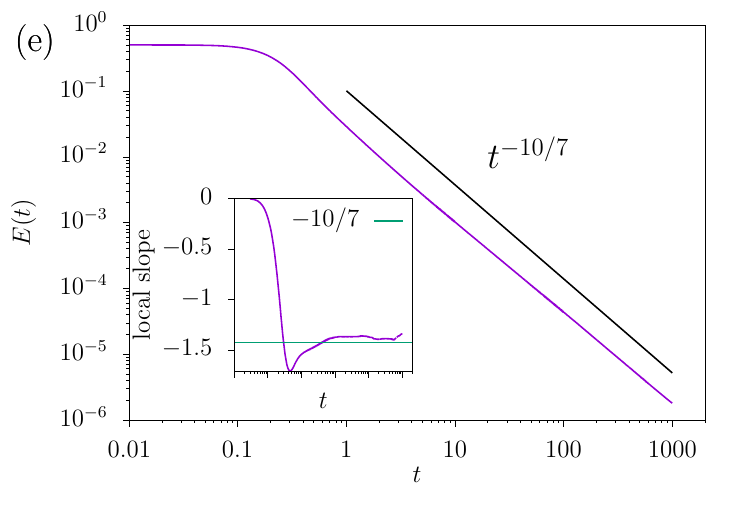}
    \includegraphics[width=0.35\columnwidth]{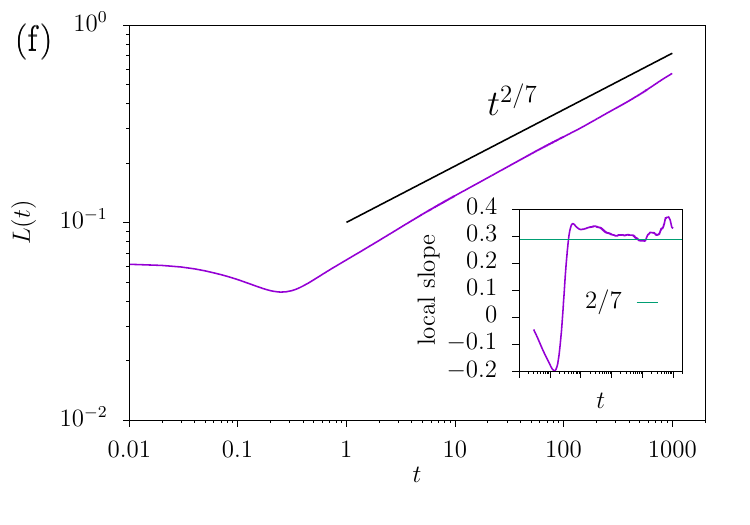} 
    } 
    \caption{(a) Log-log plots of the energy spectrum $E(k,t)$ versus the wavenumber $k$ at representative values of the time in the range $0 \leq t \leq 820$; (b) and (c) display log-log plots versus $t$ of the total energy $E(t)$ and the integral length scale $L(t)$, respectively; (d), (e), and (f) are, respectively, the hyperviscous-DNS counterparts of (a), (b), and (c).}
    \label{fig:p4} 
\end{figure}

To validate the use of hyperviscosity in a DNS of freely decaying turbulence, 
we consider the initial energy spectrum
\begin{align}
E(k, t=0) \propto k^4 \exp\left[-2\left(\frac{k}{k_p}\right)^2\right]\,,
\label{k4init}
\end{align}
with the peak wavenumber set to $k_p = 40$, which has been studied via DNSs of the 3DNSE with $\upbeta=1$, i.e., conventional viscosity [see, e.g., \cite{ishida2006decay,panickacheril2022laws}]. The phases of the velocity Fourier modes are uniformly distributed random variables, between $0$ and $2\pi$, independently and identically distributed (i.i.d.) for each velocity component\footnote{Here we do not take an average with respect to those initial phases, but present results for one realization of these phases.}. We set the initial total energy to $1/2$, namely, $E(t=0) = 1/2$;
furthermore, we take $\nu_1 = 2.0\times 10^{-4}$, in the viscous DNS [$\upbeta = 1$] or $\nu_2 = 8.97 \times 10^{-9}$ in our hyperviscous DNS [$\upbeta = 2$].
In the top row of Fig.~\ref{fig:p4} we present the results from our viscous DNS: Fig.~\ref{fig:p4} (a) shows log-log plots of the energy spectrum $E(k,t)$ versus the wavenumber $k$ at representative values of the time in the range $0 \leq t \leq 820$; Figs.~\ref{fig:p4} (b) and (c) display log-log plots versus $t$ of the total energy $E(t)$ and the integral length scale $L(t)$, respectively. Figures~\ref{fig:p4}(d), (e), and (f) are, respectively, the hyperviscous-DNS counterparts of Figs.~\ref{fig:p4} (a), (b), and (c). By comparing the plots in the top row of Fig.~\ref{fig:p4} with their counterparts in the bottom row, we see that the hyperviscous DNS yields cleaner scaling regions than the viscous DNS. 
In particular, the results for $E(t)$ and $L(t)$ [their precise definitions are given in Appendix \ref{sec:app_ns_int}] are closer to the expectations $E(t) \propto t^{-10/7}$ and  
$L(t) \propto t^{2/7}$ [based on the arguments given in \cite{1941-kolmogorov-b, comte-bellot_use_1966,1978-tatsumi--mizushima}].
This implies that, with limited spatial resolution, the DNS with hyperviscosity provides us a good method for studying freely decaying turbulence in the 3DNSE. Therefore, we use DNSs, with hyperviscosity ($\upbeta = 2$), to probe non-self-similar decay of $E(t)$ that is associated with multiple power-law regions in the initial 
energy spectrum $E(k,t=0)$ [cf. Section~\ref{subsec:res-burg-power-law} for 1D Burgulence].

Both our viscous and hyperviscous DNSs lead to substantial scaling ranges in Figs.~\ref{fig:p4} (b), (c), (e), and (f). However, they also indicate that the permanence of the large eddies (PLE)
breaks down in the following sense: if we follow the prefactor $A$ of the $k^4$ part in $E(k, t) = A k^4$, then we find that $A$ depends on time in both Figs.~\ref{fig:p4} (a) and (d).
This PLE is a key assumption in the early phenomenological treatment of energy decay in 3DNSE turbulence [see, e.g., \citet{comte-bellot_use_1966,1978-tatsumi--mizushima}\footnote{This reference also explored smaller scales, than the ones we consider here, and suggested a Kolmogorov-type spectrum (at those scales), which was subsequently revised by \cite{1980-frisch--schertzer}.}];
this breakdown can be understood by the non-conservation of the Loitsiansky invariant [as argued  by \citet{kida1997lagrangian} within a closure calculation].
In contrast, if we start with an initial spectrum $E(k,0) \sim k^2$ , this sort of the breakdown of the PLE is not observed, with both
the viscous and hyperviscous DNSs, reflecting the conservation of the associated Birkhoff-Saffman invariant [see, e.g., \cite{davidson2012freely,panickacheril2022laws}].

We note, in passing, that, in both Figs.~\ref{fig:p4} (a) and (d), the spectra $E(k,t)$ rise above the initial spectrum $E(k,t=0)$; this is the 3D NSE counterpart of the Gurbatov phenomenon [see \cite{1997-gurbatov--toth}].

\subsubsection{Two-power-law initial energy spectrum for Navier-Stokes turbulence}
\label{subsubsec:2PLNS}

We next consider an initial energy spectrum with the following two-power-law form:
\begin{align} 
    E(k, t = 0) \propto \begin{cases}
                            k^{n_1}   & 1 \le k \le k_1   \\
                            k^{n_2}  \exp[-2 (k / k_p)^2]  & k_1 \le k\,.
                        \end{cases}
 \label{eq:twopow}
\end{align}
Given our study of 1D Burgulence in Section~\ref{subsubsec:tworange}, we anticipate that the decay of $E(t)$ will not be  self-similar with the initial condition~\eqref{eq:twopow}.
In particular, $E(t) \propto t^{-\alpha}$ does not hold for all times; at intermediate times it crosses over from one power-law form to another [this can be viewed as an example of intermediate 
asymptotics in the sense of \cite{barenblatt1972self,barenblatt1996scaling}]. In our hyperviscous [$\upbeta = 2$] DNS, we use $512^3$ grid points and
we set $k_1 = 10$ and $k_p = 60$.

As in Section~\ref{subsubsec:2PLNS}, the phases of the Fourier coefficients of the initial velocity are taken to be uniformly distributed independent random variables between $0$ and $2\pi$. 
The initial total energy $E(t=0)=1/2$, the kinematic hyperviscosity is set to $\nu_2 = 1.66 \times 10^{-8}$, and for the exponents of the initial spectrum~\eqref{eq:twopow} we use the representative values $(n_1, n_2) = (1.5, 3.0)$.  [Results for other pairs $(n_1,n_2)$ are given in Appendix~\ref{subsec:appD2}.]
In this case, the \textit{na\"ive} prediction\footnote{The na\"ive theory, going back to \cite{1941-kolmogorov-b}, says that, if $E_0(k) \sim k^n$, then $E(t)\sim t^{F(n)}$, with $F(n)=\tfrac{-2(n+1)}{(n+3)}$ and $L(t) \sim t^{G(n)}$, with $G(n)=\tfrac{2}{(n+3)}$.
Therefore, if $n=3$, then $E(t)\sim t^{-4/3}$ and, if $n=1.5$, then $E(t)\sim t^{-10/9}$; the corresponding growth of the integral scale is, respectively, $L(t) \sim t^{1/3}$
and $L(t) \sim t^{4/9}$.} is an initial decay region with $E(t) \sim t^{-4/3}$ followed by another one with $E(t) \propto t^{-10/9}$,
the first because of the  $k^{3}$ part of $E_0(k)$  and the second arising from the $k^{1.5}$ part; the corresponding na\"ive power-law-growth regions in the integral scale are $L(t) \sim t^{1/3}$
and $L(t) \sim t^{4/9}$, respectively. 

\begin{figure}
 \centerline{%
 \includegraphics[width=0.35\columnwidth]{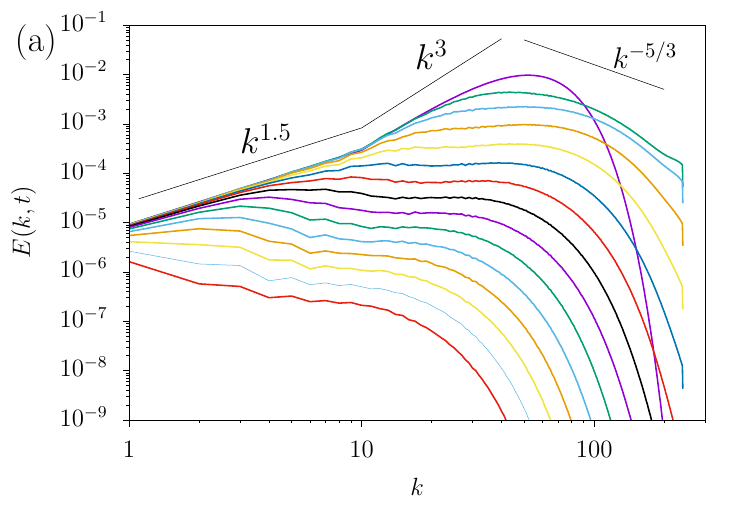} 
 \includegraphics[width=0.35\columnwidth]{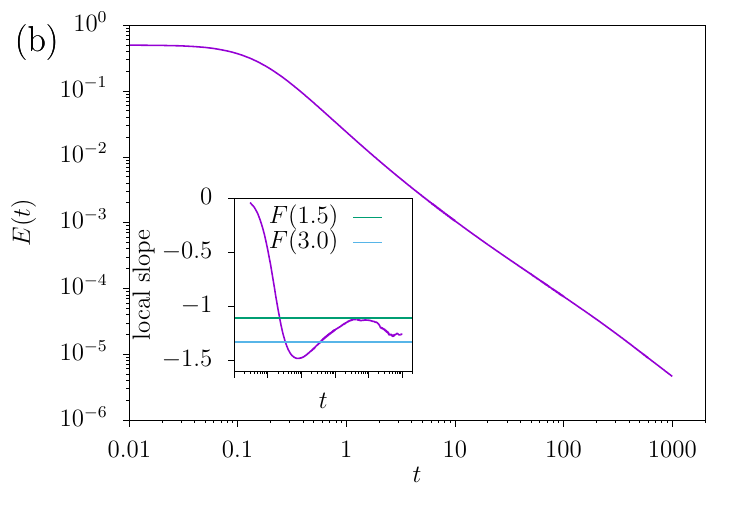}
 \includegraphics[width=0.35\columnwidth]{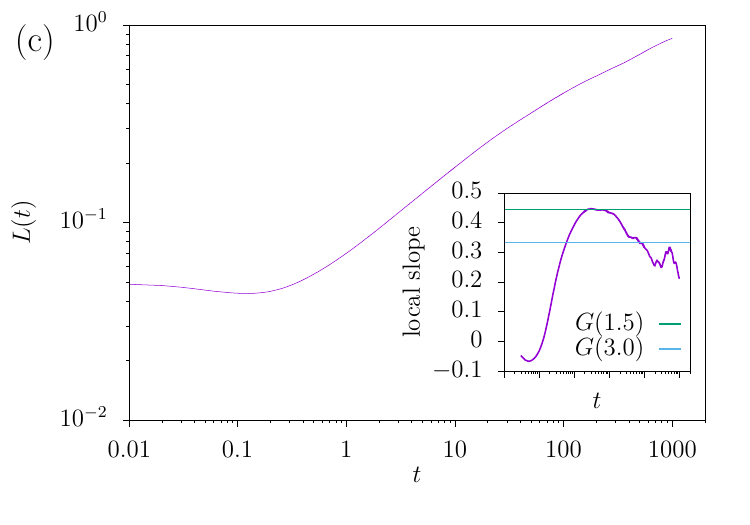}
 }
 \caption{Results for the decay of energy from our DNS for the hyperviscous 3DNSE with a two-power-law  intial energy $E(k, t=0) \propto k^{1.5}$ and $E(k, t=0) \propto k^3$, i.e., $(n_1, n_2) = (1.5,~3)$ in Eq.~\eqref{eq:twopow}. (a) Log-log plots of the energy spectrum $E(k,t)$ versus the wavenumber $k$ at representative values of the time in the range $0 \leq t \leq 820$; (b) and (c) display log-log plots versus $t$ of the total energy $E(t)$ and the integral length scale $L(t)$, respectively; in (b) and (c) logarithmic local slopes are given in the 
 insets, where $F(n) = -2 (n + 1)/(n + 3)$ is the na\"ive prediction for the power-law exponent for the decay of $E(t)$, if $E(k, t=0) \sim k^n$ and $G(n) = 2/(n + 3)$ is the corresponding  prediction for the growth exponent of the integral scale $L(t)$.}
    \label{p3ov2_3hv}
\end{figure}

We present the results of our hyperviscous DNS [$\upbeta = 2$] in Fig.~\ref{p3ov2_3hv}: Fig.~\ref{p3ov2_3hv} (a) shows log-log plots of the energy spectrum $E(k,t)$ versus the wavenumber $k$ at representative values of the time in the range $0 \leq t \leq 820$; Figs.~\ref{p3ov2_3hv} (b) and (c) display log-log plots versus $t$ of the total energy $E(t)$ and the integral length scale $L(t)$, respectively. To uncover possible power-law regimes in the log-log plots of the total energy $E(t)$ and $L(t)$, in Figs.~\ref{p3ov2_3hv} (b) and (c), respectively, we plot
logarithmic local slopes in the insets of these figures. In the inset of Fig.~\ref{p3ov2_3hv} (b), we see that this slope 
first goes below the na\"ive prediction, for the decay exponent at early times, namely, $-4/3$, and then approaches  the na\"ive prediction, for the decay exponent at late times, namely, $-10/9$. 
After this crossover region, the log-slope inset shows a narrow plateau around $-10/9$ and finally departs 
from it when the peak of the energy spectrum reaches the smallest wavenumber [cf. the spectrum $E(k,t)$ in Fig.\ref{p3ov2_3hv} (a)]; this departure occurs when
the second power-law regime $k^{1.5}$ is lost as $E(k,t)$ evolves in time. The decay of $E(t)$ is somewhat consistent with the na\"ive 
prediction, if we interpret the first turn-over as a plateau around the na\"ive decay-exponent value $-4/3$ for the $k^{3}$ part in the initial energy spectrum $E(k,t=0)$. 
However the integral scale $L(t)$ does not show a turn-over near the first exponent $1/3$ for $k^{3}$, but exhibits a plateau at the second na\"ive growth exponent $4/9$ for the $k^{1.5}$
part in the initial energy spectrum $E(k,t=0)$.

Furthermore, the extents in time of the plateaux in the logarithmic local slopes in $E(t)$ and $L(t)$ [in the insets of Figs.~\ref{p3ov2_3hv} (b) and (c)] do not coincide well with
each other. The departure from the na\"ive expectation $E(t) \sim t^{-10/9}$ occurs between $100 < t < 200$, as can be surmised from the temporal evolution of $E(k,t)$ in Figs.~\ref{p3ov2_3hv} (a).
Specifically, the energy spectrum at $t = 102$, which is the fourth curve 
from below in Fig.\ref{p3ov2_3hv} (a), does not have the $k^{1.5}$ part 
in the low-wavenumber region. In contrast, the departure from the na\"ive prediction $L(t) \sim t^{4/9}$
occurs about one decade earlier than its counterpart for $E(t)$. This may be caused by the bottleneck
effect~\citep{frisch2008hyperviscosity}, which is enhanced by the hyperviscosity, as can be seen from the nearly flat regions of the energy 
spectra at intermediate times.

In summary, then, our DNSs with hyperviscosity [$\upbeta =2$] have helped us to unveil the non-self-similar decay of $E(t)$ and growth of $L(t)$
for the two-power-law initial energy spectrum~\eqref{eq:twopow}. The crossover from one decay or growth exponent to another is subtle and 
may be viewed as an example of intermediate asymptotics \textit{\`a la} \cite{barenblatt1972self} and \cite{barenblatt1996scaling}.
To uncover these crossovers completely is a challenging numerical problem because  high spatial resolution is required to achieve sufficient scale separation between
different power-law regimes in the wavenumber $k$, and we must carry out very long runs.

\section{Conclusions}
\label{sec:conc}

We have discussed freely decaying turbulence in 1D Burgulence and 3D Navier-Stokes (NS) turbulence. Our studies have been designed to explore how different types of initial conditions that lead to non-self-similar temporal decay of the energy $E(t)$ and, hitherto unanticipated, large-scale multifractality.

 We have first investigated the decay of the energy $E(t)$ in 1D Burgulence, for a fractional Brownian motion (FBM) initial potential, with Hurst exponent $H$. We have then given the first rigorous proof that $E(t)\sim (t/t_*)^{-\frac{2-2H}{2 - H}}$, with $t_*$ is any positive reference time; furthermore, we have established the boundedness of $E(t)$ for all $t > 0$, a nontrivial result given that the intial datum is an FBM. Next, we have introduced a new type of FBM that we call an oblivious fractional Brownian motion (OFBM$_H$), with Hurst exponent $H$. We have proved that 1D Burgulence, with an OFBM$_H$ initial potential $\varphi_0(x)$, exhibits intermittency and large-scale bifractality or multifractality, which we have uncovered via the exponents $s(m)$ that follow from $\langle|u(x,t)|^{m}\rangle \sim t^{s(m)}$  [see Eq.~\eqref{eq:smdef}]. Multifractality is proved to occur if $H$ changes from one oblivious interval to another [see Section~\ref{subsec:OBHM}]. We expect that OFBMs will have applications in other fields of physics, chemistry, biology, and finance; we will explore this in future work.
 
 We have provided the first rigorous proof of genuine multifractality for turbulence in a nonlinear hydrodynamical partial differential equation (PDE);
 the specific PDE we consider is the 1D Burgers equation. We emphasize that the large-scale multifractality we have uncovered is non-universal, in as much as it depends on the initial condition; by contrast, conventional small-scale multifractlity [see, e.g., \cite{1995-frisch-book}]
 is universal. Multifractality has been proven in the Kraichnan model for passive-scalar advection [see, e.g.,~\cite{falkovich2001particles}]; 
 however, in this passive-scalar problem, the advection-diffusion equation is \textit{linear} and the statistics of the advecting velocity field are specified. Earlier studies of 1D Burgulence have obtained the exponents $\zeta_p$ analytically, but these results have always led to bifractality [see, e.g.,~\cite{vergassola1994burgers}, \cite{frisch2002burgulence}, and \cite{bec2007burgers}]. A mathematical proof of multifractality in 3D Navier-Stokes turbulence remains a challenging open question.

We  have then explored non-self-similar decay via DNSs of freely decaying 1D Burgulence, with the following initial data: 
\begin{itemize}
\item(A) $\varphi_0(x)$ a multifractal random walk, for which we have developed a spatially periodic generalisation of the multifractal random walk of ~\cite{bacry2001multifractal}, which crosses over to an FBM for lengths greater than a prescribed crossover scale $\mathcal{L}$ [see Section~\ref{subsec:mfracinit}].  The decay [growth] of $E(t)$ [$L(t)$] is non-self-similar; the multifractality of the initial condition persists, but, at very long times, the energy decays with the power-law exponent associated with the simple scaling for a fractional Brownian motion with Hurst exponent $H$ [see Section~\ref{subsec:brownian-motion}];  indeed, by tuning the value of $\mathcal{L}$, the crossover from small-scale to large-scale multifractality becomes feasible~\footnote{We note that if $\mathcal{L}_B$ is the linear size of the domain, of course, $ \mathcal{L} < \mathcal{L}_B$. Eventually, we are interested in the limit $\mathcal{L}_B \to \infty$, which can be taken in the following two ways: (i) $\mathcal{L}_B \to \infty$ with $\mathcal{L}$ held fixed $\mathcal{L}/\mathcal{L}_B \to 0$, so we only have small-scale multifractality; (ii) both $\mathcal{L}_B$ and $\mathcal{L}$ tend to infinity, such that the ratio $\mathcal{L}/\mathcal{L}_B$ goes to a finite, nonzero constant, so we can make $\mathcal{L}$ large enough to get large-scale multifractality; indeed, by tuning the value of this ratio, the crossover from small-scale to large-scale multifractality becomes feasible.}.
\item (B) Initial energy spectra $E_0(k)$ with one or more power-law regions, as a function of the wavenumber $k$, which lead, respectively, to self-similar and non-self-similar decay of $E(t)$ with time. If any one of these power-law exponents is greater than $1$, then the evolving spectrum exhibits a Gurbatov-type effect [see \cite{1997-gurbatov--toth}] with $E(k,t)$ rising above $E(k,0)$ in some ranges of $k$ and $t$. The logarithmic corrections associated with the Gurbatov effect [see Appendix~\ref{sec:app_burg}] are not easy to obtain in a DNS that has limited spatial and temporal resolution.
\end{itemize}

We have then extended these to the 3D viscous and hyperviscous NS equations. Our hyperviscous DNSs enable us to obtain $E(t)$, for initial energy spectra $E_0(k)$, with either one power law or two power laws. The former leads to self-similar decay of $E(t)$ and the latter to non-self-similar decay of $E(t)$ and the corresponding growth of $L(t)$.  
The evolution of the energy spectra, the decay of $E(t)$, and the growth of $L(t)$ are qualitatively similar to their counterparts in the 1D Burgers case discussed in Section~\ref{subsec:res-burg-power-law}.
We also obtain Gurbatov-type phenomena [see Appendix~\ref{sec:app_burg} for the 1D Burgulence counterpart].

Earlier suggestions of non-self-similar decay of $E(t)$ are based on the EDQNM closure. In particular, \cite{eyink2000free} have suggested that, with a steep power-law region in the initial spectrum $E_0(k)$, a Gurbatov-type non-self-similar decay could occur in the 3DNSE. The EDQNM study of \cite{2012-meldi-sagaut}, with two-power-law initial energy spectra $E_0(k)$,  also yields non-self-similar decay of $E(t)$ and Gurbatov-type phenomena. It is interesting to note that closure schemes, e.g., EDQNM, can capture the non-self-similar decay of $E(t)$. Of course, such closures cannot capture either small-scale or large-scale multifractality.

We end with a discussion of the possibility of investigating -- theoretically, numerically, and experimentally -- large-scale multifractality in freely decaying or forced, statistically steady NS and MHD [see, e.g., \cite{kalelkar2004decay}] turbulence. 
In our discussion of large-scale multifractality in Section~\ref{subsec:OBHM}, we have worked with the Lagrangian variable $a$. Therefore, when we try to look for signatures of large-scale multifractality in experiments or numerical studies, a Lagrangian framework might well prove to be useful. Our study of freely decaying 3D NS turbulence in Section~\ref{sec:ns-hyper-vis} has, so far, used an Eulerian description. In future work we will extend this by tracking Lagrangian particles. In Burgulence, Lagrangian particles get trapped at shocks, so we must take this into consideration [see, e.g., \cite{de2023dynamic} and \cite{de2024uncovering}].

It was shown by~\cite{frisch1975possibility} [for a recent overview see~\cite{alexakis2018cascades}], that there are good reasons to believe that an injection at intermediate wavenumbers of magnetic helicity could  drive an inverse cascade of magnetic helicity. It is important to investigate under which conditions this inverse cascade might display large-scale intermittency.

We end with suggestions for experiments that might be performed to examine large-scale multifractality in freely decaying turbulence. The natural way to design such experiments would be to begin with earlier studies of decaying turbulence in wind tunnels with fractal grids [see, e.g., \cite{krogstad2011freely} and \cite{valente2011decay}] and then generalise them using \textit{multifractal grids}. The simplest realization of such grids could employ the algorithm that we have used in Section~\ref{subsec:mfracinit} to obtain a multifractal initial condition for the initial potential in the 1D Burgers equation, where the crossover length $\mathcal{L}$ can be tuned to move from small-scale to large-scale multifractality. It could well turn out that Lagrangian measurements might be best suited to uncover large-scale multifractality.


\backsection[Funding]{ UF, KK, RP, and TM were partially supported by the French ministry of education.  TM is funded by Grants-in-Aid for Scientific Research KAKENHI (C) No.~19K03669 from JSPS. UF is also grateful to the Universit\'e C\^ote d'Azur for support. RP and DR thank
the Anusandhan National Research Foundation (ANRF), the Science and Engineering Research Board (SERB), and the National Supercomputing Mission (NSM), India, for support,  and the Supercomputer Education and Research Centre (IISc), for computational resources.}

\backsection[Declaration of interests]{The authors report no conflict of interest.}

\appendix
\section{}
\label{sec:AppA}

\subsection{Proof of the boundedness of $E(t)$ for $t > 0$}
\label{subsec:AppA}

In Sec.~\ref{sec:finite_energy} we had outlined the proof of the boundedness of $E(t)$ for $t > 0$. We give below the details of this proof for any finite time $t$, e.g., at $t=1$.

In Eq.~\eqref{eq:finite-energy-1}, we had for an arbitrary fixed  $U\ge 0$
\begin{equation}
\langle u^2(0,1)\rangle\leq U^2 {\mathcal P}\left( |u(0,1)|\leq U \right) + \sum_{k=1}^\infty{(2^kU)^2\mathcal P\left(2^{k-1}U < |u(0,1)|\leq 2^kU\right)}\,,
\label{eq:finite-energy-1A}    
\end{equation}
with $\mathcal P$ the probability distribution corresponding of the random initial condition $\varphi_0(a)=W_H(a)$, namely, the fractional Brownian walk with Hurst exponent $H$. We had divided the range of values of $|u(0,1)|$ into the initial interval $[0,U]$ and the sequence of intervals $[2^{k-1}U,\; 2^kU]\,, \rm{for} \; {\rm{integers}}\; k\geq 1$. We had used, for $|u(0,1)|$, an upper bound $U$, in the initial interval, and the bounds $2^kU$, in the intervals with  $k\geq 1$ . The squares of these upper bounds appear in the estimate~\eqref {eq:finite-energy-1A}, in the corresponding interval.
Recall that the Lagrangian coordinate $a=a_{x,t}$, corresponding to the location $x$ at time $t$, gives the location at $t=0$ corresponding to the maximum of $\left(\varphi_0(a) - \frac{(x-a)^2}{2t}\right)$ [see the max formula~\eqref{eq:max}]; and $a=x-tu(x,t)$, for a fixed $t$, is the inverse of the Lagrangian map from $a$ to $x$. 
In particular, if $t=1$, then the estimate $2^{k-1}U < |u(0,1)|\leq 2^kU$ implies that the Lagrangian coordinate $a$,
corresponding to $x=0$ at time $t=1$, satisfies the same estimate $2^{k-1}U <|a|=|u(0,1)| \leq 2^kU$. Since $\varphi(0,1)= \varphi_0(a) - a^2/2$, and $\varphi(0,1)$ corresponds to maximizing over all $a$, we have $\varphi_0(a) - a^2/2\geq \varphi(0,0)=0$. Hence, $\varphi_0(a)\geq a^2/2> (2^{k-1}U)^2/2$. 
It follows that, for the last term in Eq.~\eqref{eq:finite-energy-1A}, 
\begin{equation}
\begin{aligned}
\mathcal P\left(2^{k-1}U< |u(0,1)|\leq 2^kU\right)  \leq \mathcal P\left(\max_{-2^kU\leq a \leq 2^kU}{\varphi_0(a)}\geq \frac{1}{2}(2^{k-1}U)^2\right) \\
\leq \mathcal P\left(\max_{-2^kU\leq a \leq 0}{\varphi_0(a)}\geq \frac{1}{8}2^{2k}U^2\right) + \mathcal P\left(\max_{0\leq a \leq 2^kU}{\varphi_0(a)}\geq \frac{1}{8}2^{2k}U^2\right)\\
 = 2\mathcal P\left(\max_{0\leq a \leq 2^kU}{\varphi_0(a)}\geq \frac{1}{8}2^{2k}U^2\right).   
\end{aligned}
\label{eq:finite-energy-2}
\end{equation}
Using the exact scaling invariance of $\varphi_0(x)$, we have
\begin{equation}
\mathcal P\left(\max_{0\leq a \leq 2^kU}{\varphi_0(a)}\geq \frac{1}{8}2^{2k}U^2\right)=\mathcal P\left(\max_{0\leq a \leq 1}{\varphi_0(a)}\geq \frac{1}{(2^kU)^H}\frac{1}{8}2^{2k}U^2\right).
\label{eq:finite-energy-3}
\end{equation}

We now exploit the following result of~\cite{1978-piterbarg-prisyazhnyuk}:
\begin{equation}
\mathcal P\left(\max_{0\leq a \leq 1}{\varphi_0(a)}\geq \mathcal{M}\right)\sim C_0\mathcal{M}^\mathfrak{h}\int_\mathcal{M}^\infty{e^{-x^2/2}dx}\,,
\label{eq:finite-energy-4}    
\end{equation}
where $\mathfrak{h}=\max\{(1/H-2)\,, 0\}$. 
This result provides the exact asymptotic behaviour of the probability that the maximum of the fractional Brownian motion, on the interval $[0,1]$, exceeds 
the large level $\mathcal{M}>0$. In fact, we just need an estimate from above for the probability $\mathcal{P}\left(\max_{0\leq a \leq 1}{\varphi_0(a)}\geq \mathcal{M}\right)$.  It follows from the 
relation~\eqref {eq:finite-energy-4} that there exists $\bar{\mathcal{M}}$ such that, for all $\mathcal{M}\geq \bar{\mathcal{M}}$, the following estimate holds:
\begin{equation}
\mathcal P\left(\max_{0\leq a \leq 1}{\varphi_0(a)\geq \mathcal{M}}\right)\leq C_1\mathcal{M}^\mathfrak{h}e^{-\mathcal{M}^2/4}\,,
\label{eq:finite-energy-5} 
\end{equation}
where $C_1=3C_0\sqrt{\pi}$. 
If we denote 
\begin{equation}
M_k=\frac{1}{8(2^kU)^H}2^{2k}U^2=\frac{1}{8}U^{2-H}2^{(2-H)k}\,, 
\label{eq:finite-energy-7}
\end{equation} 
and assume that $U\geq \frac{1}{2}(8\bar{\mathcal{M}})^{1/(2-H)}$, then, for all $k\geq 1$, we have $M_k\geq\bar{\mathcal{M}}$.
Then, using Eqs.~\eqref {eq:finite-energy-1A}, \eqref {eq:finite-energy-2}, \eqref {eq:finite-energy-3}, \eqref {eq:finite-energy-5}, we obtain
\begin{equation}
 \langle u^2(0,1)\rangle\leq U^2 + C_1\sum_{k=1}^\infty{(2^kU)^2 M_k^\mathfrak{h}e^{-M_k^2/4}}\,. 
 \label{eq:finite-energy-6}
\end{equation}
It follows that the series in Eq.~\eqref {eq:finite-energy-6} converges, given that the term $e^{-M_k^2/4}$ dominates $2^{2k}M_k^\mathfrak{h}$.
Hence, boundedness of energy at time $t=1$ is established.

\section{}
\label{sec:AppB}

\subsection{Oblivious Fractional Brownian Motion}
\label{subsec:AppB}

In this Appendix we give the details of the construction of initial potentials  that lead to large-scale multifractality in freely decaying 1D Burgulence,
which we had discussed briefly in Section~\ref{subsec:OBHM}. First we will construct initial potentials that lead to large-scale \textit{bifractality}; then we will generalise this construction to obtain an initial potential that leads to genuine large-scale multifractality.

Consider a sequence of independent identically distributed intervals (iid) in the Lagrange variable $a$. We assume that the length $l$ of the intervals has the probability distribution function (PDF) $p(l)$, where 
$p(l)\sim l^{-\gamma}$ as $l\to \infty$, with the tail exponent $\gamma>1$. We are interested in random initial potentials with stationary increments,
so we must ensure  that the point process, corresponding to the end points of the intervals, is translationally invariant. This can be achieved by the following procedure: 
We start with some large negative $a=-L$, and then begin adding iid intervals, sampled according to the PDF $p(l)$ in the positive direction. 
In the limit $L\to \infty$, the starting point $a=-L$ plays no role, so we obtain, in this limit, a translationally invariant point field of the endpoints of the intervals. Although the above construction is conceptually correct, it is not easy to implement numerically. Below we describe a better way to achieve our goal of constructing a
translationally invariant point field of the endpoints of the intervals. It can be shown that, for a fixed non-random point, the distribution of the length of the interval containing this point is different from the PDF $p(l)$. Indeed, it is more probable that long intervals contain a given point; the corresponding PDF is proportional to 
$lp(l)$. 

Now, the construction of the translationally invariant point field can be described as follows. We first sample the length $l_0$ of the interval $\Delta_0$, containing the origin $a=0$, using its PDF, which is proportional to $lp(l)$. Then we sample the location of the origin uniformly within the interval of length $l_0$, i.e.,
we choose $\Delta_0=[-\epsilon, l_0-\epsilon]$, where $\epsilon$ is distributed uniformly in $[0, l_0]$. 
Next, we add intervals $\Delta_{-i}, \; i>0$ and $\Delta_i, \; i>0$ to the left and to the right of $\Delta_0$. The length of each interval $\Delta_i, \; i\neq 0$ is an independent random variable with the distribution given by the PDF $p(l)$; this construction can be carried out only if the exponent $\gamma>2$; otherwise, $\int_0^\infty{lp(l)dl}=+\infty$ and a probability distribution with the PDF proportional to $lp(l)$ does not exist.

We now construct the initial potential $W(a), \, a\in \mathbb R^1$.  In each of the intervals $\Delta_i, \; i \in \mathbb{Z}$, we choose an independent realization of 
a Fractional Brownian Motion (FBM) with the Hurst exponent $H$. Notice that these FBMs are not extended beyond $\Delta_i$. For the interval $\Delta_0$, we assume that the FBM starts at the origin; for all other $\Delta_i$, we assume that it starts at 
the leftmost point of $\Delta_i$ for positive $i$, and at the rightmost point of $\Delta_i$ for negative $i$.
Our potential $W(a)$ must be continuous, so we next move the
FBMs inside $\Delta_{-1}$ and $\Delta_1$ vertically, so that the values at the end points of $\Delta_0$ are matched [see Fig.~\ref{fig:orw_schematic}].
We repeat this matching process for the intervals $\Delta_{-i}$ and $\Delta_i$ consequently for $i= 2, \, 3, \dots$. The process constructed above is exactly the one that we call an Oblivious Fractional Brownian Motion  with the Hurst exponent $H$ ($\text{OFBM}_H$).

\subsection{Large-scale bifractality}
\label{subsec:AppBbifrac}

{\bf {Case A ($0<H<1/2$).}} In the case $0<H<1/2$  we are  interested in the averages of the inverse powers of the speed, i.e., $\langle|u(x,t)|^{m}\rangle$ for negative $m$. 
Consider first the contribution from the typical events, when the intervals $\Delta_i$ are not anomalously long. To estimate the variance of $W(L)$,
as $|L|\to \infty$, we note that the average length of intervals $\Delta_i$ is finite because $\gamma>2$ . Hence, the total
length of the union of intervals $\Delta_i, \, i\in [-n,n]$ is of the order of $n$. The FBMs inside each one of the intervals  $\Delta_i$ are independent, so the 
variance $\langle W^2(L)\rangle$ is of the order of
\begin{equation}
    \label{variance_W(a)A}
 \sum_{0\leq i \leq n} {|\Delta_i|^{2H}},
\end{equation}
where $|L|$ and $n$ are of the same order. It can be shown that the PDF for $|\Delta|^{2H}$ is
\begin{equation}
\frac{p(l^{1/2H})l^{1/2H}}{2Hl}\,,\;{\rm{which\;decays\; as}} \;\; \sim \frac{1}{l^{1+(\gamma-1)/2H}}, \;\;{\rm{as}} \;\; l\to\infty\,.
\label{eq:PDFDeltaA}
\end{equation}

Below we will use the following well-known asymptotic formula [see, e.g., ~\cite{feller1991introduction}] for sums of positive independent identically distributed (iid)
random variables with heavy-tailed PDFs. Let $\xi_i, \, i \in \mathbb{N}$ be positive iid random variables, with their PDF decaying as $\xi^{-(1+\tau)}$. Then
\begin{equation}
    \label{asymptotic_heavy_tailsA}
\sum_{1\leq i\leq n}{\xi_i}\sim 
\begin{cases}
n^{\frac{1}{\tau}}, \ \ \text{if} \ \  \tau<1\\
n\log{n}, \ \ \text{if} \ \ \tau=1\\
n, \ \ \text{if} \ \  \tau>1 .
\end{cases}
\end{equation}
Note that the condition $\tau > 1$ corresponds to the case when $\langle\xi\rangle$ is finite.
We now use the asymptotic relation~\eqref{asymptotic_heavy_tailsA},  with $\tau = (\gamma-1)/2H$. We are considering the case $0 < H <1/2$, so we get $\tau = (\gamma-1)/2H>1$, and hence, 
\begin{equation}
    \label{asymptotic_sum_H2A}
\sum_{0\leq i\leq n} {|\Delta_i|^{2H}}\sim n\,;
\end{equation}
since the total length is also of the order $n$, the relation~\eqref{asymptotic_sum_H2A} implies that $\langle W^2(L)\rangle$, namely,
the variance of $W(L)$, scales as $|L|$ as $|L| \to \infty$. This relation allows us to find the asymptotic behaviour of the Lagrangian coordinate $a=L(t)$, corresponding to space-time location $(x=0,t)$. Note that the order of $L(t)$ is determined by the relation
\begin{equation}
    \label{L(t)}
t\,\left(\frac{L(t)}{t}\right)^2\sim \sqrt{\langle W^2(L(t))\rangle} \sim \sqrt{|L(t)|}\,,
\end{equation}
from which it follows that
\begin{equation}
    \label{L(t) + u(0,t)}
|L(t)| \sim t^{\frac{2}{3}}, \, \, \, |u(0,t)|=\frac {|L(t)|}{t}\sim t^{-\frac{1}{3}}\,,
\end{equation}
and finally we get
\begin{equation}
    \label{main_event_1A}
|u(0,t)|^{m}\sim t^{-\frac{m}{3}},
\end{equation}
which gives the order of the contribution to $\langle|u(0,t)|^{m}\rangle, \, m>-1$ coming from the \textit{typical events} mentioned above. 

Next we consider the case when the interval $\Delta_0$ is so long that the velocity at the origin is determined by the FBM with fixed Hurst exponent $H$ \textit{inside this interval}. 
We denote by $q_H(u,1)$ the PDF for the velocity $u(0,1)$ at time $t=1$. It is easy to show that $q_H(u,1)$ is a continuous function
that tends to a positive constant as $u\to 0$. This PDF decays rapidly as $|u| \to \infty$, namely,
\begin{equation}
\label{eq:qHA}
-\log{q_H(u,1)}\sim \frac{1}{4}|u|^{4-2H}\,.
\end{equation}
To see why Eq.~\eqref{eq:qHA} holds, we again use the variational principle~\eqref {eq:max}. Let $u$ be a large positive velocity. If we denote by $P_\epsilon(u)$ the probability that $(1-\epsilon)u\leq~u(0,1)\leq~(1+\epsilon)u$, then
\begin{equation}
P_\epsilon(u)\leq \mathcal P\left(\max_{0\leq a \leq (1+\epsilon)u}{\varphi_0(a)}\geq \frac{(1-\epsilon)^2u^2}{2}\right)\,.
\label{eq:estimate_1}    
\end{equation}
In fact, it is easy to see that $P_\epsilon(u)$ and $\mathcal P\left(\max_{0\leq a \leq (1+\epsilon)u}{\varphi_0(a)}\geq \frac{(1-\epsilon)^2u^2}{2}\right)$ are of the same order.
From the scaling invariance of $\phi_0(a)$ we obtain
\begin{equation}
\mathcal P\left(\max_{0\leq a \leq (1+\epsilon)u}{\varphi_0(a)}\geq \frac{(1-\epsilon)^2u^2}{2}\right)=\mathcal P\left(\max_{0\leq a \leq 1}{\varphi_0(a)}\geq \frac{1}{2}\frac{(1-\epsilon)^2u^2}{((1+\epsilon)u)^H}\right)\,.
\label{eq:iquality_2}    
\end{equation}
Using the asymptotic relation~\eqref{eq:finite-energy-4} we get
\begin{equation}
- \log{P_\epsilon(u)}\sim \frac{(1-\epsilon)^4}{4(1+\epsilon)^{2H}}u^{4-2H}\,,
\label{eq:estimate_3}
\end{equation}
and thence, in the limit $\epsilon \to 0$, we obtain Eq.~\eqref{eq:qHA}. The case of large negative values of $u(0,1)$ can be considered in a similar way.
 
We now use the exact scaling relation 
\begin{equation}
u(x,t)=u(x,1)t^{-\alpha(H)}\,,\;\; {\rm{where}}\;\; \alpha(H)= (1-H)/(2-H)\,,
\label{eq:alphaA}
\end{equation}
which should be understood in the distributional sense\footnote{That is, having the same probabilistic law in the sense used in Eqs.~\eqref{eq:lambda-invariance} and ~\eqref{eq:velocity-at-time-t-in-terms-of-velocity-at-time-unity} for conventional fractional Brownian motion.},
to get the following expression for the PDF of the velocity field $u(0,t)$ at time $t$:
\begin{equation}
    \label{pdfattimet}
  q_H(u,t)=t^{\alpha(H)}q(ut^{\alpha(H)},1).  
\end{equation}
It follows that 
\begin{equation}
    \label{eq:negative_moments}
  \langle |u(0,t)|^{m}\rangle=\int_{-\infty}^{\infty}{|u|^{m}q_H(u,t)du}=t^{\alpha(H)m}\int_{-\infty}^{\infty}{|u|^{m}q(u,1)du},  
\end{equation}
which, in particular, implies that negative moments $\langle |u(0,t)|^{m}\rangle$ are finite for $-1<m<0$. Also, since the speed
$|u(0,t)|$ scales like $t^{-\alpha(H)}$, we have 
\begin{equation}
\label{eq:asyL}
|L(t)|\sim t^{1-\alpha(H)}=t^{1/(2-H)}\,.
\end{equation}
If $|\Delta_0|$ is much larger than $|L(t)|$, then, with large probability, $L(t)$ will be inside $\Delta_0$. It is easy to see
that, in this case, the speed $|u(0,t)|$ will, indeed, scale like 
$t^{-\alpha(H)}$. To estimate the probability of such an event we note that the PDF of the length of $\Delta_0$ is given by $p_0(l)=\frac{1}{M}lp(l)$, where $M=\int_0^{\infty}lp(l)dl$. Thus, we have
\begin{equation}
    \label{eq:long_Delta_0}
  \mathcal {P}(|\Delta_0|>t^{\frac{1}{2-H}}) =\frac{1}{M}\int_{t^{\frac{1}{2-H}}}^\infty{lp(l)dl}\sim \frac{\gamma-2}{M}\frac{1}{t^{\frac{\gamma-2}{2-H}}}\,, 
\end{equation}
whence it follows that the contribution to $\langle |u(0,t)|^{m}\rangle$, coming from such an event with a long interval $\Delta_0$, is 
\begin{equation}
\label{main_event_2}
 |u(0,t)|^{m}\sim t^{-\alpha(H)m-(\gamma-2)/(2-H)}\,.
\end{equation}

In order to establish multifractality, we are interested in the behaviour of the scaling exponent $s(m)$ that is defined by the scaling relation
\begin{equation}
    \langle|u(0,t)|^{m}\rangle \sim t^{s(m)}\,.
    \label{eq:smdefA}
\end{equation}
From the Eqs.~\eqref{main_event_1A} and \eqref{main_event_2}, we conclude that 
\begin{equation}
    s(m)\geq \max{[\{-\frac{m}{3}\}\,, \, \{-\alpha(H)m-(\gamma-2)/(2-H)\}]}\,.
    \label{eq:smformulaA}
\end{equation}
Note that  $\alpha(H)-(\gamma-2)/(2-H)>1/3$, for the tail exponent in the range  $2<\gamma<2+ \frac{1-2H}{3}$, so we can use Eq.~\eqref{eq:alphaA} to show
that $s(m)$ is not a linear function of $m$.
In summary, this result is obtained as follows: (a) Typical events provide the dominant contribution in the case of small $|m|$, so we conclude that the scaling exponent $s(m)$ behaves as $-\frac{m}{3}$ in the limit when $m\to 0$; (b) by contrast, the rare events [see Eqs.~\eqref{eq:negative_moments} and \eqref{eq:long_Delta_0}] lead to the contribution $t^{-\alpha(H)m-(\gamma-2)/(2-H)}$, so $s(m)\geq~-\alpha(H)m-(\gamma-2)/(2-H)$. 
Hence, for such values of $\gamma$ and for $m$ close enough to $-1$, we have $s(m)>-\frac{m}{3}$. It follows that $s(m)$
is not a linear function which is a manifestation of the multifractal
nature of $\langle|u(0,t)|^{m}\rangle \sim t^{s(m)}$ for large $t$. 
The terms $-\frac{m}{3}$ and $-\alpha(H)m-(\gamma-2)/(2-H)$ provide the dominant contributions in the whole range 
of the scaling exponent $m>-1$, so $s(m)=\max{\{-\frac{m}{3}, \, -\alpha(H)m-(\gamma-2)/(2-H)\}}$. Hence, for $2<\gamma<2+ \frac{1-2H}{3}$ we have 
\begin{eqnarray}
    s(m)&=&-m/3\;\; {\rm{for}} \;\; m\geq m_A(H,\gamma)\equiv-\frac{3(\gamma-2)}{1-2H}\,;\nonumber\\
    s(m)&=&-\alpha(H)m-(\gamma-2)/(2-H) \;\; {\rm{for}} -1<m<m_A(H,\gamma)\,. 
    \label{eq:smbifractalA}
\end{eqnarray}
This is an example of bifractal scaling, because $s(m)$ is a piecewise linear function of $m$; later we will show how to generalise the OFBM$_H$, by introducing a range of Hurst exponents $H$, to obtain genuine multifractality.

{\bf Case B ($1/2<H<1,\; \tau= (\gamma-1)/2H >1$).} Our construction and analysis in the second case, with $1/2<H<1$, is very similar to our discussion for Case A above.
Again, we consider $\text{OFBM}_H$ and assume that $\gamma>2$ and $\tau=(\gamma-1)/2H>1$; this inequality is satisfied if $\gamma>1+2H$. Since $\tau>1$, all
asymptotic behaviours in the main probability event remain the same as in Case A, so $|u(0,t)|$ scales as $t^{-1/3}$. Our estimates in the case of a long $\Delta_0$ are also unchanged. The speed $|u(0,t)|$ scales as $t^{-\alpha(H)}$ if $|\Delta_0|\sim t^{1/(2-H)}$. Hence, the contribution to $s(m)$ remains the same as in Case A, namely, 
$-\frac{m}{3}$ and $-\alpha(H)m-(\gamma-2)/(2-H)$. The only difference is that we now have $1/2<H<1$, so $\alpha(H)<1/3$. It follows that, in the case of long $\Delta_0$, the speed  $|u(0,t)|$ is larger than in the main probability event. Hence, there exists $m_B(H,\gamma)>0$ such that $s(m)=-\alpha(H)m-(\gamma-2)/(2-H)$ gives the dominant contribution for $m>m_B(H, \gamma)$, while the term $-\frac{m}{3}$ dominates for $-1<m \leq m_B$. An easy calculation gives $m_B(h,\gamma)=\frac{3(\gamma-2)}{2H-1}$; and
since $\gamma>1+2H$, the exponent $m_B(h,\gamma)$ is greater than 3. Finally, we get
\begin{eqnarray}
    s(m)&=&-m/3\,, \;\; {\rm{for}} \;\; -1 < m \leq m_B(H, \gamma) \,;\nonumber\\
    s(m)&=&-\alpha(H)m-(\gamma-2)/(2-H) \,,\;\; {\rm{for}} \;\; m > m_B(H, \gamma)\,. 
    \label{eq:smbifractalB}
\end{eqnarray}
Again, this is an example of bifractal scaling.

{\bf Case C ($1/2<H<1;\; \tau= (\gamma-1)/2H <1$).} 
In Case B above, we assumed $1/2<H<1$ and $\tau>1$. Let us now consider the last possible case when $1/2<H<1$ and $\tau<1$, so $2<\gamma<1+2H$. Our analysis 
in the case of a long interval $\Delta_0$ remains unchanged; again, the contribution to $s(m)$ remains $-\alpha(H)m-(\gamma-2)/(2-H)$.
However, the asymptotic analysis for the main probability event is different. Since $\sum_{i=1}^n{|\Delta_i|^{2H}}\sim n^{1/\tau}$ and
$L(t)\sim \sum_{i=1}^n{|\Delta_i|}\sim n$, we have
\begin{equation}
    \label{L(t)_case_3A}
t\left(\frac{L(t)}{t}\right)^2\sim \sqrt{\langle W^2(L(t))\rangle} \sim \sqrt{n^{\frac{1}{\tau}}} \sim L(t)^{\frac{1}{2\tau}}\,.
\end{equation}
It follows that
\begin{equation}
    \label{L(t) + u(0,t)A}
|L(t)| \sim t^{\frac{2\tau}{4\tau-1}}=t^{\frac{\gamma-1}{2\gamma-2-H}}, \, \, \, |u(0,t)|=\frac {|L(t)|}{t}\sim t^{-\frac{\gamma-1-H}{2\gamma-2-H}}.
\end{equation}
We finally get that the contribution to $s(m)$ is equal to $-m \frac{\gamma-1-H}{2\gamma-2-H}.$ We
compare the two contributions $-\alpha(H)m-(\gamma-2)/(2-H)$ and  $-m \frac{\gamma-1-H}{2\gamma-2-H}$; then a simple calculation gives the following expression for the threshold $m_C(H,\gamma)$:
\begin{equation}
    \label{thresholdA}
m_C(H,\gamma)=\frac{2\gamma-2-H}{H}.
\end{equation}
Hence, $s(m)=-m \frac{\gamma-1-H}{2\gamma-2-H}$, for $-1<m\leq m_C(H,\gamma)$; and $s(m)=-\alpha(H)m-(\gamma-2)/(2-H)$, for $m>m_C(H,\gamma)$.

In summary, we obtain bifractal behaviour with
\begin{eqnarray}
    s(m)&=&-m \frac{\gamma-1-H}{2\gamma-2-H}\,,\;\; {\rm{for}} \;\; -1 < m \leq m_C(H, \gamma) \,;\nonumber\\
    s(m)&=&-\alpha(H)m-(\gamma-2)/(2-H) \,,\;\; {\rm{for}} \;\; m > m_C(H, \gamma)\,. 
    \label{eq:smbifractalA}
\end{eqnarray}

Note that the energy corresponds to the exponent $m=2$. For this value of $m$, in the first two cases considered above, the dominant contribution to $s(m)$ comes from the term $-\frac{m}{3}$. Hence, the energy decays as $t^{-2/3}$. The threshold $m_C(H,\gamma)$ is:
\begin{eqnarray}
m_C(H,\gamma)&\leq& 2 \;\; {\rm{ in}} \;\; C_1\equiv \{(H,\gamma): 2/3<H<1,\;\; 2<\gamma\leq 3H/2 +1\}\,;\nonumber \\
m_C(H,\gamma) &>& 2 \;\; {\rm{ in}} \;\; C_2 \equiv \{(H,\gamma): 1/2<H<1, \;\max{\{2, 3H/2+1\}}<\gamma < 1+2H\}.\nonumber\\ 
\end{eqnarray}
The areas $C_1$ and $C_2 = C\setminus C_1$ are shown in Fig.~\ref{fig:CasesABC} in Section~\ref{subsec:OBHM}. 
Therefore, the energy decays as follows: 
\begin{eqnarray}
E(t)&\sim& t^{-(\gamma-2H)/(2-H)}\,,\;\; {\rm{for}}\;\; (H,\gamma) \in C_1\,; \\
E(t)&\sim& t^{-(1-H/(2\gamma -2-H))}\,,\;\;{\rm{for}}\;\; (H,\gamma) \in C_2\,.
\label{eq:EdecayA}
\end{eqnarray}
The exponent $(\gamma-2H)/(2-H)\leq 1/2$, when $(H,\gamma) \in C_1$, whereas, if $(H,\gamma) \in C_2$, the exponent $1/2<1-H/(2\gamma -2-H)<2/3$. Note that in all three cases there is also a subdominant contribution to $E(t)$ with a faster decay in the limit $t\to \infty$.

\subsection{Large-scale multifractality}
\label{subsec:AppBmfrac}

{\bf{Genuine Large-scale Multifractality}}

In all three Cases A, B, and C considered above, the exponent $s(m)$ consists of two different pieces that are linear in $m$, so
we have bifractal scaling.
We now generalize the OFBM$_H$, which we used in Cases A-C above, to build an initial condition that leads to genuine large-scale multifractality. The crucial idea is to allow the Hurst exponent $H$ to vary, and then use the construction of the OFBM$_H$ with an $H$-dependent tail exponent $\gamma=\gamma(H)$. 
We provide the details for Case B. 

We proceed as in Case B above, by choosing $H_0$, $\gamma_0$, and $m_0$ such that  $1/2 < H_0 <1, \; \gamma_0 > (1+2H_0), \; {\rm{and}}\; m_0 > m_B(H,\gamma)$. We then sample $H$ uniformly from the interval $[H_0-\epsilon,H_0+\epsilon]$, where $\epsilon$ is small and positive.
We also use the tail exponent $\gamma=\gamma(H)$ and set $\gamma(H_0)=\gamma_0$. The exact dependence of $\gamma$ on $H$ will be specified below. Given the continuity of $m_B(H,\gamma)$, we can say that, for some small $\delta>0$, which depends on $\epsilon$, the whole interval $[m_0-\delta, m_0+\delta]$ will be
above the  threshold $m_B(H,\gamma(H))$, for all $H\in [H_0-\epsilon,H_0+\epsilon]$. Then, for all $m\in [m_0-\delta, m_0+\delta]$, we have: 
\begin{equation}
    \label{s(m)_multifractalA}
s(m)=\max_{H\in [H_0-\epsilon,H_0+\epsilon] }{s_{H,\gamma(H)}(m)}=
\max_{H\in [H_0-\epsilon,H_0+\epsilon] }\left\{-\alpha(H)m-\frac{\gamma(H)-2}{2-H}\right\}\,.
\end{equation}
We now choose $\gamma(H)$ in such a way that, for $m=m_0$, the maximum in the above expression is attained at $H=H_0$.
Namely, we require that 
\begin{equation}
    \label{s(m)_multifractal_conditionA}
\frac{d}{dH}\left[-\alpha(H)m_0-\frac{\gamma(H)-2}{2-H}\right](H_0)=0\,;\; \, \frac{d^2}{dH^2}\left[-\alpha(H)m_0-\frac{\gamma(H)-2}{2-H}\right](H_0)<0\,.
\end{equation}
An easy calculation shows that the first condition is satisfied  if 
\begin{equation}
    \label{s(m)_multifractal_condition_1A}
\frac{d\gamma}{dH}(H_0)=\frac{m_0-\gamma_0+2}{2-H_0}\,.
\end{equation}
The second condition requires that 
\begin{equation}
    \label{s(m)_multifractal_condition_2A}
\frac{d^2\gamma}{dH^2}(H_0)> 0.
\end{equation}
We can now set
\begin{equation}
    \label{gamma(H)A}
\gamma(H)=\gamma_0 + \frac{m_0-\gamma_0+2}{2-H_0}(H-H_0) + A (H-H_0)^2,
\end{equation}
where $A>0$ is an arbitrary positive constant. To find $s(m)$
for $m\in [m_0-\delta,m_0+\delta]$, we have to find $H(m)$ by solving a quadratic equation
\begin{equation}
    \label{quadratic_eqA}
\frac{d\gamma}{dH}(H)= \frac{m_0-\gamma_0+2}{2-H_0} + 2A(H-H_0)=\frac{m-\gamma(H) +2}{2-H}\,,
\end{equation}
then find $\gamma(m)=\gamma(H(m))$, and, finally, substitute 
$H=H(m), \, \gamma=\gamma(m)$ into the expression $s(m)= -(1-H)/(2-H)m- (\gamma-2)/(2-H)$. 

If we use, as an example, the values $H_0=3/4, \, \gamma_0=3, \, m_0=7$, we can proceed as follows.
We first choose $\epsilon=1/2$ which means that the Hurst exponent $H$ is sampled within a maximum possible interval $1/2<H<1$.
We shall use the suggested expression \eqref{gamma(H)A} and choose
$A=16$. After substituting the values $H_0=3/4, \, \gamma_0=3, \, m_0=7, A=16$, we get $\gamma(H)=3+6(4H-3)/5 + (4H-3)^2$. Note that
this choice of $\gamma(H)$ corresponds to area $B$ for all $1/2<H<1$ [see Fig.~\ref{fig:CasesABC} in Section~\ref{subsec:OBHM}]. Differentiating $s(m;H)= -(1-H)/(2-H)m- (\gamma(H)-2)/(2-H)$
with respect to $H$, we get that $ds(m;H)/dH$ vanishes only if $H$
satisfies the following condition:
\begin{equation}
    \label{s(m)_multifractal_condition_1HA}
m= \frac{d\gamma}{dH}(H)(2-H) + \gamma(H) -2\,.
\end{equation}
Substituting $\gamma(H)=3+6(4H-3)/5 + (4H-3)^2$ in Eq.~\eqref{s(m)_multifractal_condition_1HA}, we get the quadratic equation $H^2-4H+\frac{32+m}{16}=0$ with two solutions $H=2\pm \sqrt{32-m}/4$. Since we are interested in the interval $H\in (1/2, 1)$, only the solution $H_{-}=2 - \sqrt{32-m}/4$ is of interest to us. We can check that $1/2<H_{-}<1$ for $-4<m<16$. Given that $ds(m;H)/dH>0$ for $1/2<H<H_{-}$ and
$ds(m;H)/dH<0$ for $H_{-}<H<1$, we get $\max_{H\in [1/2,1]}s(m;H)=
s(m; H_{-})$, when $-4<m<16$. In the case $m\geq 16$, the maximum of $s(m;H)$, over the interval $H\in [1/2, 1]$, is attained at $H=1$. 
Note that we only consider the case when $m>-1$; otherwise, the average speed raised to the power $m$ diverges. Substituting $H_{-}=2 - \sqrt{32-m}/4$, 
we get the following result for a contribution to $s(m)$ coming from long intervals $\Delta_0$:
\begin{equation}
    \label{s(m)_Long_Intervals-1A}
s(m)=\frac{224}{5}-m -8\sqrt{32-m}, \, m\in (-1,16)
\, \, {\text {and}} \, \, s(m)=-\frac{16}{5}, \, m \geq 16\,.
\end{equation}
We also have to take into account a contribution $-m/3$ coming from the  main probability event. It can be shown that the 
term $-m/3$ is the dominant one for $-1<m<24(\sqrt{5}-1)/5$. This leads to the following final answer for $s(m)$ [see Fig.~\ref{fig:sm_plot} in Section~\ref{subsec:OBHM}]:
\begin{equation}
    \label{s(m)_final_app}
s(m)=\begin{cases}
			-\frac{m}{3}, & \, -1<m\leq \frac{24(\sqrt{5}-1)}{5} \\
           \frac{224}{5}-m -8\sqrt{32-m}, & \, \frac{24(\sqrt{5}-1)}{5}<m<16 \\
           -\frac{16}{5}, &\, m\geq 16.  
		 \end{cases}
\end{equation}
Clearly, Eq.~\eqref{s(m)_final_app} implies genuine multifractality because $s(m)$ has truly nonlinear dependence on $m$, and it is not just a combination of different linear functions of $m$ (as, e.g., in Eq.~\eqref{eq:smbifractalA}).

\section{}
\label{sec:app_burg}

\subsection{Energy decay in 1D Burgulence with multifractal initial data}
\label{subsec:app_burg1}

\subsubsection{Initial data for the velocity: MRW with $H=0.5$}

We considered freely decaying turbulence in the 1D inviscid Burgers equation~\eqref{eq:burg}, with the multifractal-random-walk (MRW) initial condition obtained using Eqs.~\eqref{eq:mfin1}-\eqref{eq:mfin4} with $H=1/2$
for the initial potential $\varphi_0$ [cf. Type A initial data in \cite{1992-she--frisch}]. We now consider such decay with MRWs for the \textit{initial velocity} [cf. Type B initial data in \cite{1992-she--frisch}].
In our numerical studies, which use Eqs.~\eqref{eq:max} and ~\eqref{eq:burgvel}, we discretize the system with $N=2^{14}$ points. 

\begin{figure} 
    \centerline{
        \includegraphics[width=0.5\columnwidth]{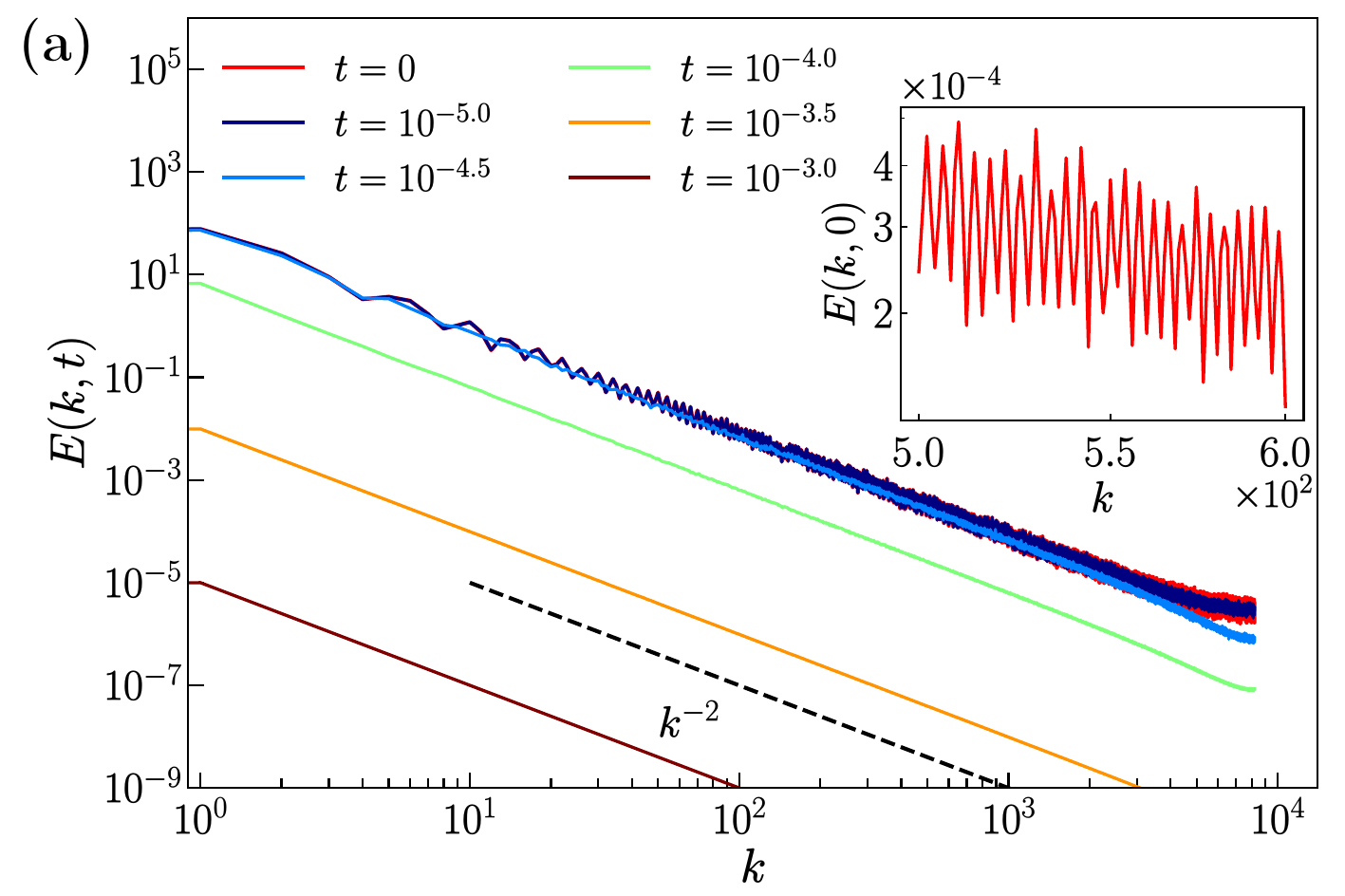} 
        \includegraphics[width=0.5\columnwidth]{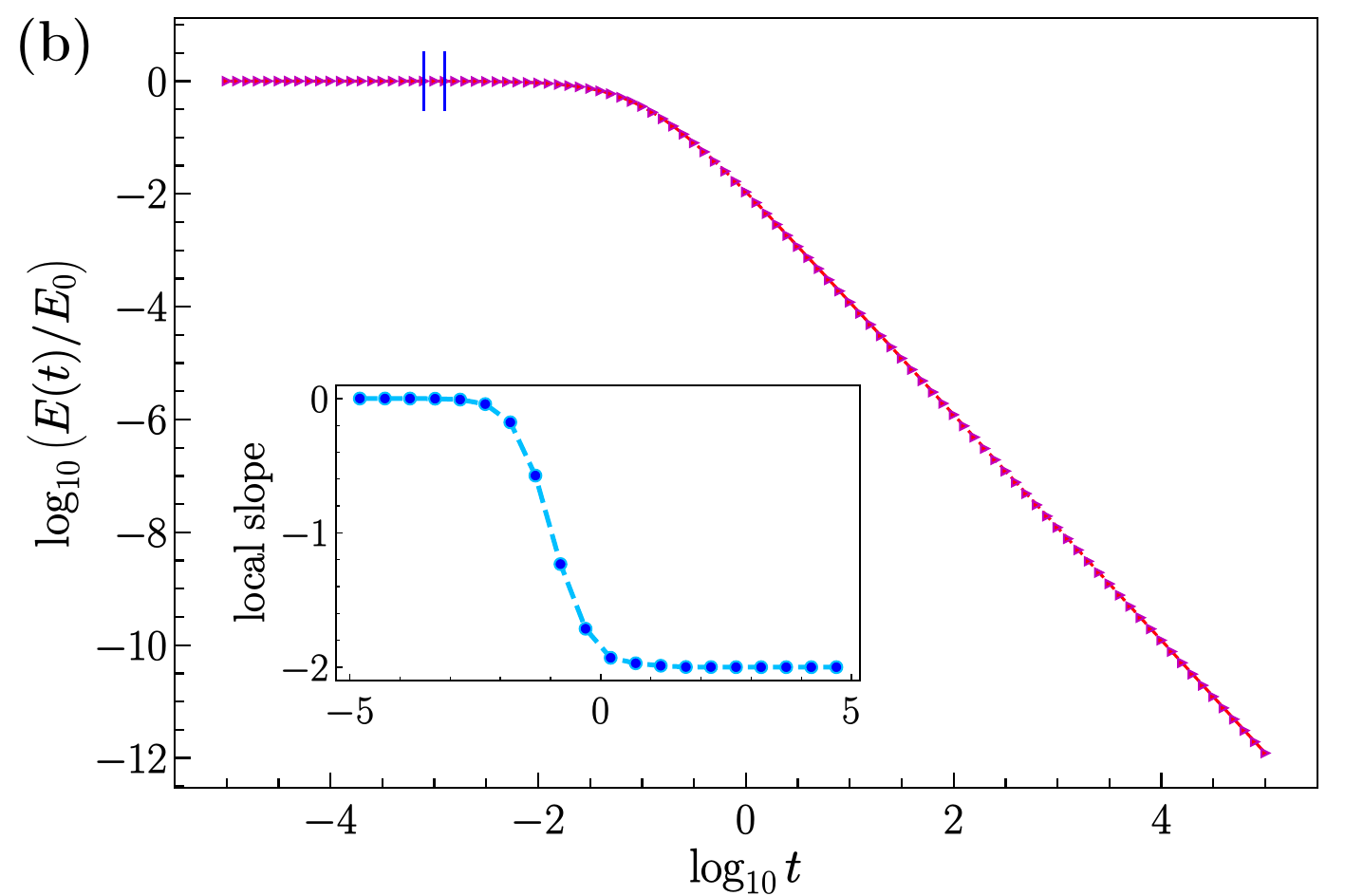}
    }
    \caption{Log-log plots of (a) the energy spectrum $E(k,t)$ versus the wavenumber $k$, at some representative values of $t$ [the inset shows the initial spectrum $E(k,0)$ in detail in the wavenumber range $5\times 10^2\leq k\leq 6 \times 10^2$] and (b) the scaled total energy $E(t)/E(t=0)$ versus the time $t$ for the  multifractal initial condition [Eqs.~\eqref{eq:mfin1}-~\eqref{eq:mfin4}], for the velocity, with Hurst exponent $H=1/2$; the inset shows a plot of the local slope. We compute structure functions in Fig.~\ref{fig:sfunmfrw} at the point in time that lies at the centre of the interval indicated by the blue vertical lines. }
    \label{fig:mrw-a1-ende}
\end{figure}

In Fig.~\ref{fig:mrw-a1-ende} (a) we show log-log plots of the energy specturm $E(k,t)$ versus the wave number $k$ at different representative times $t$; at early times 
$0 \leq t \lesssim 10^{-5}$, this spectrum is not of a simple, power-law form because of the multifractal initial condition for the velocity; however, for $10^{-4.5} \lesssim t$, the spectrum has the power-law form $E(k,t) \sim k^{-2}$ because of the formation of shocks. The decay of the total energy $E(t)$ is shown in the log-log plot of Fig.~\ref{fig:mrw-a1-ende} (b); the temporal decay does not have a single-exponent, power-law form for $0 \leq t \lesssim 1$; however, at later times, it shows the power-law decay $E(t) \sim t^{-2}$, once the integral length scale becomes comparable to the system size.  We compute the order-$p$ velocity structure functions $S_p(\ell,t) \equiv [u(x+\ell,t) - u(x,t)]^p$ and plot it versus the separation $\ell$ [see the log-log plot in Fig.~\ref{fig:mrw-a1-ende} (c)]; we obtain the multiscaling exponents $\zeta_p$, which follow from the power-law form $S_p(\ell,t) \sim \ell^{\zeta_p}$ for $\ell$ in the pink-shaded region in  Fig.~\ref{fig:mrw-a1-ende} (c). We use a local-slope analysis [Fig.~\ref{fig:mrw-a1-ende} (d)] to extract these exponents, which we plot versus the order $p$ in  Fig.~\ref{fig:mrw-a1-ende} (e) at  $t=0$ (red curve)  and $t=10^{-3}$ (blue curve). We observe that multifractality is present at $t=10^{-3}$ (see Fig.~\ref{fig:mrw-a1-ende}), insofar as $\zeta_p$ is a nonlinear function of $p$. 

\begin{center}
	\begin{figure}
            \centering
		\includegraphics[width=0.9\linewidth]{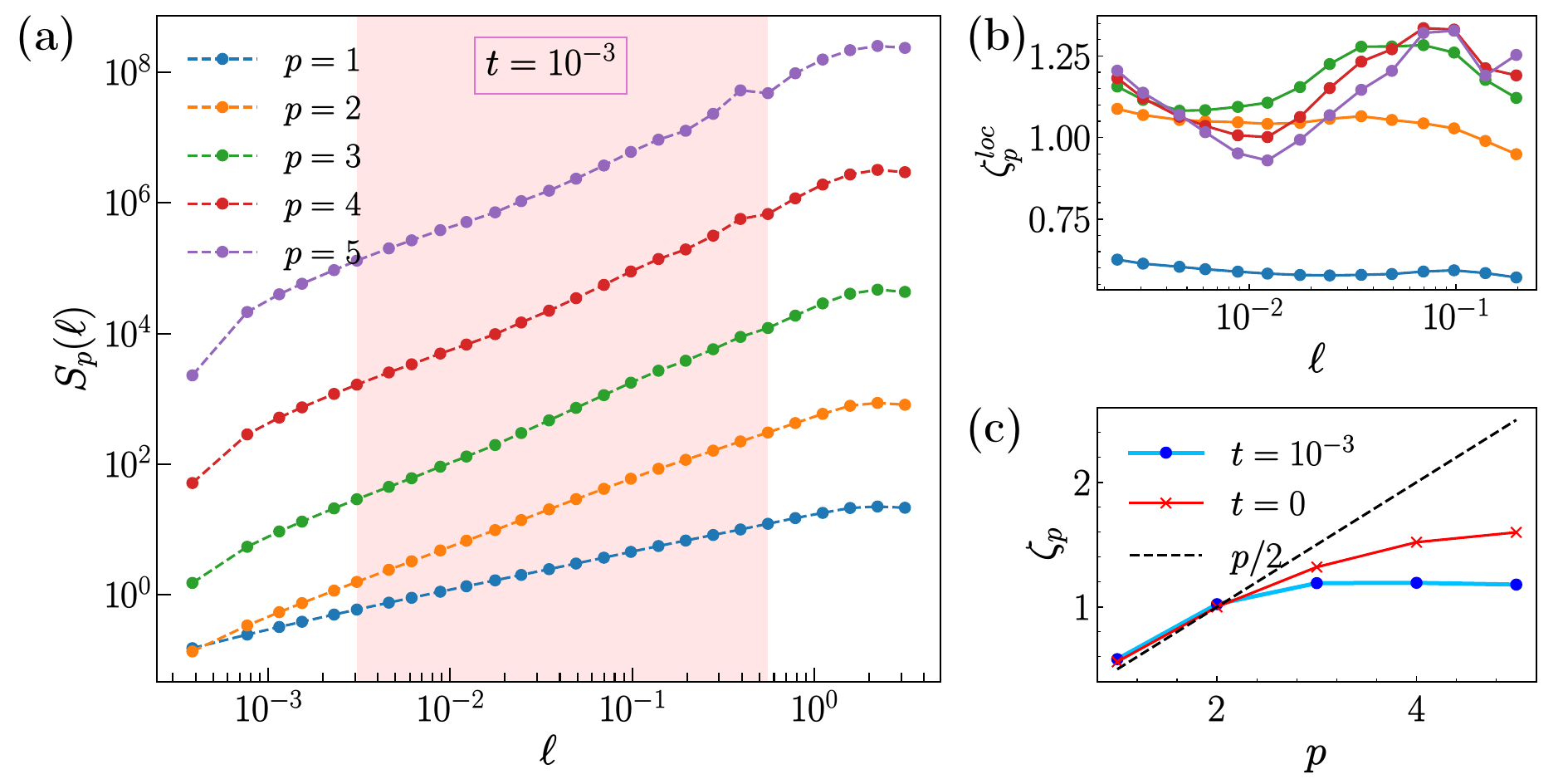}
		\caption {Plots for the  multifractal initial condition, for the velocity, with Hurst exponent $H=1/2$ at $t=10^{-3}$: (a) Log-log plots versus $\ell$ of the structure functions of order $p=1, \ldots, 5$. (b) Plots of $\zeta_p^{loc}$, obtained from local slopes of the structure functions in (a), versus $\ell$. (c) Plots of $\zeta_p$ versus $p$ (in blue) at $t=10^{-3}$; the red curves show $\zeta_p$ (in red) for the multifractal random walk of~\cite{bacry2001multifractal}.}
         \label{fig:sfunmfrw}
	\end{figure}

\end{center}

\begin{figure} 
    \centerline{
        \includegraphics[width=0.5\columnwidth]{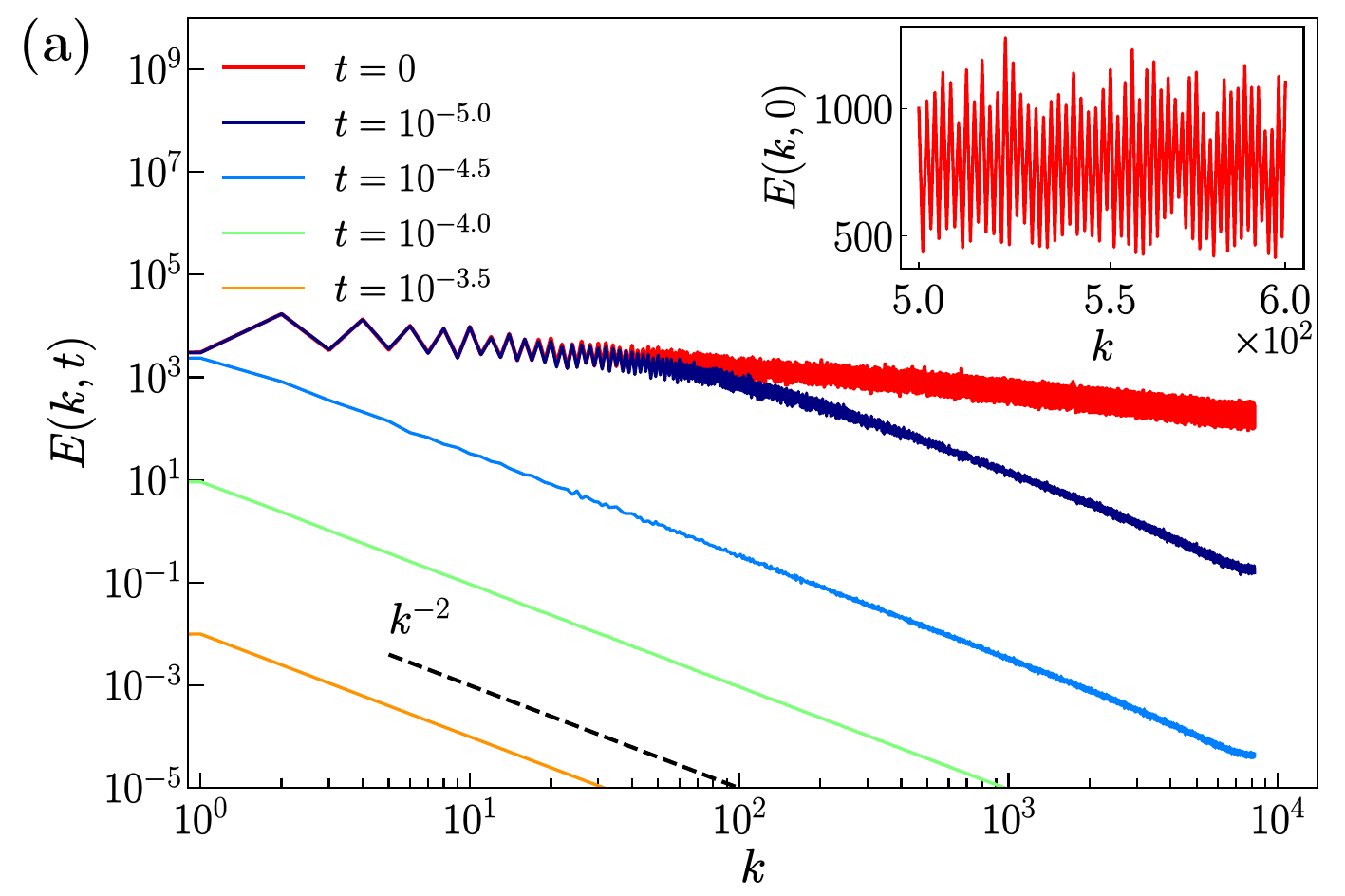} 
        \includegraphics[width=0.5\columnwidth]{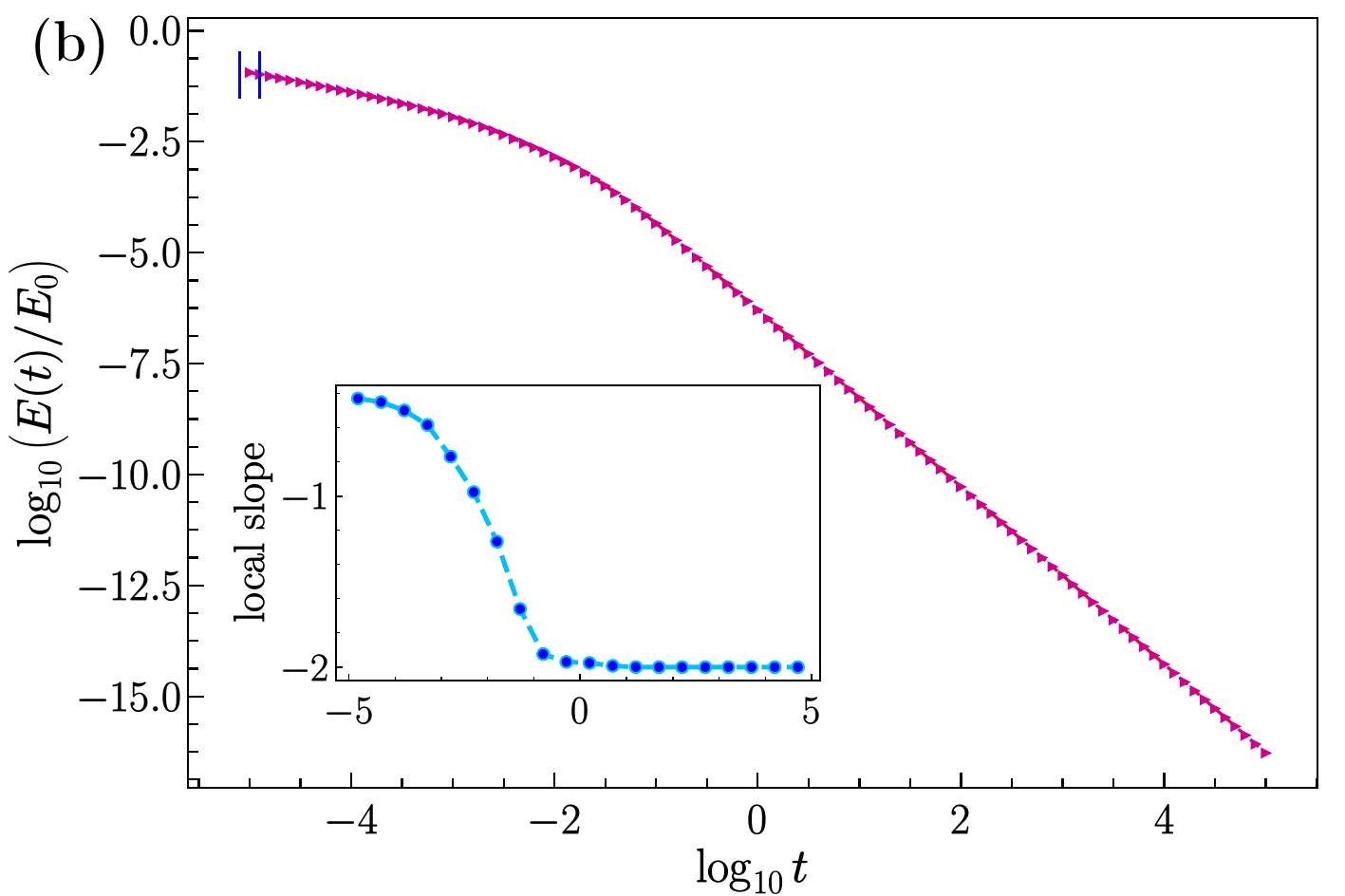}
    }
    \caption{Log-log plots of (a) the energy spectrum $E(k,t)$ versus the wavenumber $k$, at some representative values of $t$ [the inset shows the initial spectrum $E(k,0)$ in detail in the wavenumber range $5\times 10^2\leq k\leq 6 \times 10^2$] and (b) the scaled total energy $E(t)/E(t=0)$ versus the time $t$ for the  multifractal initial condition [Eqs.~\eqref{eq:mfin1}-~\eqref{eq:mfin4}], for the potential $\varphi_0$, with Hurst exponent $H=0.75$; the inset shows a plot of the local slope. We compute structure functions in Fig.~\ref{fig:sfunmfrwphi-1p5} at the point in time that lies at the centre of the interval indicated by the blue vertical lines. }
    \label{fig:mrw-a1p5-ende-phi}
\end{figure}
\begin{center}
	\begin{figure}
            \centering
		\includegraphics[width=0.9\linewidth]{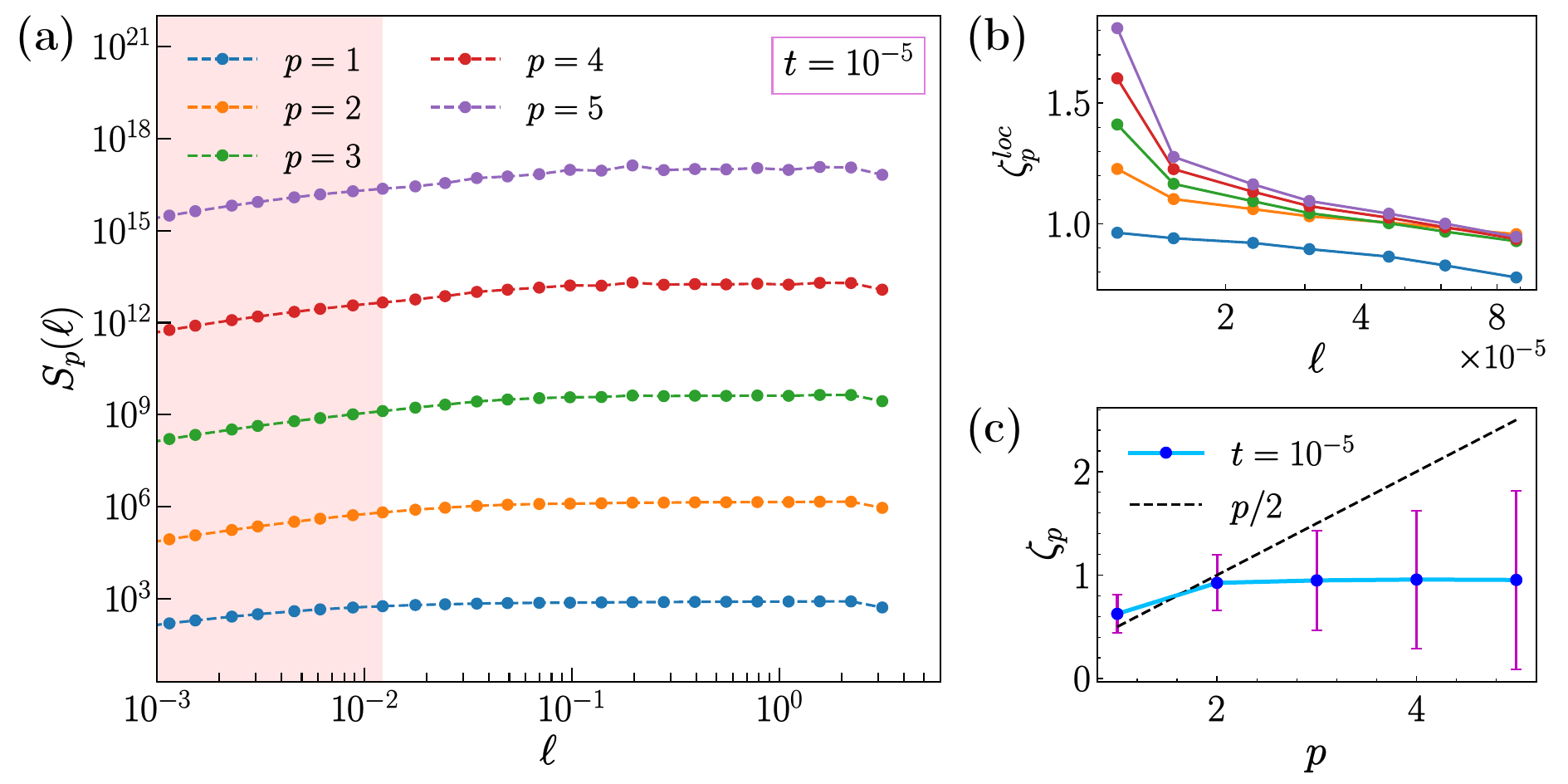}
		\caption {Plots for the  multifractal initial condition, for the potential $\varphi_0$, with Hurst exponent $H=0.75$ at $t=10^{-5}$: (a) Log-log plots versus $\ell$ of the structure functions of order $p=1, \ldots, 5$. (b) Plots of $\zeta_p^{loc}$, obtained from local slopes of the structure functions in (a), versus $\ell$. (c) Plots of $\zeta_p$ versus $p$ (in blue) at $t=10^{-5}$.}
        \label{fig:sfunmfrwphi-1p5}
	\end{figure}
\end{center}

\begin{figure} 
    \centerline{
        \includegraphics[width=0.5\columnwidth]{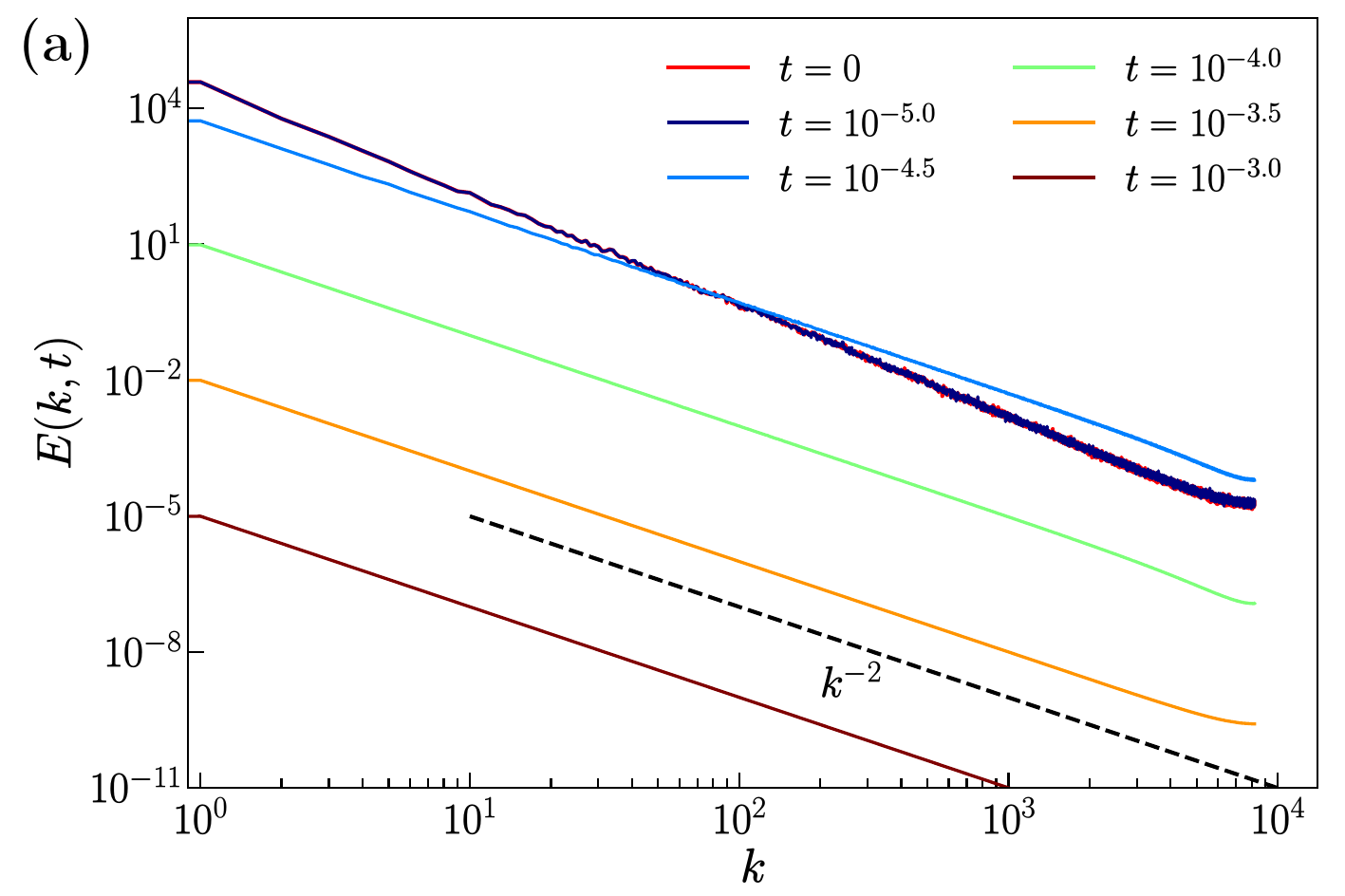} 
        \includegraphics[width=0.5\columnwidth]{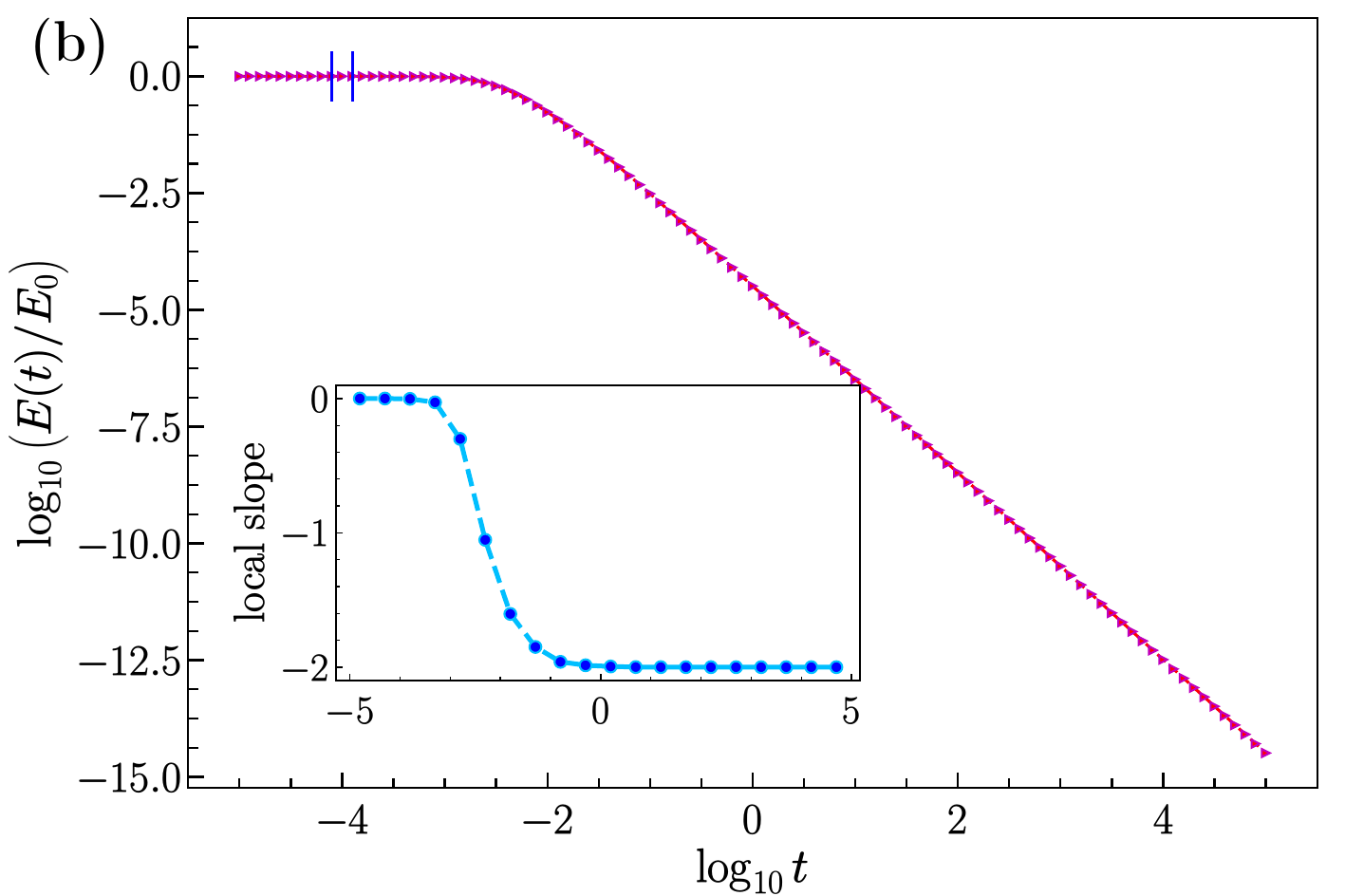}
    }
  \label{fig:sf-1p5}
  \label{fig:sf-1p5}
    \caption{Log-log plots of (a) the energy spectrum $E(k,t)$ versus the wavenumber $k$, at some representative values of $t$, and (b) the scaled total energy $E(t)/E(t=0)$ versus the time $t$ for the  multifractal initial condition [Eqs.~\eqref{eq:mfin1}-~\eqref{eq:mfin4}], for the velocity, with Hurst exponent $H=0.75$; the inset shows a plot of the local slope. We compute structure functions in Fig.~\ref{fig:sf-1p5} at the point in time that lies at the centre of the interval indicated by the blue vertical lines.}
    \label{fig:mrw-a1p5-ende}
\end{figure}
\begin{center}
	\begin{figure}
            \centering
		\includegraphics[width=0.9\linewidth]{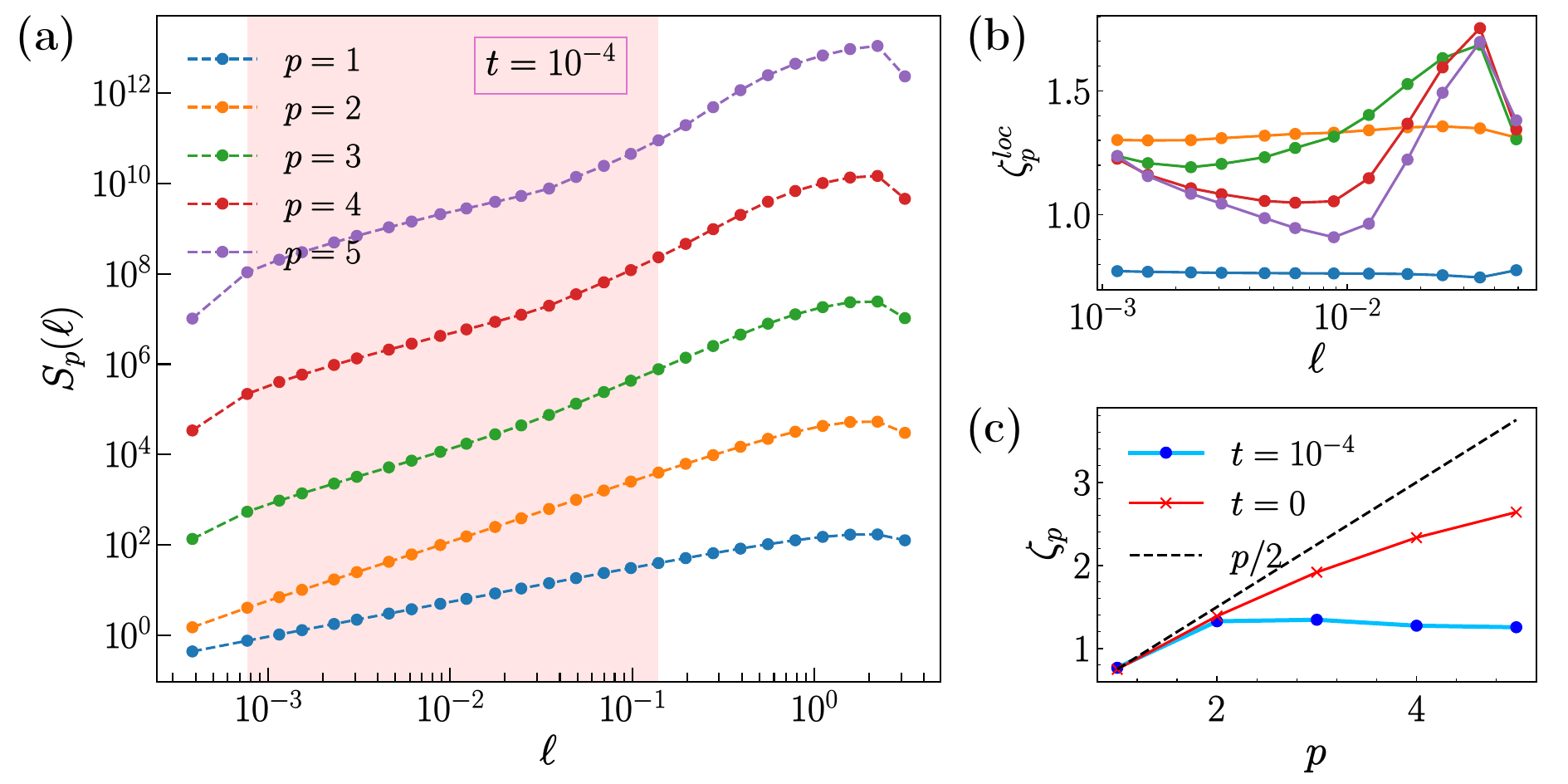}
		\caption{Plots for the  multifractal initial condition, for the velocity, with Hurst exponent $H=0.75$ at $t=10^{-4}$: (a) Log-log plots versus $\ell$ of the structure functions of order $p=1, \ldots, 5$. (b) Plots of $\zeta_p^{loc}$, obtained from local slopes of the structure functions in (a), versus $\ell$. (c) Plots of $\zeta_p$ versus $p$ (in blue) at $t=10^{-4}$; the red curves show $\zeta_p$ (in red) for the multifractal random walk of \cite{bacry2001multifractal}.}
  \label{fig:sf-1p5}
	\end{figure}
\end{center}

\subsubsection{Initial data for the potential and velocity: MRW with $H=0.75$}

In addition, we consider multifractal initial conditions, with $\tilde{B}_n$, increments in  the (periodised) fractional Brownian motions with Hurst exponent $H=0.75$, to construct the following sequence of random numbers:
\begin{equation}
    \tilde{A}_n = \tilde{B}_n e^{\Omega_n}, \quad 1 \leq n \leq N.
\end{equation}
Then we consider the sequence 
\begin{equation}
	\tilde{A}'_n = \tilde{A}_n - \frac{1}{N}\sum_{i=1}^{N} \tilde{A}_i\,,
\end{equation}
from which we obtain a multifractal random walk using $A'_n$ as follows:
\begin{equation}
	\tilde{M}_n = 
	\begin{cases}
	0, & \quad  n=1 \\
	\sum_{i=1}^{n-1} \tilde{A}'_i,  & \quad n > 1 .
	\end{cases}
\end{equation}
For the  multifractal initial condition for the potential $\varphi_0$, with Hurst exponent $H=0.75$, we present plots for energy spectra, energy decay, structure functions, and exponents in Figs.~\ref{fig:mrw-a1p5-ende-phi} and \ref{fig:sfunmfrwphi-1p5}.
For the  multifractal initial condition for the velocity, with Hurst exponent $H=0.75$, we present plots for energy spectra, energy decay, structure functions, and exponents in
Figs.~\ref{fig:mrw-a1p5-ende} and \ref{fig:sf-1p5}.

\subsection{Energy decay in 1D Burgulence with power-law initial energy spectra and connections to the Gurbatov effect}
\label{subsec:app_burg2}

We begin with some well-known results [see, e.g., \cite{1997-gurbatov--toth}] for the case when the initial (average) energy spectrum $E_{0}(k)$ is
\begin{equation}
    E_{0}(k) =  \left\langle  \left| \widetilde{u}_{k}(0) \right|^{2} \right\rangle  = A  \, \mathcal{E}(k) \, \exp \! \left[ - 2 k^{2}/k^{2}_{c} \right] , 
\end{equation} 
where $\left\langle \cdot \right\rangle $ is the average over realizations. The energy spectrum $E(k,t)$ at time $t$ is defined as
\begin{equation}
	E(k,t) := \left\langle \left| \widetilde{u}_{k}(t) \right|^{2} \right\rangle.
\end{equation}
Furthermore, we define the (average) energy 
\begin{equation}
	E(t) := \left\langle  \int_{ - \infty}^{ \infty }  \text{d} k \left| \widetilde{u}_{k}(t) \right|^{2} \right\rangle,
\end{equation}
and the integral length scale  
\begin{equation}
	L(t) := \frac{1}{E(t)}\displaystyle \left\langle    \int_{-\infty}^{\infty} \text{d} k \, k^{-1} \left| \widetilde{u}_{k}(t) \right|^{2}     \right\rangle.
\end{equation}
The single-power-law case  
\begin{equation}
	\mathcal{E}(k) = |k|^{n} \ ,
\end{equation}
where the exponent $n$ satisfies $-1 < n < 2 $, was studied by \cite{1997-gurbatov--toth} in detail. We recall here that, for $-1 < n < 1$, the energy decay is self-similar with 
\begin{equation}
	E(t) \propto  t^{ e_{E}} \ , \qquad \text{ where } \ \ e_{E} = \displaystyle \frac{- 2 (n +1)}{n+3}   \ .
	\label{eq:expe}
\end{equation}
The integral length scale $L(t)$ increases with time as
\begin{equation}
	L(t) \propto t^{e_{I}} \ , \qquad \text{ where } \ \ e_{I} = \displaystyle \frac{2}{n+3} \ .
	\label{eq:expi}
\end{equation}

For $1 < n < 2$, we encounter the Gurbatov phenomenon,  namely, $E(k,t) > E_0(k)$ for wavenumbers $k \leq K(t) \sim 1/L(t)$.
This leads to non-self-similar decay (growth) of the total energy (integral length scale) because of the following logarithmic corrections [see \cite{1997-gurbatov--toth} and \cite{roy2021steady}]:
\begin{eqnarray}
    E(t)  &\sim&  t^{-1}\sigma_\psi \ln^{-1/2}(t/t_{nl})\,; \;\; L(t) \sim t^{1/2} \sigma_\psi^{1/2}\ln^{-1/4}(t/t_{nl})\,.\nonumber \\
{\rm{Here,}}\;\;\sigma_\psi^2  &\equiv&  \langle \int_{-\infty}^{\infty} dk k^{-2} |\tilde{u}_k(0)|^2\rangle \,; \;\; t_{nl} \equiv \frac{\sigma_\psi}{\int_{-\infty}^\infty dk E_0(k)}\,.
    \label{eq:burggurbdecay}
\end{eqnarray}

If we use a single sharp peak in the initial energy spectrum $E_0(k)$, with the passage of time $t$, the spectrum $E(k,t)$ develops a $k^2$ part at small $k$ and a $k^{-2}$ part at large $k$   [see Fig. 8.2 in \cite{roy2021steady}], the former because of Proudman-Reid-type beating interactions [discussed for 3D NS turbulence in Section~\ref{sec:ns-hyper-vis}] and the latter because of the development of shocks. The total energy $E(t)$ shows the power-law decay $\sim t^{-1}$ at intermediate times (by analogy with the single-power-law case discussed above), because of the $k^2$ part in $E(k,t)$; this crosses over to a decay of the form  $\sim t^{-2}$ at large times, when the integral length scale, which grows with $t$, becomes comparable to the system size.

\bigskip
\noindent

\subsection{Initial data with energy spectra that have two power-law spectral ranges}
\label{subsec:app_burg3}

In Section~\ref{subsubsec:tworange} we presented results where $E_0(k)$ has two power-law spectral ranges are present in the initial spectrum [see Eqs.~\eqref{eq:E0k2p1} and \eqref{eq:power-law} in Section~\ref{subsubsec:tworange}]. Here we consider three more such cases with different combinations of power-laws for the two spectral ranges. We recall that the general form for a composite two-range initial spectrum is written in terms of the power-laws $\mathcal{E}_1(k)$ and $\mathcal{E}_2(k)$ with exponents $n_1$ and $n_2$ respectively [see Eqs.~\eqref{eq:E0k2p1} and \eqref{eq:power-law} in Section~\ref{subsubsec:tworange}]. We consider the following three cases by considering different values for the exponents $n_1$ and $n_2$:
\begin{itemize}
    \item Case Ib: $n_1 = 0.25, n_2 = 0.75$.
    \item Case Ic: $n_1 = 1.50, n_2 = 0.5$.
    \item Case Id: $n_1 = 1.25, n_2 = 1.75$.
\end{itemize}
We describe our results for these cases below.
\begin{figure}
    \centerline{
        \includegraphics[width=0.333\textwidth]{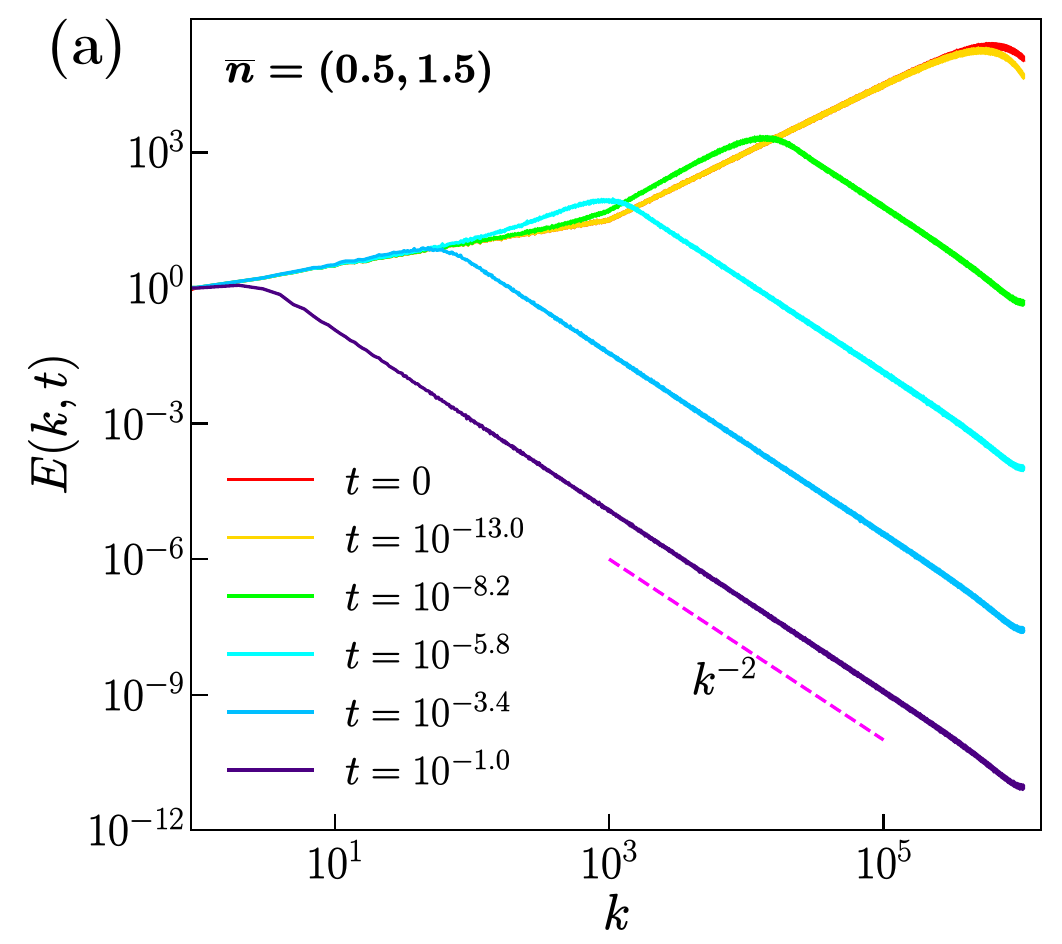}
        \includegraphics[width=0.333\textwidth]{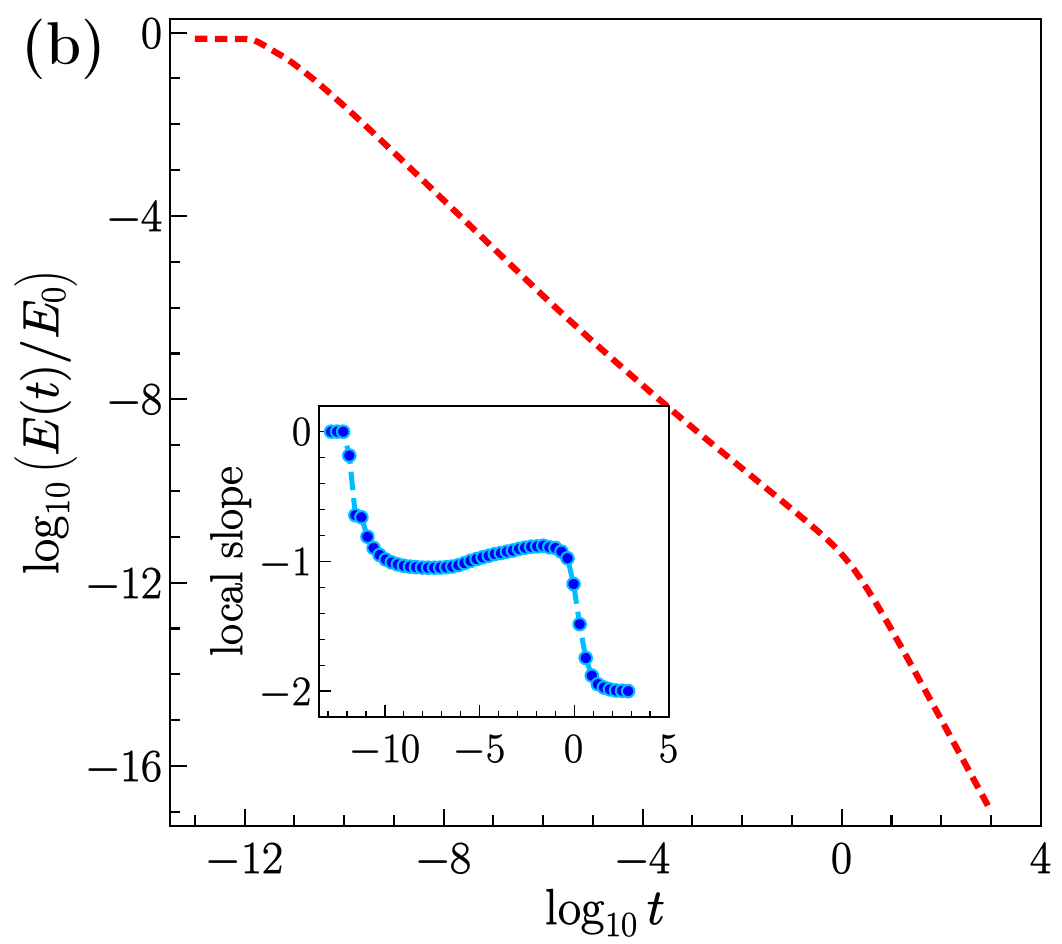}
        \includegraphics[width=0.333\textwidth]{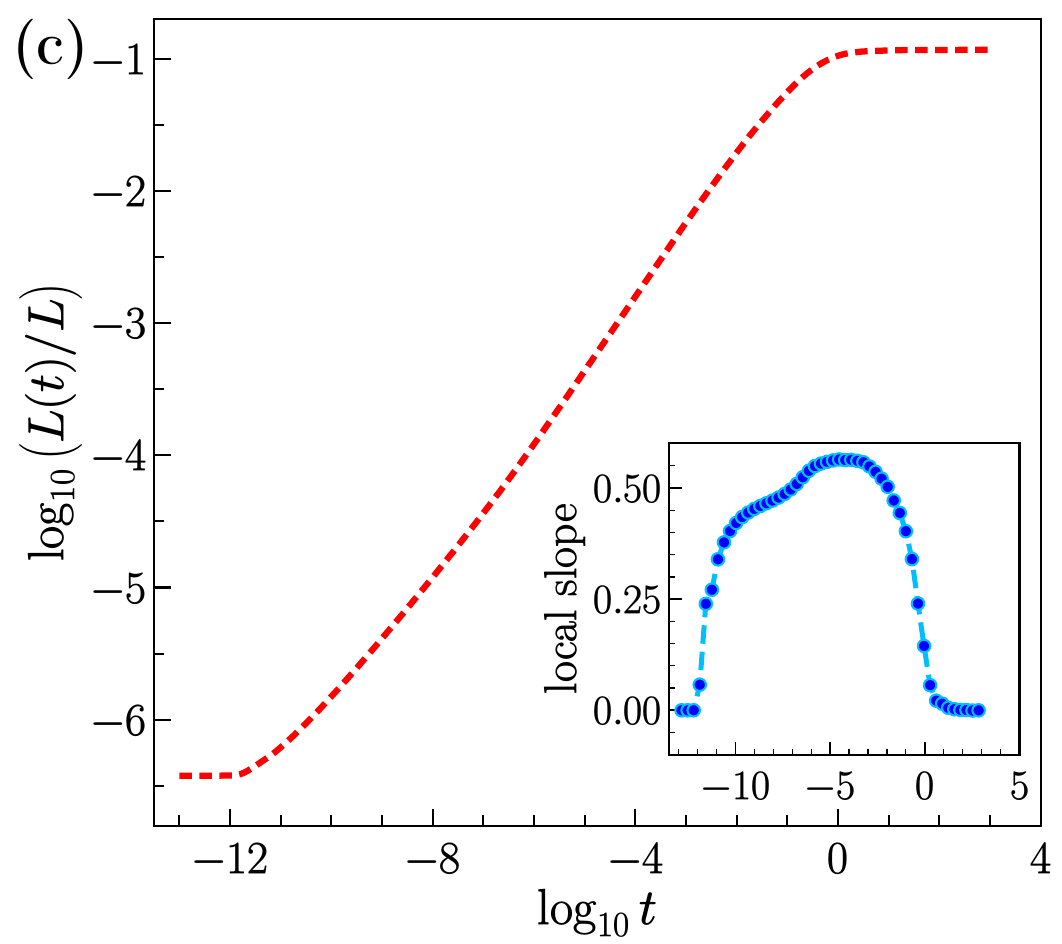}
    }
    \centerline{
        \includegraphics[width=0.333\textwidth]{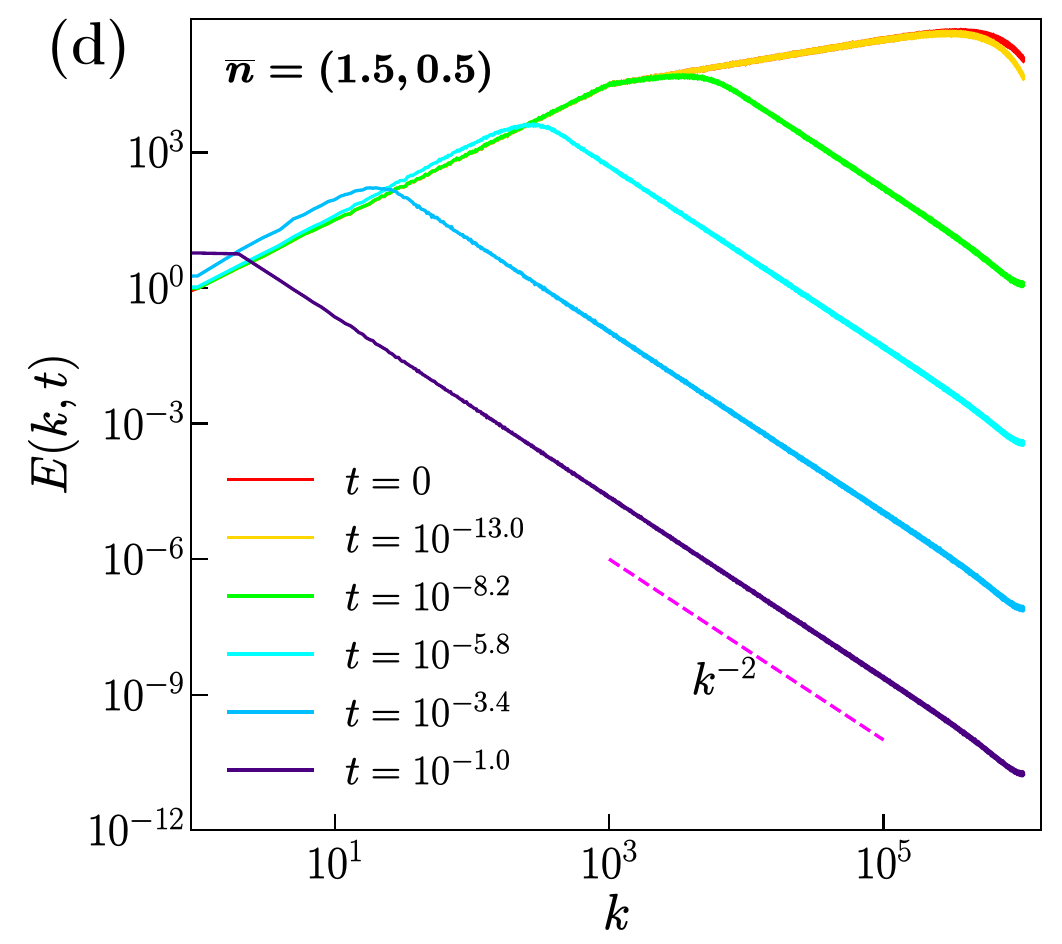}
        \includegraphics[width=0.333\textwidth]{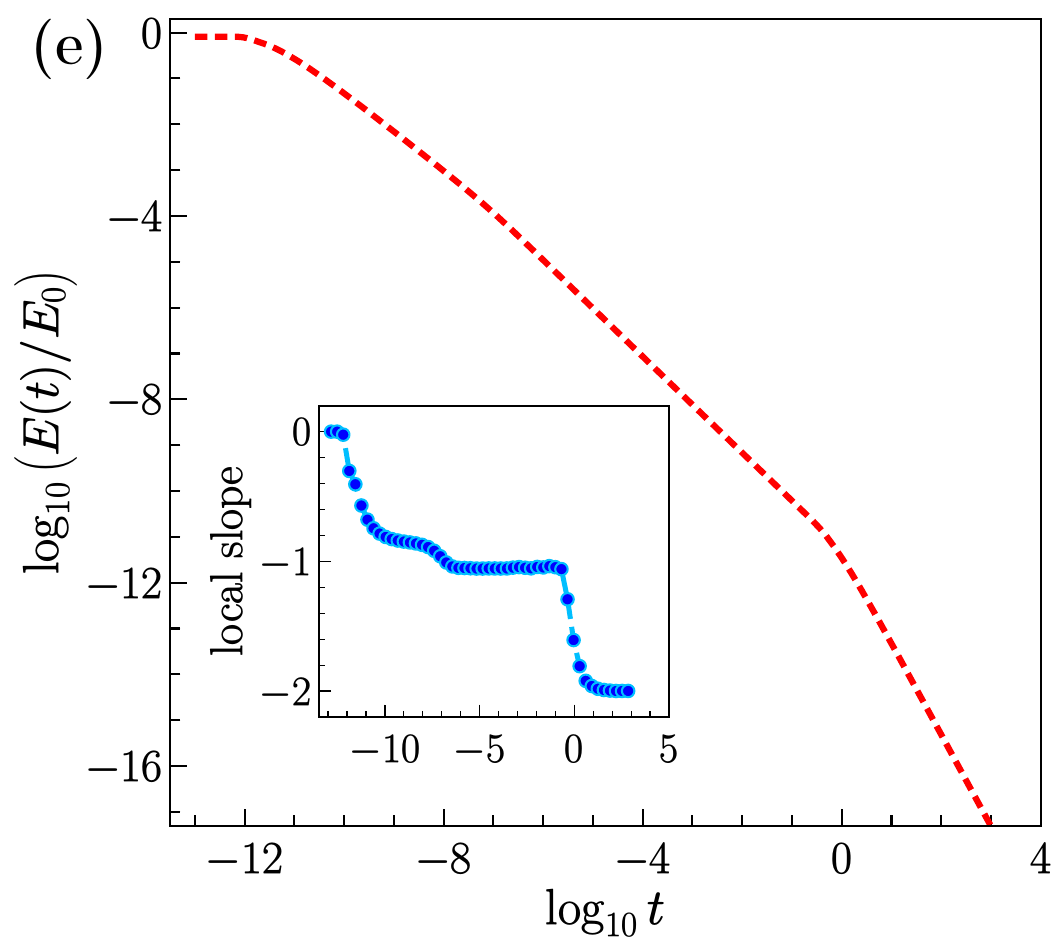}
        \includegraphics[width=0.333\textwidth]{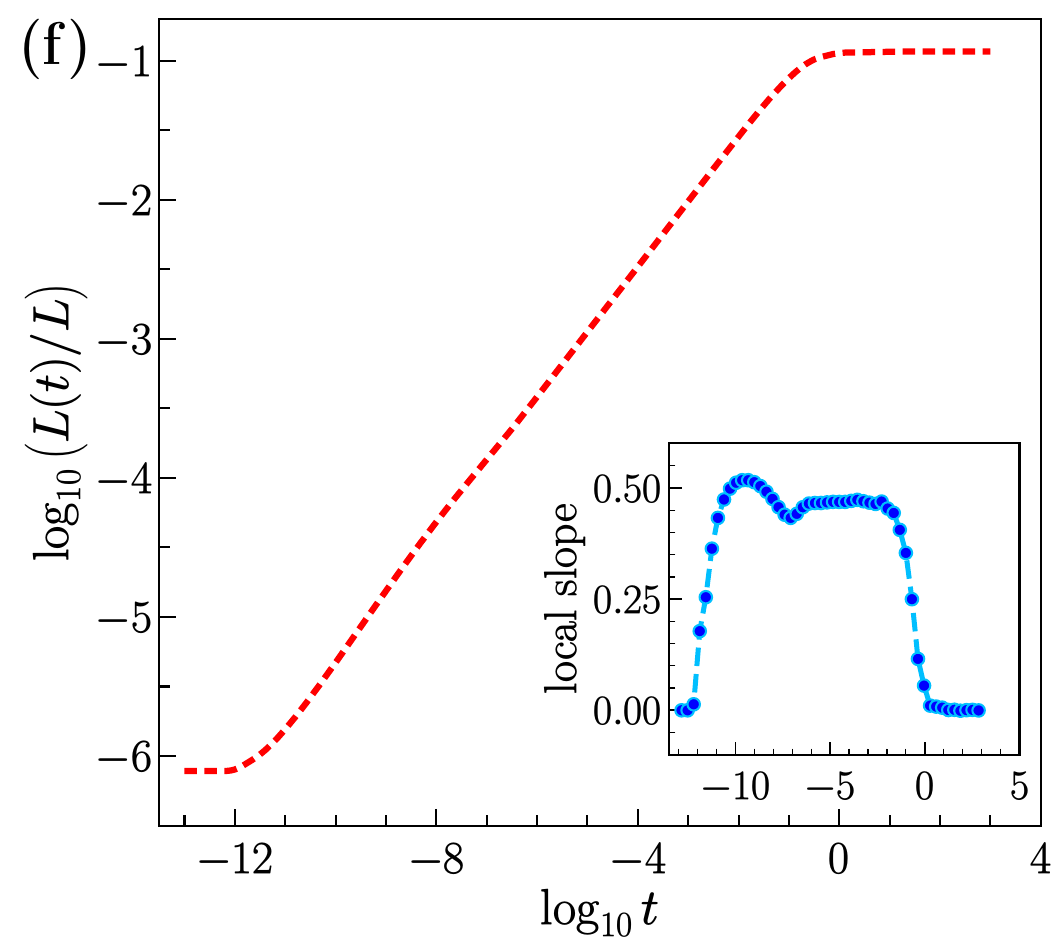}
    }
    \centerline{
        \includegraphics[width=0.333\textwidth]{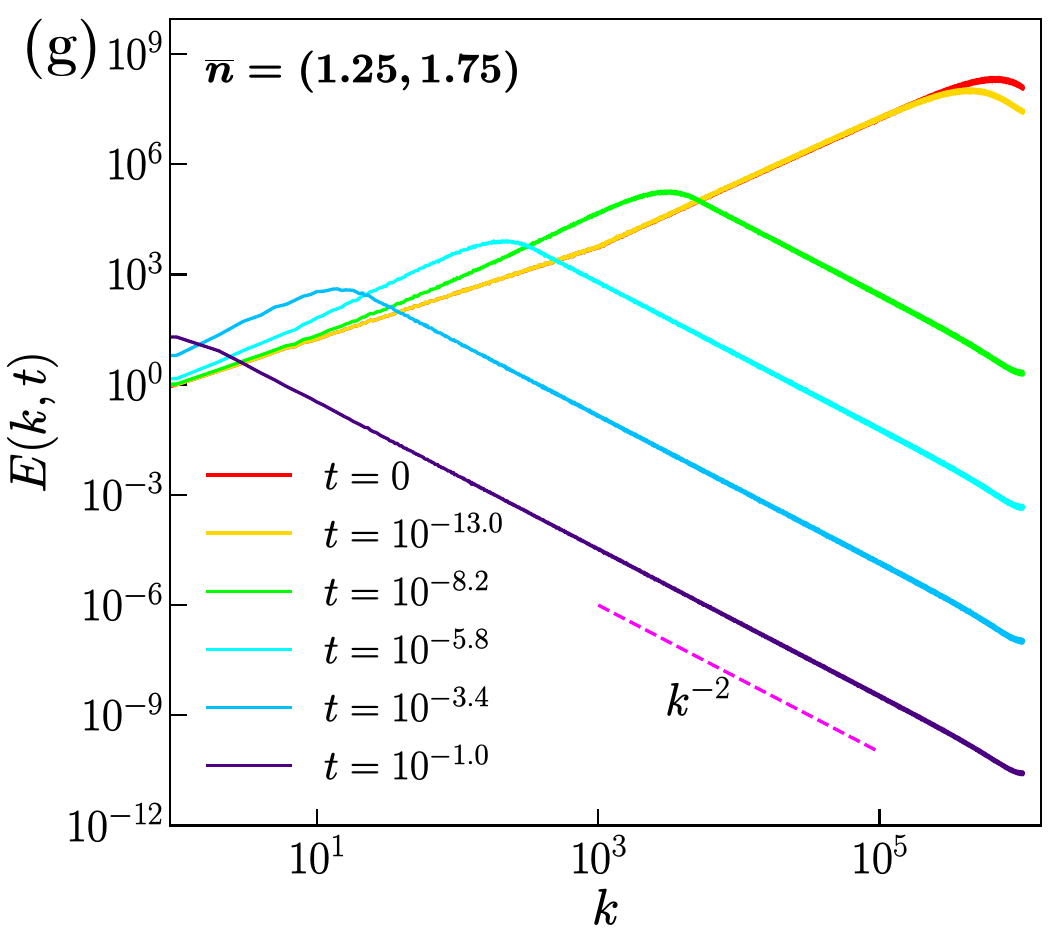}
        \includegraphics[width=0.333\textwidth]{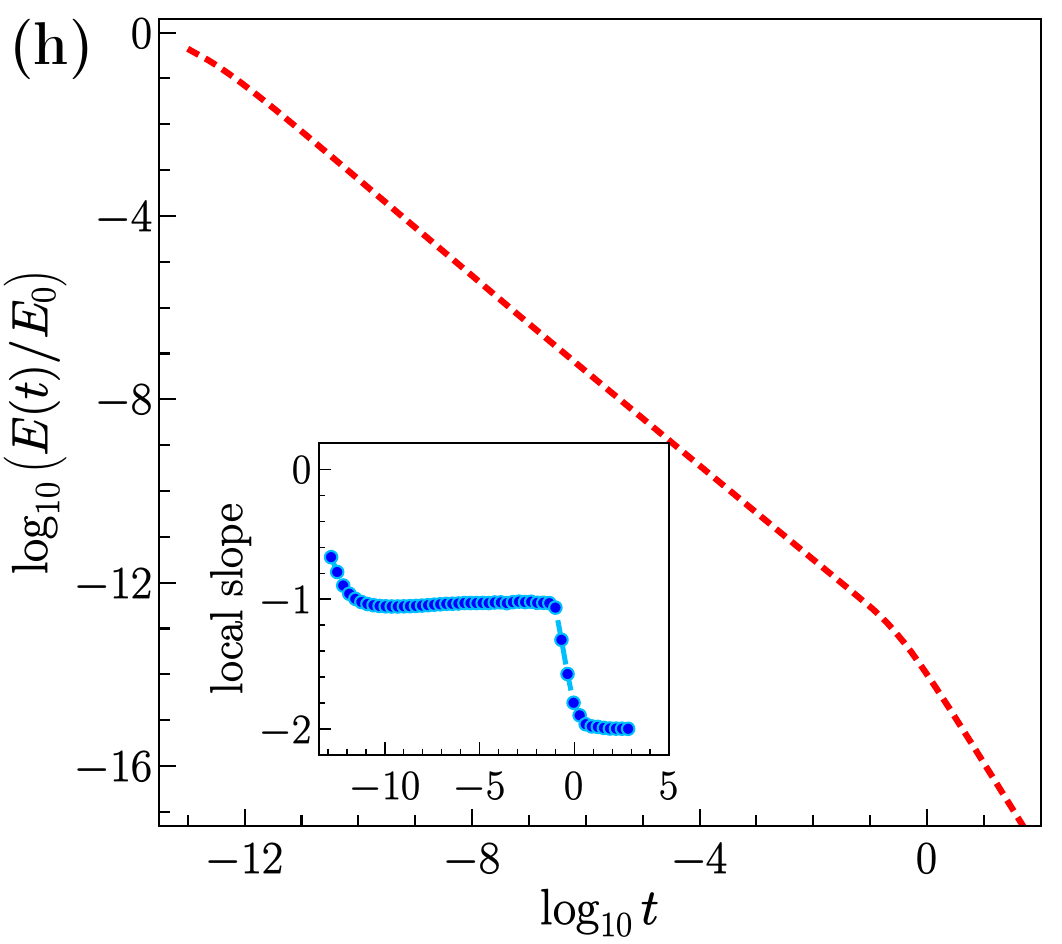}
        \includegraphics[width=0.333\textwidth]{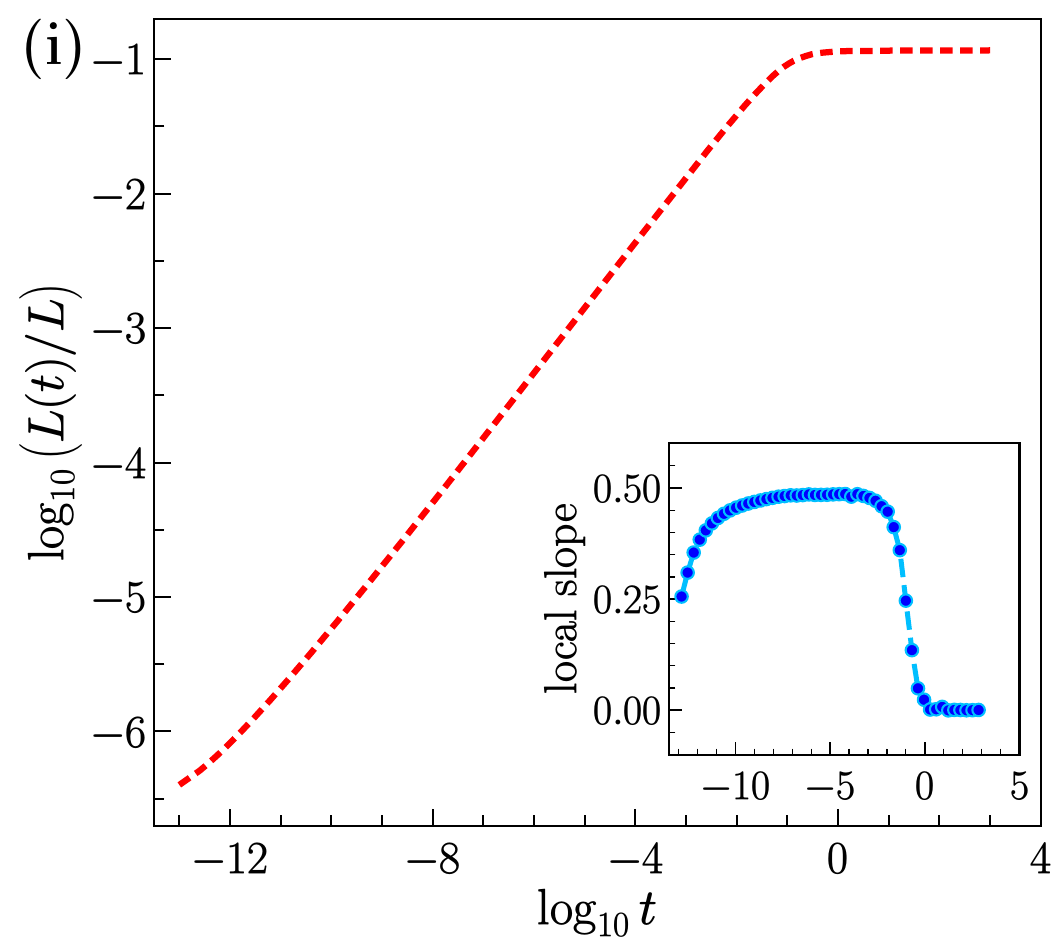}
    }
    \caption{Log-log plots for (a) the energy spectrum $E(k,t)$ versus the wavenumber $k$ at representative times $t$, (b) the decay of the total energy $E(t)$ with time $t$, and (c) the growth of integral length scale  $L(t)$ with time $t$ for case Ib, with a two-power-law initial energy spectrum [see Section~\ref{subsubsec:fourrange} where Fig.~\ref{fig:caseI_burg} gives the analogous plots for case Ia]; the insets show local slopes that can be used to estimate the decay and growth exponents in (b) and (c), respectively. (d)-(f) are the counterparts of (a)-(c) for case Ic;
    (g)-(i) are the counterparts of (a)-(c) for case Id.}
    \label{fig:caseII_extra}
\end{figure}
\bigskip\noindent

\textbf{\underline{Case \textbf{Ib}:} } As in Case Ia [Section~\ref{subsubsec:tworange}], there is exactly one peak at $k_{p}(t)$. Depending on $k_{p}(t)$, the spectrum has different behaviours as observed in Fig.~\ref{fig:caseII_extra}(a). When $k_{1} < k_{p}(t) < k^{2}_{1}$, $E(k,t)$ is the same as $E_{0}(k)$ for some $k'$ with $k< k' < k_{1}$; also, there is a spectral interval $k'' < k < k'''$ with $k'<k''$ and $k'''< k_{p}(t)$, where $E(k,t) \sim k^{2}$ with $E(k,t)> E_{0}(k)$. A continuous curve bridges these two spectral regions. This is reminiscent of the Gurbatov phenomenon
discussed in \cite{1997-gurbatov--toth}. When $k_{p}(t)<k_{1}$, i.e., in the low-wavenumber region, the peak height diminishes with time until the second spectral range, where the spectrum was proportional to $k^{2}$, vanishes. Eventually, the spectrum becomes such that $E(k,t) \leq E_{0}(k)$ for all $k$. Thus the memory of the initial spectral range with $n_{2} =1.5$ lingers in the system for a long time. 

The decay of $E(t)$ shows two main regimes. In our DNS, the first regime occurs approximately for $t \in [10^{-8}, 10^{-3}]$, whereas the second regime is seen for $ 10^{0} \lesssim t \lesssim 10^{3}$. Although we observe $E(t) \sim t^{-1} $ in the first regime, the power-law decay in the second regime is not very clear. Only towards the end of the second regime do we observe that $E(t)\sim t^{ - 0.9} $ (see the inset in Fig.~\ref{fig:caseII_extra}(b)). The exponents for the growth of $L(t)$ are not clear (see Fig.~\ref{fig:caseII_extra}(c)).

\bigskip
\noindent
\textbf{\underline{Case \textbf{Ic}:} } There is a single peak at $k_{p}(t)$ in the spectrum just as in the case Ib. Again the decay of $E(t)$ shows two power-law regimes. However, this case appears to be simpler than case IIb as we can see by comparing in the plots of $E(k,t)$ in Figs.~\ref{fig:caseII_extra}(a) and (d). When $k'< k_{p}(t) <k''$ the decay is exactly like the single-power-law case with $n = 0.5$. The exponent for the decay of $E(t)$ is $\simeq -0.9$. But, for $ k_{p}(t) < k' $, we observe the Gurbatov phenomenon, i.e., the behaviour resembles the single-power-law case 
with $n = 1.5$ [see Figs.~\ref{fig:caseII_extra}(e) and (f)].

\bigskip\noindent
\textbf{\underline{Case \textbf{Id}:} } The evolution of $E(k,t)$ exhibits the Gurbatov phenomenon [Fig.~\ref{fig:caseII_extra}(g)]. The decay forms for $E(t)$ and the growth of
for $L(t)$ are similar to their counterparts for a single-power initial spectrum $E_0(k) \sim k^n$ for $1<n<2$ [see \cite{1997-gurbatov--toth}]. The exponents for the decay of the energy decay and the growth of the integral length scale are roughly $-1.0$ and $0.5$ [Figs.~\ref{fig:caseII_extra}(h) and \ref{fig:caseII_extra}(i)], but they should have logarithmic corrections
[as discussed in \cite{1997-gurbatov--toth}]. 

\subsection{Initial data with energy spectra that have four power-law spectral ranges}
\label{subsec:app_burg4}

In Section~\ref{subsubsec:fourrange} we presented results for $E_0(k)$ with four power-law spectral ranges [see Eq.~\eqref{eq:four-power} in Section~\ref{subsubsec:fourrange} for case IIa]. The exponents $n_i$ in Eq.~\eqref{eq:four-power} are $n_1=n_3 = 1.5, n_2 = n_4 = -1.5$ in case IIb. We describe our main observations for this case below.

\begin{figure}
    \centerline{
        \includegraphics[width=0.32\textwidth]{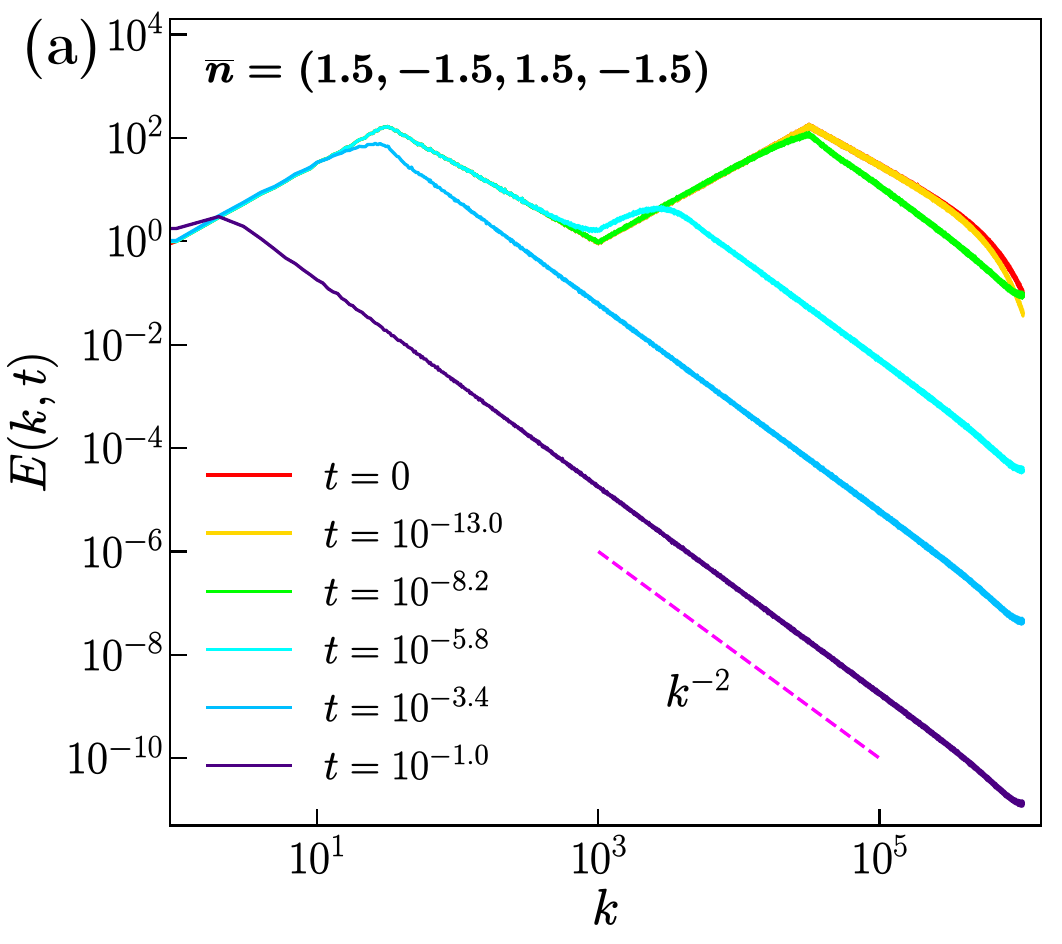}
        \includegraphics[width=0.32\textwidth]{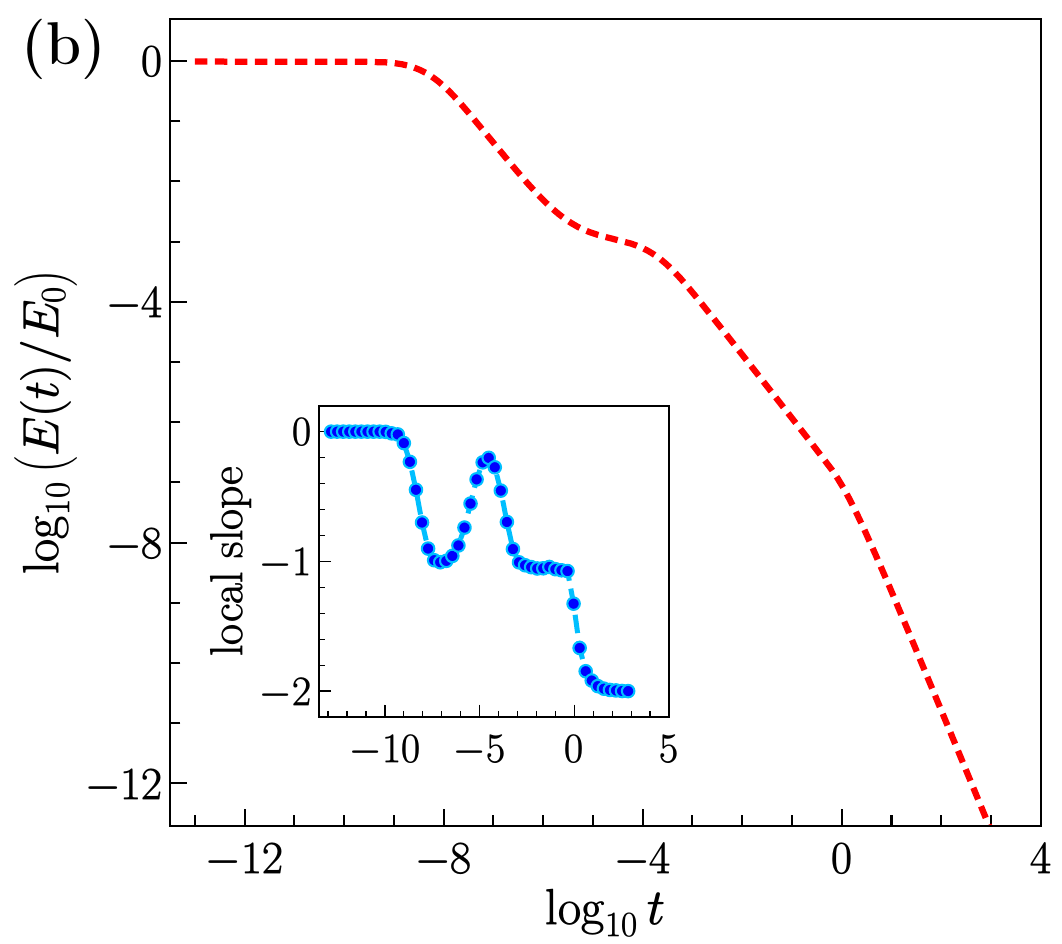}
        \includegraphics[width=0.32\textwidth]{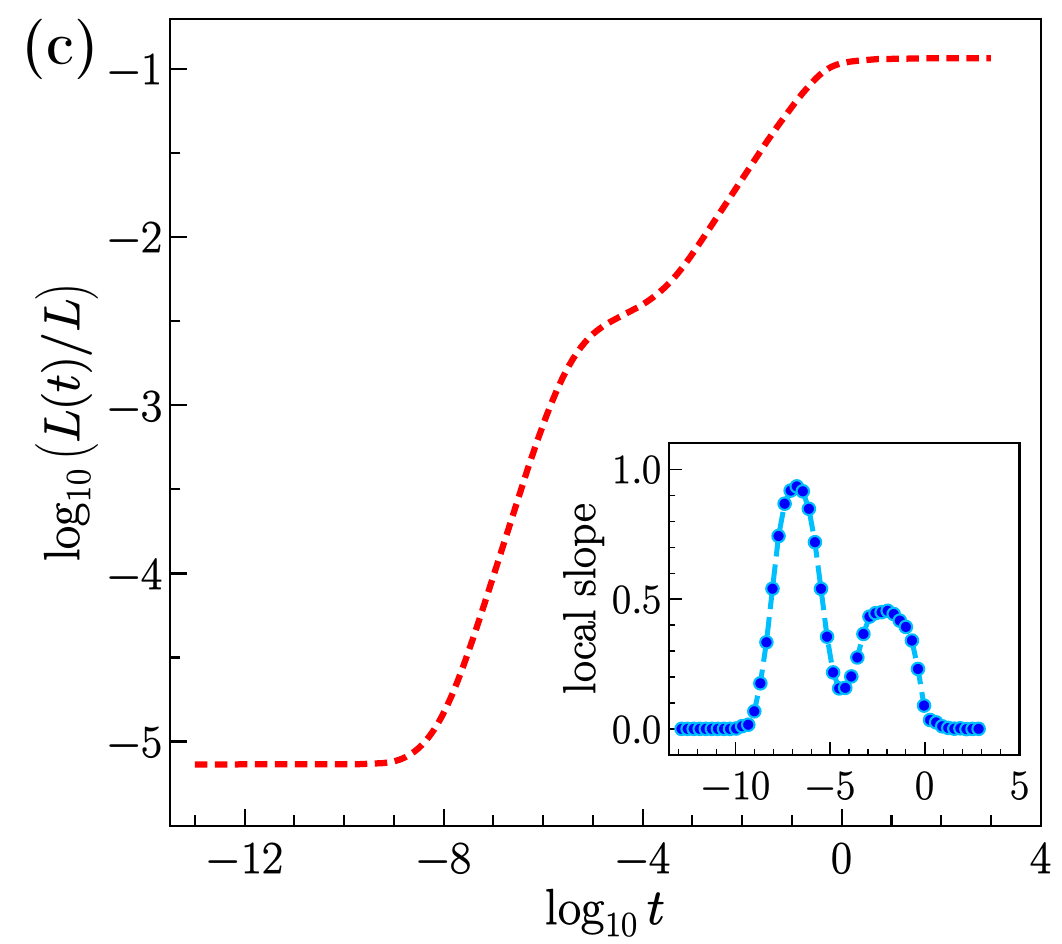}
    }
    \caption{Log-log plots for (a) the energy spectrum $E(k,t)$ versus the wavenumber $k$ at representative times $t$, (b) the decay of the total energy $E(t)$ with time $t$, and (c) the growth of integral length scale  $L(t)$ with time $t$ for case IIb, with a four-power-law initial energy spectrum [see Section~\ref{subsubsec:fourrange} and Fig.~\ref{fig:caseII_burg} for case IIa]. The insets show local slopes that can be used to  estimate the decay and growth exponents in (b) and (c), respectively.}
    \label{fig:caseIII_extra}
\end{figure}

We consider, at time $t$, the peak in the spectrum at $k_{p}(t)$. When $k_{p}(t)$ is in the first and the third spectral ranges [see Fig.~\ref{fig:caseIII_extra}(a)], we observe that some part of the spectrum rises above the initial spectrum. The local slope corresponding to the exponent for the energy decay is approximately $-1.0$ [Fig.~\ref{fig:caseIII_extra}(b)]. The exponent for the growth of the integral length scale remains constant at approximately $0.5$ only when $k_{p}(t)$ is in the first spectral range [Fig.~\ref{fig:caseIII_extra}(c)]. When the peak is in the third spectral range, the exponent rises close to the value $1$ [see the inset in Fig.\ref{fig:caseIII_extra}(c)]. We also observe a short period of slowing down of the energy decay and the growth of the integral-length scale-growth when $k_{p}(t)$ passes through the junction of the second and the third spectral ranges.

Thus, we have considered different types of initial spectra in cases I-II where the energy decay can have complicated dependences on time $t$ in different temporal ranges. By carrying out this study, we have gone beyond the results for simple spectral ranges, presented in the study by \cite{1997-gurbatov--toth}. For similar studies on the Navier-Stokes turbulence, the reader is referred to \cite{2012-meldi-sagaut}, which treats this decay using closure theory, and the numerical studies in Section~\ref{sec:ns-hyper-vis}.

\section{\label{sec:app_ns}}

\subsection{Navier--Stokes: Definitions of the integral quantities\label{sec:app_ns_int}}
\label{subsec:appD1}

For our Navier--Stokes studies in Section~\ref{sec:ns-hyper-vis} 
the total energy $E(t)$ and the energy spectrum $E(k, t)$ are related to the Fourier series for the velocity field as follows:
\begin{eqnarray}
\vec{u}(\vec{x}, t) &=& \sum_{\vec{k}} \hat{\vec{u}}(\vec{k}, t) \e^{\i \vec{k}\cdot\vec{x}}\,;\nonumber \\
 E(t) &=& \sum_{\vec{k}} \frac{1}{2}|\hat{\vec{u}}(\vec{k}, t)|^2 = \frac{1}{(2\pi)^3}\int \frac{1}{2} |\vec{u}(\vec{x}, t)|^2 d\vec{x}\,;
\label{defEtot} \\
 E(k, t) &=& \sum_{\substack{\vec{k} \\ k \le |\vec{k}| < k + \Delta k}} \frac{1}{2}|\hat{\vec{u}}(\vec{k}, t)|^2 \frac{1}{\Delta k}\,.
\label{defspc} 
\end{eqnarray}
The carets denote spatial Fourier transforms, we take $\Delta k = 1$, and we use the following standard definitions of the root-mean-square (rms) velocity, the integral length, and the large-scale turnover time:
\begin{equation}
 u_{\rm rms}(t) = \left(\frac{2 E(t)}{3}\right)^{1/2}\,;\;\; 
 L(t) = \frac{\pi}{2 u^2_{\rm rms}} \sum_{k = 1}^{k_{\rm max}} k^{-1} E(k, t) \Delta k\,;\;\;
 \tau(t) = \frac{L(t)}{u_{\rm rms}(t)}\,.
 \label{eq:urmsLT}
\end{equation}

Here $k_{\rm max}$ is the largest integer that lies below the truncation wavenumber $\sqrt{2}N/3$.

\subsection{Navier-Stokes: Decay of the two-power-law spectrum \label{sec:app_ns_decay}}
\label{subsec:appD2}

Studies of the decay of $E(t)$ in freely decaying 3DNS turbulence have a long history. For a summary see Sec. 1.1 lines 69-131 and \cite{panickacheril2022laws}. We summarise some more results below:

(A) If we use a single sharp peak in the initial spectrum $E_0(k)$, the decay of $E(t)$ is qualitatively similar to that in 1D Bugulence [see Appendix~\ref{subsec:app_burg2}]. In particular, $E(t)$ is expected to decay with a single power law (self-similar decay), but the measurements of the decay exponent are spread over a considerable range. Furthermore, the spectrum $E(k,t)$ develops a $k^4$ part at small $k$ and a $k^{-5/3}$ part at large $k$, the former because of the \cite{1954-proudman-reid} beating interaction and the latter because of the evolution towards a K41 spectrum. These regimes show up clearly in  our DNSs, especially for the hyperviscous NS equation, but they are not reported here because  we concentrates on intial data for which the initial spectrum $E_0(k)$ has power-law regions (see (B) and (C) below). Of course, there are corrections to the K41 spectrum because of small-scale multifractality [see, e.g., ~\cite{1995-frisch-book}, \cite{parisi1985multifractal}, and \cite{ray2008universality}].

(B) A single power law in $E_0(k) \sim k^n$, with $ 2 < n < 4$, cut off at large $k$: The decay of $E(t)$ is qualitatively similar to that in the 1D Burgulence: In particular, $E(t)$ is expected to decay with a single power law (self-similar decay), but the measurements of the decay exponent are spread over a considerable range [see, e.g., \cite{panickacheril2022laws,2012-meldi-sagaut}.] The cases $n=4$ and $n=2$ yield different powers as expected [see, e.g., \cite{panickacheril2022laws}]. We have discussed our results for the case $n=4$ in detail in Section~\ref{sec:ns-hyper-vis}]. Note also the study by \cite{biferale2003decay} of decaying anisotropic turbulence; their `` $\ldots$ initial conditions are taken from the stationary ensemble of a forced random Kolmogorov flow''.

(C) Two or more power-law regimes in $E_0(k)$, cut off at large $k$, as discussed in detail in Section~\ref{sec:ns-hyper-vis}. In short, the evolution of the energy spectra, the decay of $E(t)$, and the growth of $L(t)$ are qualitatively similar to their counterparts in the 1D Burgulence [see Section~\ref{subsubsec:tworange}]. 

We study two additional cases of energy decay in the hyperviscous 3DNSE of the type that we have studied in Section~\ref{subsubsec:2PLNS}.

\begin{figure}
    \centerline{
    \includegraphics[width=0.35\columnwidth]{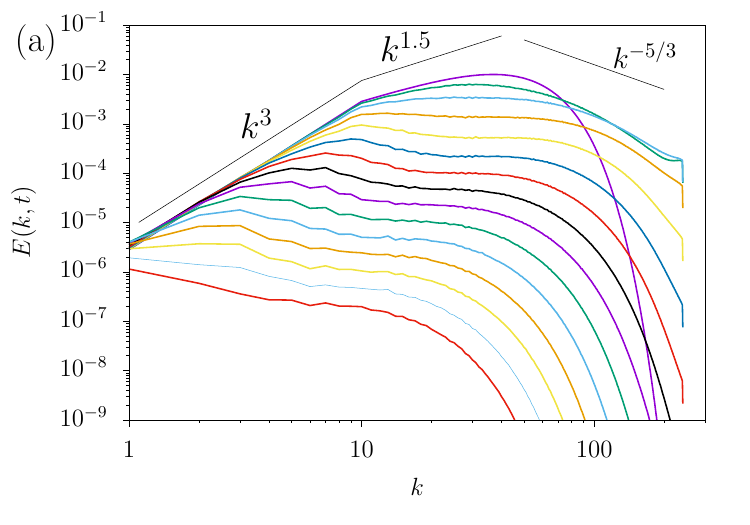} 
    \includegraphics[width=0.35\columnwidth]{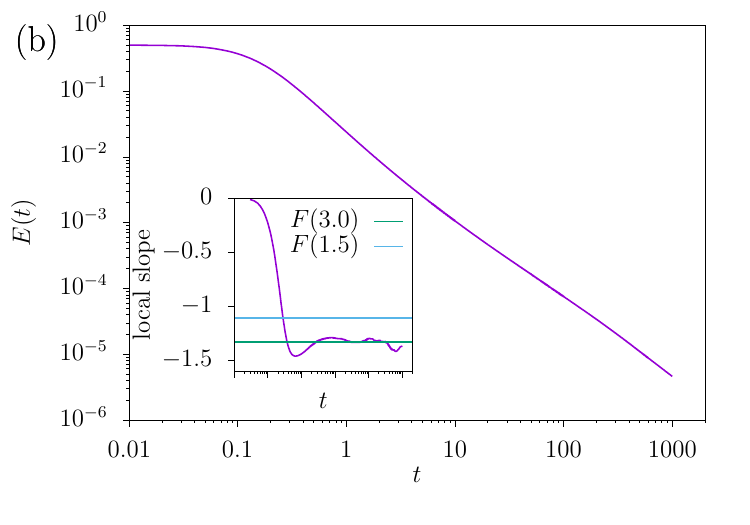}
    \includegraphics[width=0.35\columnwidth]{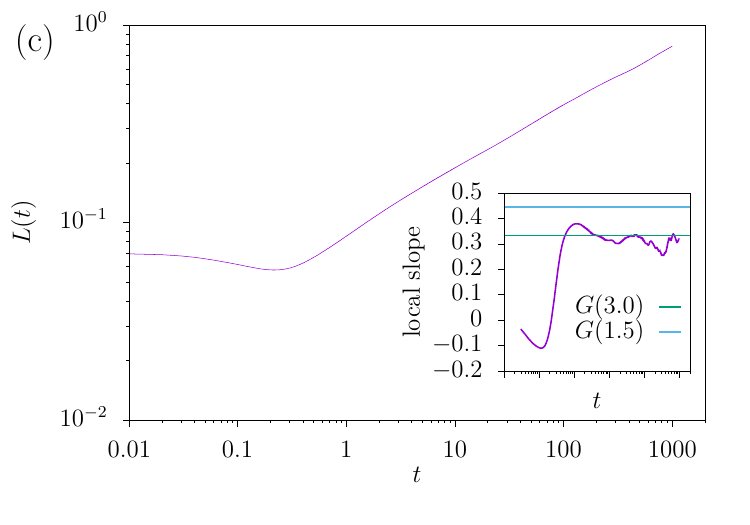}}
    \caption{The same as  the caption of Fig.~\ref{p3ov2_3hv}  in Section~\ref{subsubsec:2PLNS}, but with $(n_1, n_2) = (3, 1.5)$ in Eq.~\eqref{eq:twopow}.} 
    \label{p3_3ov2hv}
\end{figure}

We first consider the pair of spectral exponents $(n_1, n_2) = (3, 1.5)$ with
$k_1 = 60$ in Eq.~\eqref{eq:twopow} and $\nu_2 = 1.66 \times 10^{-8}$.
Our results are displayed in Fig.~\ref{p3_3ov2hv}, which is the counterpart of Fig.~\ref{p3ov2_3hv} in Section~\ref{subsubsec:2PLNS}. The logarithmic local slopes of $E(t)$ and $L(t)$, shown in the insets of Figs.~\ref{p3_3ov2hv} (b) and (c), support our qualitative conclusion in Fig.~\ref{p3ov2_3hv} of Section~\ref{subsubsec:2PLNS}: Energy decays is non-self-similar, but, at large times, the energy-decay exponent is determined by the $k^{n_1} = k^3$ part in the low-$k$ region of the energy spectrum. Here, we do not observe a decay exponent that is close to the na\"ive expectation for the $k^{n_2}$ part in the initial energy spectrum. 
With this pair of values $(n_1, n_2) = (3, 1.5)$, we do not observe a mismatch
of the time windows for the plateaux in the logarithmic local slopes of $E(t)$ and $L(t)$ [unlike what we found for the pair $(n_1, n_2) = (1.5, 3)$ in Fig.~\ref{p3ov2_3hv} of Section~\ref{subsubsec:2PLNS}].

\begin{figure}
    \centerline{%
     \includegraphics[width=0.35\columnwidth]{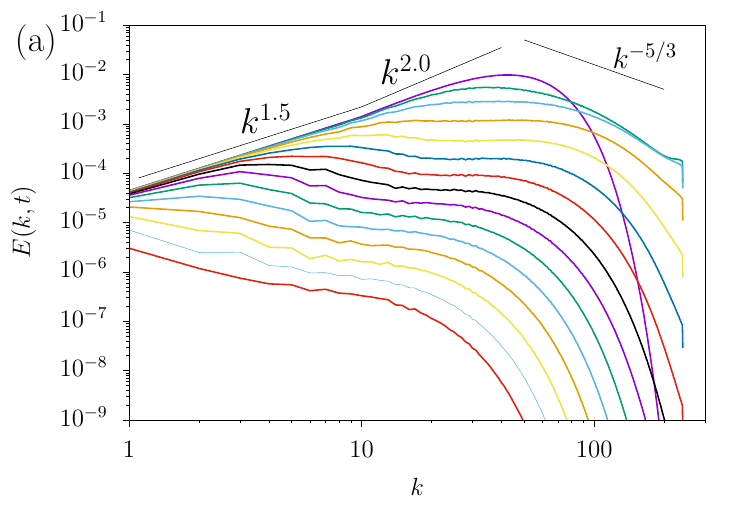} 
     \includegraphics[width=0.35\columnwidth]{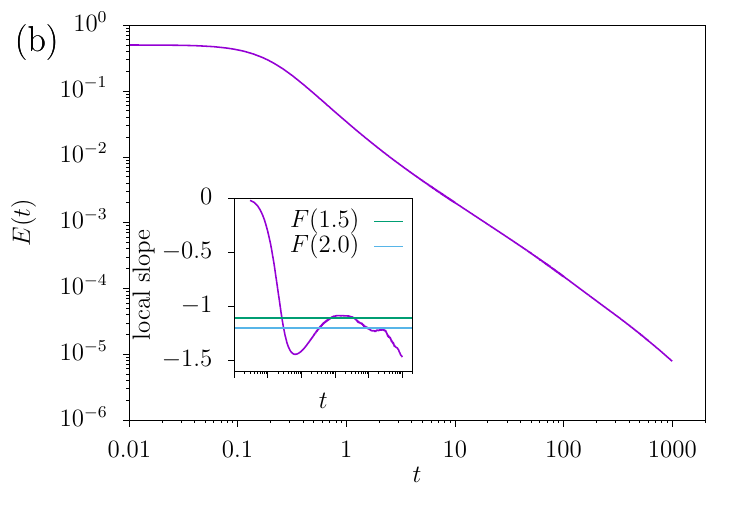}
     \includegraphics[width=0.35\columnwidth]{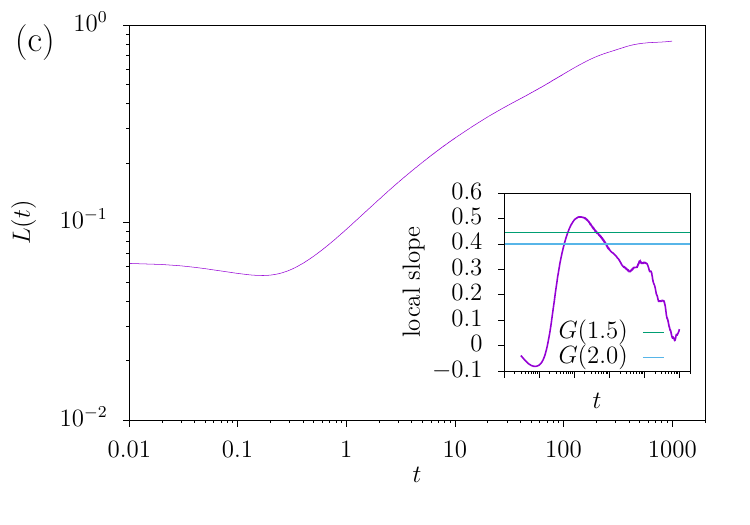}}
    \caption{The same as the caption of Fig.~\ref{p3ov2_3hv} in Section~\ref{subsubsec:2PLNS}, but with $(n_1, n_2) = (1.5, 2)$.} 
    \label{1p5_2hv}
\end{figure}

Finally we consider $(n_1, n_2) = (1.5, 2)$ with $k_1 = 60$ in Eq.\eqref{eq:twopow}, with $n_1$ quite close to $n_2$; we
set $\nu_2 = 1.32 \times 10^{-8}$. Our DNS results are depicted in  Fig.~\ref{1p5_2hv},  which is the counterpart of Fig.~\ref{p3ov2_3hv} in Section~\ref{subsubsec:2PLNS}. 
The logarithmic local slope of $E(t)$, shown  in the inset of in  Fig.~\ref{1p5_2hv} (b), is similar to its counterpart in Fig.~\ref{p3ov2_3hv} in Section~\ref{subsubsec:2PLNS}. 
$(n_1, n_2) = (1.5, 3)$ shown in Fig.~\ref{p3ov2_3hv} (b). It is interesting that the two small plateaux
agree with the na\"ive predictions for energy-decay exponents for initial energy spectra with $k^{1.5}$ and $k^{2}$ power-law forms. 
However, the local slope of the integral scale does not have a well-developed plateau, hence, it is at variance with
the na\"ive predictions for the growth exponents for $L(t)$. 
Therefore, with this pair, $(n_1, n_2) = (1.5, 2)$, the decay of $E(t)$ is non-self-similar; and it does not support the dominance of 
the $k^{n_1}$ part of the energy spectrum at large times. 

In summary, our hyperviscous DNS [$\upbeta = 2$] elucidates the subtle dependence of the decay of $E(t)$ on the powers $n_1$ and $n_2$ that characterise the initial energy spectrum
in Eq.\eqref{eq:twopow}.

\bibliographystyle{jfm}
\bibliography{all-refs}

\end{document}